\documentclass[12pt]{iopart}
\usepackage{iopams}  

\usepackage{xcolor}
\usepackage{graphicx}
\usepackage{tikz}
\usetikzlibrary{decorations.pathmorphing, patterns,shapes}
\usepackage{pgfplots}

\expandafter\let\csname equation*\endcsname\relax

\expandafter\let\csname endequation*\endcsname\relax

\usepackage{amsmath}
\usepackage{mathrsfs}
\usepackage[normalem]{ulem}

\usepackage[T1]{fontenc}

\usepackage{float}
\usepackage{tikz}
\usetikzlibrary{decorations.pathmorphing}
\usetikzlibrary{arrows.meta}

\usepackage{subcaption}

\newcommand*{\del}{\partial}

\newcommand*{\sigmab}{\bar\sigma}
\newcommand*{\mub}{\bar\mu}

\begin{document}
\title{Non-linear scattering of plane gravitational and electromagnetic waves}

\author{B Camden$^1$, C Stevens$^1$, J Forbes$^2$}

\address{${}^1$School of Mathematics and Statistics, University of Canterbury, Christchurch 8041, New Zealand}
\address{${}^2$School of Physical and Chemical Sciences, University of Canterbury, Christchurch 8041, New Zealand}

\ead{breanna.camden@pg.canterbury.ac.nz, john.forbes@canterbury.ac.nz, chris.stevens@canterbury.ac.nz}

\begin{abstract}
Fully non-linear, plane-symmetric exact solutions of the Einstein equations describing the scattering of gravitational and electromagnetic waves have existed for many years. For these closed-form solutions to be found, idealized wave profiles such as the Dirac delta and Heaviside theta functions must be assumed. Although pathological in that future curvature singularities generically occur, these exact solutions give useful insights into the non-linear features of the scattering process. Only a limited number of exact solutions exist and this leaves many other physically-motivated scattering situations without a non-linear description. The aim of this paper is to shed light on these unexplored cases. This is achieved through numerical solutions of the Friedrich-Nagy initial boundary value problem for the Einstein equations coupled to the source-free Maxwell equations in plane symmetry. Interesting non-linear scattering effects are presented for a variety of electromagnetic and gravitational wave profiles not currently described by an exact solution. Implications for electromagnetic wave observations are investigated through analyzing the time-delay and frequency shift imparted on the electromagnetic wave through the non-linear scattering with a gravitational wave of a given strain. Simple arguments supported by the numerical solutions suggest dramatic effects on the radiation may be observable.
\end{abstract}
\maketitle

\section{Introduction}
The Friedrich-Nagy system \cite{friedrich1999initial}, a well-posed Initial Boundary Value Problem (IBVP) formulation of the Einstein equations, is ideal for studying non-linear gravitational wave interactions. The constraint-preserving boundary conditions, coupled with the ingoing and outgoing gravitational waves being system variables, allows for arbitrary wave profiles to be imposed directly and easily. Further, if the space-time between the waves is an exact solution of the Einstein equations before collision, which is commonly the case, then one does not need to solve constraints to impose initial data. One can therefore experiment with arbitrary wave profiles by simply changing the boundary data.

Despite being very well suited to the study of vacuum space-times with arbitrary cosmological constant where one wants precise control over the gravitational radiation, the Friedrich-Nagy formulation has had very little attention from the numerical relativity community. In fact, the only numerical implementation in the literature is \cite{frauendiener2014numerical,frauendiener2021can}, where the authors successfully implemented the formulation for the first time, with the simplifying assumption of plane symmetry. The numerical viability of the system was confirmed for the cases of colliding gravitational plane waves in vacuum with a vanishing and a positive cosmological constant respectively.

Fully non-linear, exact solutions of the Einstein field equations describing plane gravitational waves have been studied extensively over the years. Of most interest to the present work in vacuum, the Khan-Penrose \cite{khan1971scattering} and Nutku-Halil \cite{nutku1977colliding} solutions describe the scattering of two impulsive plane gravitational waves with colinear and non-colinear polarisation respectively. In electrovacuum, Griffith's solution \cite{griffiths1975colliding} describing the scattering of an impulsive plane gravitational wave and a shock plane electromagnetic wave, and the Bell-Szekeres solution \cite{bell1974interacting} describing the scattering of two shock plane electromagnetic waves. One can see a common theme here: there is a head-on collision of two plane waves whose wave profiles are always either impulsive, namely a Dirac delta function, or shock, namely a Heaviside theta function. These specific wave profiles allow analytical methods to be employed to obtain closed-form solutions. Generalizing these solutions to include a non-vanishing cosmological constant or more general wave profiles seem to be outside the tractability of the analytical methods previously employed successfully. On the other hand, it is completely trivial for the numerical implementation of the Friedrich-Nagy formulation \cite{frauendiener2014numerical,frauendiener2021can} to alter the cosmological constant or wave profiles. This makes it an excellent choice as a means to continue the study of fully non-linear scattering of gravitational and electromagnetic waves beyond current analytical capabilities.

The non-linear scattering of gravitational and electromagnetic waves is largely unexplored outside of plane-symmetric exact solutions. This is in part due to the popularity of formulations stemming from the ADM formulation, like BSSN, which are the stock-standard systems for simulating compact binary mergers. Although being extremely successful handling these situations, they are not well suited to scattering problems. This is because they have a complicated characteristic structure whereby the characteristic analysis of a timelike boundary is difficult and they do not represent gravitational waves explicitly. These features make the construction of a wellposed IBVP framework very difficult, let alone in such a way that is easily amenable to scattering experiments. 

As such, there are many interesting open questions regarding the non-linear scattering process: What happens in situations not described by exact solutions? How do the wave polarisations and profiles affect the scattering? When do the non-linear effects become significant? Are there any interesting features that appear in the non-linear regime? Although non-linear scattering is interesting in and of itself, there are also interesting questions about how non-linear scattering could affect electromagnetic wave observations. Does non-linear scattering induce a non-trivial time delay on the electromagnetic wave? A frequency shift? Does the polarisation change? What is the strain of the gravitational wave required to induce such features? What happens if we use realistic frequencies? Although our current formulation is a plane-symmetric toy model, the non-linear characteristics of the scattering process we present should underscore significant non-linear aspects of a realistic scattering process. Furthermore, it should illuminate key features that are crucial in inducing changes in observed electromagnetic wave signals, that could be further pursued in more general cases. There are a rich array of astrophysical scenarios in which electromagnetic waves will encounter strong gravitational waves, so there is an opportunity to begin to make predictions for the phenomenology of current and future observations.


The paper follows the same conventions as \cite{penrose1984spinors} throughout and commonly uses m$^{-1}$ to measure frequency in relativistic units with $c=G=1$. The usual Hz, which has units of inverse seconds $s^{-1}$, can be found by multiplying the frequency written in $m^{-1}$ units by $c$. When discussing the polarisation of a wave, we take this to be the real number $\alpha\in[0,1)$ in the overall phase $e^{2\pi\alpha i}$ of the wave profile. The layout is as follows: Sec.~\ref{sec:background} outlines two exact solutions: the collision of two impulsive plane gravitational waves with non-aligned polarisation and the collision of one impulsive plane gravitational wave and one shock plane electromagnetic wave with aligned polarisation. The equations are presented and generalize those of \cite{frauendiener2021can} to include a coupling to the Maxwell equations. These two cases are used to verify our numerical implementation before proceeding to more general situations. Sec.~\ref{sec:numericalimplementation} details the numerical implementation of the electovacuum equations and performs convergence tests to the two exact solutions detailed in the previous section to confirm correctness. Sec.~\ref{sec:nonlinearbehaviour} presents non-linear scattering results outside the scope of exact solutions. Sec.~\ref{sec:astro} details physical quantities that appear in the scattering scenario, derives expressions for them and explored how non-linear scattering can affect properties of the electromagnetic wave. Sec.~\ref{sec:summary} summarizes and discusses our results and lays out future work.

\section{Background and equations}\label{sec:background}
\subsection{Review of the Nutku-Halil solution}
The Nutku-Halil solution is an analytic solution of the full Einstein vacuum field equations with vanishing cosmological constant \cite{nutku1977colliding} and exemplifies several generic features of non-linear plane wave collisions. It describes the head-on collision and subsequent scattering of two impulsive plane gravitational waves, each with arbitrary polarisation, generalizing the colinear case of Khan and Penrose \cite{khan1971scattering}. The metric is represented in double null coordinates $\{u,v,x,y\}$, where $u,v$ are null and $x,y$ describe the planes of symmetry. The line element for this solution in these coordinates is
\begin{equation}\label{eq:NHLineElement}
	\text{d}s^2 = 2\frac{1-E\bar{E}}{t r w}\text{d}u\text{d}v
	- \frac{t^2}{1 - E\bar{E}}[(1-E)\text{d}x + i(1+E)\text{d}y]
	[(1-\bar{E})\text{d}x - i(1+\bar{E})\text{d}y],
\end{equation}
where 
\begin{gather}
	p := u\theta(u), \qquad
	q := v\theta(v), \qquad
	r := \sqrt{1 - p^2}, \qquad
	w := \sqrt{1 - q^2}, \nonumber \\
	t := \sqrt{1 - p^2 - q^2}, \qquad
	E := e^{-2\pi i\alpha}pw + e^{2\pi i\beta}qr,
\end{gather}
$\theta$ is the Heaviside theta function and $\alpha,\,\beta\in[0,1)$ determine the polarisations of the waves $\Psi_0$ and $\Psi_4$ respectively.

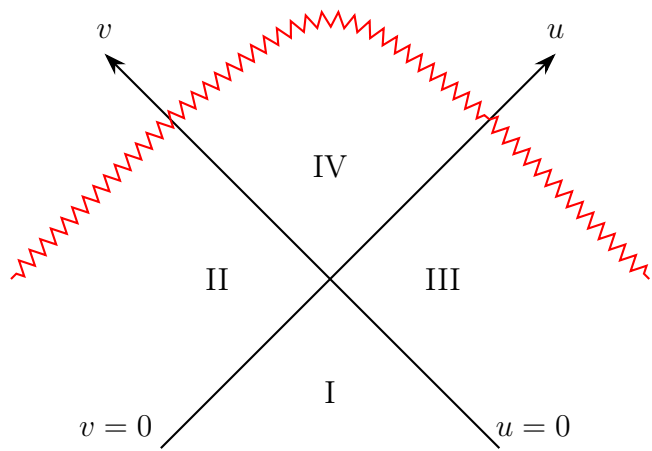
\begin{figure}[H]
	\centering
	\begin{tikzpicture}[scale=3]
		\draw[thick, -{Stealth[length=3mm, width=2mm]}] (-0.75,-0.75) -- (1,1);
		\draw[thick, -{Stealth[length=3mm, width=2mm]}] (0.75,-0.75) -- (-1,1);

		\draw[thick, red, decoration = {zigzag, segment length = 2mm, amplitude = 1mm}, decorate] (-0.7071,0.7071) -- (-1.4142,0);
		\draw[thick, red, decoration = {zigzag, segment length = 2mm, amplitude = 1mm}, decorate] (0.7071,0.7071) -- (1.4142,0);
		\draw[thick, red, decoration = {zigzag, segment length = 2mm, amplitude = 1mm}, decorate] (-0.7071,0.7071) .. controls (0,1.3) ..  (0.7071,0.7071);

		\node at (0, -0.5) {I};
		\node at (-0.5, 0) {II};
		\node at (0.5, 0) {III};
		\node at (0, 0.5) {IV};

		\node at (1, 1.1) {$u$};
		\node at (-1, 1.1) {$v$};

		\node at (0.9, -0.65) {$u=0$};
		\node at (-0.95, -0.65) {$v=0$};

	\end{tikzpicture}
	\caption{The four regions of the Nutku-Halil space-time and the singularity structure.}
	\label{fig:NHDiagram}
\end{figure}

Fig. \ref{fig:NHDiagram} presents how the Nutku-Halil space-time can be broken up into four distinct regions: the initial region I given by $u,v<0$ is described by Minkowski space-time, the two regions II and III containing one gravitational wave given by $u<0, v\geq0$ and $u\geq0,v<0$ respectively and the scattering region IV given by $u,v\geq0$. A prominent feature of exact colliding plane gravitational wave space-times is the inevitable future curvature singularity in the scattering region. For the current representation of the Nutku-Halil space-time this occurs on the surface $u^2 + v^2 = 1$. There also exist so-called \emph{fold} singularities \cite{griffiths2016colliding} that appear along $v=1$ in region II and $u=1$ in region III.
Quantities of special interest derived from this space-time are the non-vanishing Weyl scalars $\Psi_0,\,\Psi_2$ and $\Psi_4$. We present them here written in terms of a null tetrad different from \cite{chandrasekhar1984nutku} so that the wave profiles in regions II and III are $\Psi_4 = e^{2\pi i \beta}\delta(v)$ and $\Psi_0 = e^{2\pi i \alpha}\delta(u)$. This then allows a direct relationship between the wave profiles and the polarization constants $\alpha, \beta$ that we will exploit later. This correspondence is not possible when using the null tetrad of Chandrasekhar \cite{chandrasekhar1984nutku}, however, Chandrasekhar's frame exemplifies that the space-time can be written solely in terms of the relative polarization $\alpha - \beta$.

Our new tetrad is
\begin{equation}
\textbf{l} = \frac{f}{w\sqrt{t}}\text{d}v, \qquad
\textbf{n} = \frac{f}{r\sqrt{t}}\text{d}u, \qquad
\textbf{m} = \frac{(\bar{E} - 1)\text{d}x + i(1 + \bar{E})\text{d}y}{\sqrt{2}f},
\end{equation}
where 
\begin{equation}
	f := \sqrt{t^2 + 2pq\Big{(}pq - \cos[2\pi(\alpha + \beta)]rw\Big{)}}.
\end{equation}
For comparison to the numerical results of subsequent sections, we only present the Weyl scalars in the scattering region $u, v > 0$ along $u=v$. These are
\begin{align}
	\Psi_0(u,u)
	&= \frac{3u\sqrt{1 - u^2}}
	{\sqrt{1 - 2u^2}(1 - (1 + e^{2\pi i(\alpha + \beta)})u^2)^3}e^{2\pi i(2\alpha + \beta)}, \\
	\Psi_2(u,u)
	& = \frac{1 - 3(u^2 - u^4) - (1 + e^{2\pi i (\alpha + \beta)})u^6}
	{(1 - 2u^2)^{3/2}(1 - (1 + e^{2\pi i(\alpha + \beta)}u^2))^3}e^{2\pi i (\alpha + \beta)}, \\
	\Psi_4(u,u)
	&= \frac{3u\sqrt{1 - u^2}}
	{\sqrt{1 - 2u^2}(1 - (1 + e^{2\pi i(\alpha + \beta)})u^2)^3}e^{2\pi i(\alpha + 2\beta)}, \\
	\Psi_1(u,v) &= \Psi_3(u,v) = 0.
\end{align}
The Weyl scalar curvature invariants $I_1$ and $I_2$ \cite{griffiths2016colliding,penrose1984spinors}, which we define as
\begin{align}
	I_1 + i I_2 
	&:= 3\Psi_2^2 + \Psi_0\Psi_4,
\end{align}
are thus easily obtained from these expressions. As these scalars are built from curvature quantities in a frame-independent way, their divergence is a useful way to detect the presence of a physical curvature singularity.

\subsection{Review of Griffith's 1975 solution}
Griffith's 1975 solution is an analytic solution of the full Einstein vacuum field equations coupled to an electromagnetic field with vanishing cosmological constant \cite{griffiths1975colliding}. It describes the head-on collision and subsequent scattering of a plane electromagnetic shock wave and a plane impulsive gravitational wave. The metric is represented in double null coordinates $\{u,v,x,y\}$, where $u,v$ are null and $x,y$ describe the planes of symmetry. The line element describing this solution can be written in standard Szekeres form \cite{griffiths2016colliding} as
\begin{equation}\label{eq:BSLineElement}
	\text{d}s^2 = 2e^{-M}\text{d}u\text{d}v
	- e^{-U}\Big{(}e^V\text{cosh}W\text{d}x^2
    -2e^{-U}\text{sinh}W 
    + e^{-V}\text{cosh}W\text{d}y^2\Big{)},
\end{equation}
where\footnote{Note that $b$ here is $\sqrt{2}$ times the original $b$ used in \cite{griffiths1975colliding} to match the conventions we use from \cite{penrose1984spinors}.}
\begin{gather}
    e^{-U} = \cos^2\sqrt{2}bv - a^2u^2, \qquad
    e^V = \frac{1- au}{1+au}, \\
    e^{-M} = \frac{\cos\sqrt{2}bv\sqrt{1-a^2u^2}}{\sqrt{\cos^2\sqrt{2}bv - a^2u^2}}, \qquad
    W = 0.
\end{gather}
The singularity structure is similar to that described in Fig.~\ref{fig:NHDiagram} but with the curvature singularity in region IV being given by $a^2u^2 = \cos^2\sqrt{2}bv$.

The Weyl scalar curvature invariant $I_1$ is computed to be
\begin{equation}
    I_1 = \frac{48a^4b^2u^2\sin^2\sqrt{2}bv}{(1-a^2u^2)(1-2a^2u^2+\cos2\sqrt{2}bv)^3}.
\end{equation}

\subsection{The equations}
As the Friedrich-Nagy gauge in plane symmetry has already been derived and implemented numerically in \cite{frauendiener2014numerical,frauendiener2021can}, here we summarize only the relevant parts for the present study and refer the reader to these references for more details.

The line element of a space-time containing two commuting spacelike Killing vectors can be put into the form
\begin{equation}\label{eq:PWLineElement}
	\text{d}s^2 = \text{d}t^2 - 2\frac{B}{A}\text{d}t\text{d}z - \frac{1-B^2}{A^2}\text{d}z^2
	+ \frac{2}{(\xi \bar\eta - \bar\xi \eta)^2}(\eta\text{d}x - \xi\text{d}y)
	(\bar\eta\text{d}x - \bar\xi\text{d}y),
\end{equation}
where the metric functions $A, B, \xi$ and $\eta$ are functions of the coordinates $(t,z)$ chosen to describe the temporal and spatial directions orthogonal to the planes spanned by the Killing vectors. This metric yields a null tetrad
\begin{align}
	l^a\nabla_a &= \frac{1}{\sqrt{2}}\Big{(}(1 + B)\partial_t + A\partial_z\Big{)}, \\
	n^a\nabla_a &= \frac{1}{\sqrt{2}}\Big{(}(1 - B)\partial_t - A\partial_z\Big{)}, \\
	m^a\nabla_a &= \xi\partial_x + \eta\partial_y.
\end{align}
The Friedrich-Nagy system originates from the commutator, spin-coefficient and Bianchi equations \cite{penrose1984spinors}, which give equations for the derivatives of the metric, connection and curvature respectively. In order to incorporate electromagnetic waves into this system, we supplement these equations with the source-free Maxwell equations. Taking the energy momentum tensor in the Einstein field equations to be the Maxwell energy tensor, and considering the symmetric and skew components respectively, we obtain the following relationship\footnote{We neglect to explicitly represent the Infeld-van der Waerden symbols $\sigma^a_{AA'}$ in transforming tensor to spinor indices, and vice-versa.} between $\Phi_{ab}$ (proportional to the trace-free Ricci tensor) and the components of the symmetric Maxwell spinor $\phi_{AB}$
\begin{equation}
    \Phi_{AA'BB'} \equiv \Phi_{ab} = 2\varphi_{AB}\bar{\varphi}_{A'B'}.\label{eq:PhiDef}
\end{equation}

\color{black} In vacuum and with vanishing cosmological constant, the resulting evolution equations are
\begin{subequations}\label{eq:EvEqs}
	\begin{align}
		\sqrt2 \del_t A &= (\mu + \bar\mu)\,A, \label{ee:1}\\
		\sqrt2 \del_t B &= (2\rho - 2\rho' + F + \bar F) + (\mu + \bar\mu) B,\label{ee:2}\\
		\sqrt2 \del_t \rho &= 3\rho^2 + \sigma \sigmab + \rho(F + \bar F) + \Phi_{00} - \Phi_{11}, \label{ee:3}\\
		\sqrt2 \del_t \rho' &= 3\rho^{\prime2}  + \sigma' \sigmab' - \rho'(F + \bar F) - \Phi_{11} + \Phi_{22} \color{black},\label{ee:4}\\
		\sqrt2 \del_t \sigma &= 4\rho\sigma - \rho'\sigma + \rho\sigmab' + \sigma(3F - \bar F) - \Phi_{02}+ \Psi_0,\label{ee:5}\\
		\sqrt2 \del_t \sigma' &= 4\rho'\sigma' - \rho\sigma' +
		\rho'\sigmab - \sigma'(3 F - \bar F) - \Phi_{20} +
		\Psi_4,\label{ee:6}\\
		\sqrt2 \del_t \mu &= \mu^2 + \mu\bar \mu - 3 (\rho - \rho')^2 +
		(\mu + \bar \mu) (\rho + \rho') - \sigma \bar\sigma -
		\sigma'\bar\sigma'\nonumber\\ 
		& + 2 \sigma\sigma' - (\rho - \rho')(\bar F + 3F) - F^2 - F \bar F - \sqrt2 A\del_z F\nonumber\\
        &  - \sqrt2 B\del_t F - \Phi_{00} + 2\Phi_{11} - \Phi_{22}, \label{ee:7}\\[5pt]
		&\hspace{-3.9cm}(1-B) \del_t \Psi_0 - A \del_z \Psi_0 = \sqrt2 \left((2\rho - \rho' + 2 F + 2\mu)\Psi_0 + 3\sigma\Psi_2 \right)\nonumber\\
        & - \sqrt2 (\rho + F - \bar F - \mu - \mub))\Phi_{02}\nonumber\\
        & + \sqrt2(2 \sigma \Phi_{11} + \sigmab' \Phi_{00}),\label{ee:8}\\
		&\hspace{-3.9cm}(1+B) \del_t \Psi_4 + A \del_z \Psi_4 = \sqrt2 \left((2\rho' - \rho - 2 F + 2\mu)\Psi_4 + 
		3\sigma'\Psi_2\right)\nonumber\\
        & - \sqrt2(\rho' - F + \bar F - \mu + \mub))\Phi_{20}\nonumber\\
        & + \sqrt2(2 \sigma'\Phi_{11} + \sigmab \Phi_{22}) ,\label{ee:9}\\[5pt]
         (1-B)\del _t\varphi_0 - A\del _z\varphi_0 & = \sqrt2((\rho + F + \mu)\varphi_0 + \sigma\varphi_2) ,\label{ee:10}\\
        \del _t\varphi_1 & = \sqrt2(\rho + \rho')\varphi_1 ,\label{ee:11}\\
        (B + 1)\del _t\varphi_2 + A\del _z\varphi_2 &  = \sqrt2((\rho' - F + \mu)\varphi_2 + \sigma'\varphi_0),\label{ee:12}
	\end{align}
        
\end{subequations}
while the constraints take the form
\begin{subequations}\label{eq:ConstrEqs}
	\begin{align}
	0=C_1 &:= \sqrt2 A\del_z\rho - (1 - 3 B) \rho^2 - (1 - B) \sigma\sigmab + \rho (\mu + \bar\mu + 2\rho') \nonumber \\
	&\quad  + \rho B (F + \bar F) -(1-B)\Phi_{00} - (1+B)\Phi_{11},\label{ce:1}\\[4pt]
	0=C_2 &:= \sqrt2 A\del_z\rho' + (1 + 3 B) {\rho'}^2 + (1 + B) \sigma'\sigmab' - \rho' ( \mu + \bar\mu + 2\rho) \nonumber \\
	&\quad - \rho' B (F+\bar F) + (1-B)\Phi_{11} + (1+B)\Phi_{22},\label{ce:2}\\[4pt]
	0=C_3&:=\sqrt2 A\del_z\sigma + (1+B) \rho\sigmab' - 2 (1-2B)\rho\sigma + (1-B) \rho'\sigma  \nonumber\\ 
        &\quad + \sigma(3\mu - \bar\mu) + B\sigma(3F - \bar F) - (1-B) \Psi_0  - (1+B)\Phi_{02},\label{ce:3}\\
	0=C_4&:= \sqrt2 A\del_z\sigma' - (1-B) \rho'\sigmab + 2
	(1+2B)\rho'\sigma' - (1+B) \rho\sigma' \nonumber\\ 
        &\quad - \sigma'(3\mu - \bar\mu) - B\sigma'(3F - \bar F) + (1+B)
	\Psi_4 + (1-B)\Phi_{20}, \label{ce:4}\\
         0=C_5 & := \sqrt2 A\partial_z\varphi_1 +2 ((B+1)\rho ' + (B-1)\rho )\varphi_1, \label{ce:5}
\end{align}
\end{subequations}
for system variables $A,B,\mu,\rho,\rho',\sigma,\sigma',\Psi_0,\Psi_4,\phi_0,\phi_1,\phi_2$ and gauge source function $F$. It was shown in \cite{friedrich1999initial} that (in vacuum) and without any symmetry assumptions that these equations constitute a wellposed IBVP. We have a symmetric hyperbolic evolution system and the constraints propagate with maximally dissipative boundary conditions. In plane symmetry but now coupled to the electromagnetic field equations, it is straightforward to show that the constraints above still propagate, see \ref{app:CP} for the subsidiary system.

It was shown in \cite{frauendiener2021can} that the choice $F = \rho' - \rho$ leads to $\del_t B=0$ and a zero acceleration along $\partial^a_t$, which corresponds to the Gau\ss\;gauge. Although this is notorious for developing caustics, it has the property of having spatially constant curves correspond to free-falling observers. This will be useful when we calculate the strain in Sec.~\ref{sec:physicalparams}. It also makes $\partial^a_t$ and $\partial^a_z$ orthogonal by killing the $\text{d}t\text{d}z$ term in the line element, which makes comparisons to exact solutions easier. In any case, as colliding plane gravitational wave space-times inevitably contain a future curvature singularity, this gauge has been found to not cause any issues before this singularity is reached and the simulation stops.

The only partial differential equations appearing are first order advection equations for $\Psi_0$, $\varphi_0$, $\Psi_4$ and $\varphi_2$, with the characteristic speeds of the first two being $-A/(1-B)$ and the latter two $A/(1+B)$.  These reduce to $\pm A$ with our gauge choice.

As we choose $\phi_1=0$ initially, it will remain zero for all time and the constraint $C_5$ is identically satisfied.

We finally supplement our system with equations for null coordinates defined through $l^a\nabla_a u = 0 = n^a\nabla_a v$. These yield advection equations to which boundary conditions are imposed that reproduce standard Minkowski space-time null coordinates when we do not have any waves and use Minkowski space-time initial and boundary data (see \cite{frauendiener2021can} for a detailed discussion). Then defining $T:=\sqrt{2}(v+u)$ and $Z:=\sqrt{2}(v-u)$ we can produce conformal diagrams that will be useful for visualising the results.

\section{Numerical implementation}\label{sec:numericalimplementation}
Due to all functions depending only on the $t,z$-coordinates, we have a $1+1$ system. The system is evolved using the method of lines. We discretize the $z$-direction into equi-distant points with step size $\Delta z$ on an interval $[z_l,z_r]$ with $z_l < z_r$. The temporal direction is discretized into equi-distant points by marching in time with a timestep of $\Delta t = 0.2\Delta z$. The system is implemented in the Python package COFFEE \cite{doulis2019coffee} which contains all the numerical methods required for evolving our system. Spatial derivatives are approximated with Strand's fourth-order (third order near the boundary) summation-by-parts finite-difference operator \cite{strand1994summation}. A stable imposition of boundary conditions is accomplished with the simultaneous approximation term method \cite{carpenter1994time} where we take $\tau=1$ throughout. We march in time with the standard explicit Runge-Kutta 4 algorithm.

Initial data is chosen to be that of Minkowski space-time. It is noted that in our gauge with $B=0$ the initial choice of $A$ is equivalent to a rescaling of the $z$-direction.

\subsection{Emulation of exact solutions}\label{sec:emulationofexactsolns}
In order to validate the correctness of our numerical implementation we reproduce the Weyl scalar curvature invariants $I_1$ and $I_2$ of the Nutku-Halil solution and Griffith's 1975 solution. This has the benefit of not needing to relate frames in the scattering region to obtain a comparison. 

An obvious issue with reproducing these exact solutions is that impulsive wave profiles are given by the Dirac delta function and shock wave profiles are given by the Heaviside step function. To deal with these numerically, we define the functions
\begin{equation}\label{eq:deltaprofileapprox}
    \delta_N(a,t) := 
    \begin{cases} 
	    a\sin^8(bt) & 0\leq t\leq\frac{\pi}{b} \\
	    0 & \text{otherwise}
	\end{cases},
\end{equation}
where $b := 35\pi a/128$, and
\begin{equation}
    \Theta_N(h,t) := 
    \begin{cases} 
        p & t < -c \\
	    \frac{e^{h(t+c)}}{e^{ch} + e^{h (t+c)}} & -c\leq t\leq c \\
	    1+p & t > c
	\end{cases},
\end{equation}
where $c := \frac{\log|p^{-1} - 1|}{h}$ and $p := 10^{-16}$.
These have the property that $\delta_N(a,t)\rightarrow \delta(t)$ as $a\rightarrow\infty$, $\int_0^{\pi/b}\delta_N(a,t) = 1$, $\Theta_N(h,t)\rightarrow\Theta(t)$ as $h\rightarrow\infty$ and $\Theta_N(h,t <- c) = p$ and $\Theta_N(h,t >= c) = 1+p$. The shift $c$ is chosen so that $\Theta_N(h,-c) = p$ and the function smoothly deviates from 0.

In both of the exact solutions we consider here, we take initial data corresponding to the Minkowski space-time and choose $z_l = -2 = -z_r$. The non-vanishing initial conditions are essentially just a choice of the spatial coordinate $z$ (through the choice of $A$) and null coordinates $u,v$:
\begin{equation}\label{eq:standardID}
    A(0,z) = 4, \qquad
    u(0,z) = -\frac{z-z_l}{\sqrt{2}A}, \qquad
    v(0,z) = \frac{z-z_r}{\sqrt{2}A}.
\end{equation}

The final task is to relate the coordinate systems. For simplicity we look along the curve $u=v$ only. This simplifies the analysis but still serves as a satisfactory and non-trivial comparison. Equating the line element Eq.~\eqref{eq:PWLineElement} with that of the exact solution along $u=v$ yields an ODE for $u=u(t)$ that is readily found via numerical integration.

\subsubsection{The Nutku-Halil solution}
\begin{figure}[H]
    \centering
    \begin{subfigure}[t]{0.45\textwidth}
        \centering
        \includegraphics[height=5.5cm]{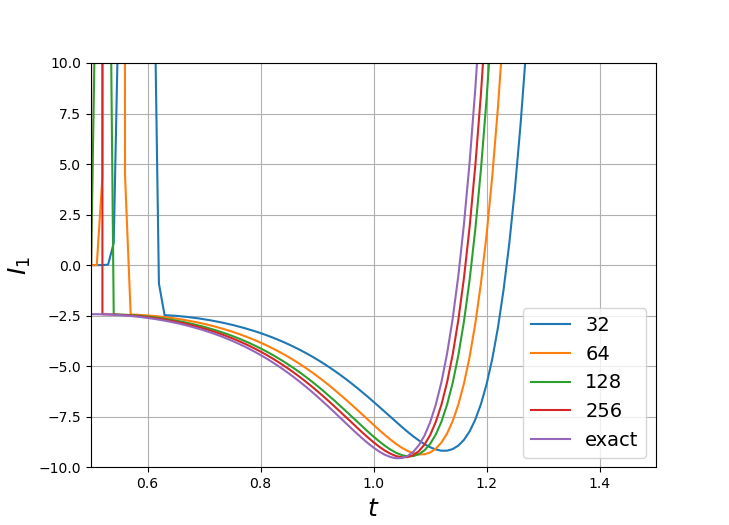}
        \caption{$I_1$}
    \end{subfigure}%
    ~ 
    \begin{subfigure}[t]{0.45\textwidth}
        \centering
        \includegraphics[height=5.5cm]{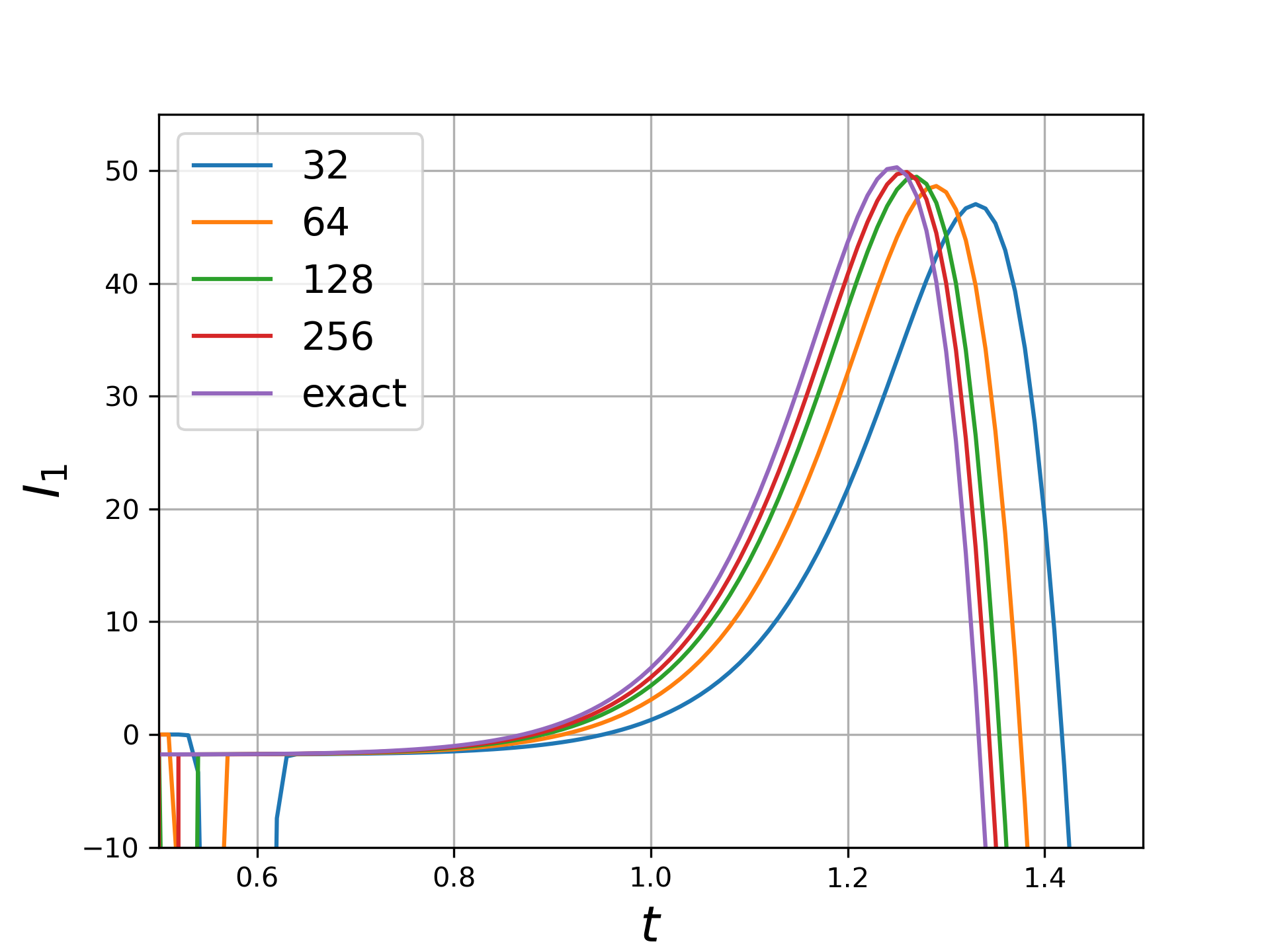}
        \caption{$I_2$}
    \end{subfigure}
    \caption{Convergence of the Weyl scalar invariants $I_1$ and $I_2$ to the Nutku-Halil solution for $\alpha=0.1$ and $\beta=0.7$ along $u=v$ for increasing wave amplitude $a$ and computational resolution. The curves from right to left represent simulations using waves of amplitude $32,64,128$ and $256$ respectively, with the left-most curve being the exact solution.}\label{fig:NHReproduction}
\end{figure}

The boundary conditions to approximate the Nutku-Halil solution are taken to be
\begin{gather} 
	\Psi_0(v, z_r) = e^{2\pi i\alpha}\delta_N(a,u),\qquad
	\Psi_4(u, z_l) = e^{2\pi i\beta}\delta_N(a,v), \nonumber  \\
    \phi_0(v, z_r) = \phi_2(u, z_l) = 0, \qquad
    u(t, z_l) = v(t, z_r) = \frac{t}{\sqrt{2}}, \label{eq:NHBCs}
\end{gather}
which together with the initial data Eq.~\eqref{eq:standardID} should reproduce the exact solution in the limit $a\rightarrow\infty$ with increasing resolution. Equating the line elements given by Eqs~\eqref{eq:NHLineElement}, \eqref{eq:PWLineElement}, and recalling our gauge choice $B=0$, yields along the curve $u=v$
\begin{equation}
	2\frac{1 - E\bar{E}}{(1 - u^2)\sqrt{1 - 2u^2}}du^2 = dt^2,
\end{equation}
which can easily be solved numerically for $u = u(t)$. Numerical integration of our IBVP given by the evolution system Eqs.~\eqref{eq:EvEqs} with initial conditions Eq.~\eqref{eq:standardID} and boundary conditions Eqs.~\eqref{eq:NHBCs} with the choice $\alpha=0.1$ and $\beta=0.7$ are performed for a selection of increasing amplitudes $a$ and resolutions of the computational domain. The resulting Weyl scalar invariants along $u=v$, corresponding to $z=0$ in this case due to the symmetry of the setup, are seen to converge to the exact solution as $a\rightarrow\infty$ and with increasing resolution, as shown in Fig.~\ref{fig:NHReproduction}.

\begin{figure}[H]
    \centering
    \includegraphics[height=5.5cm]{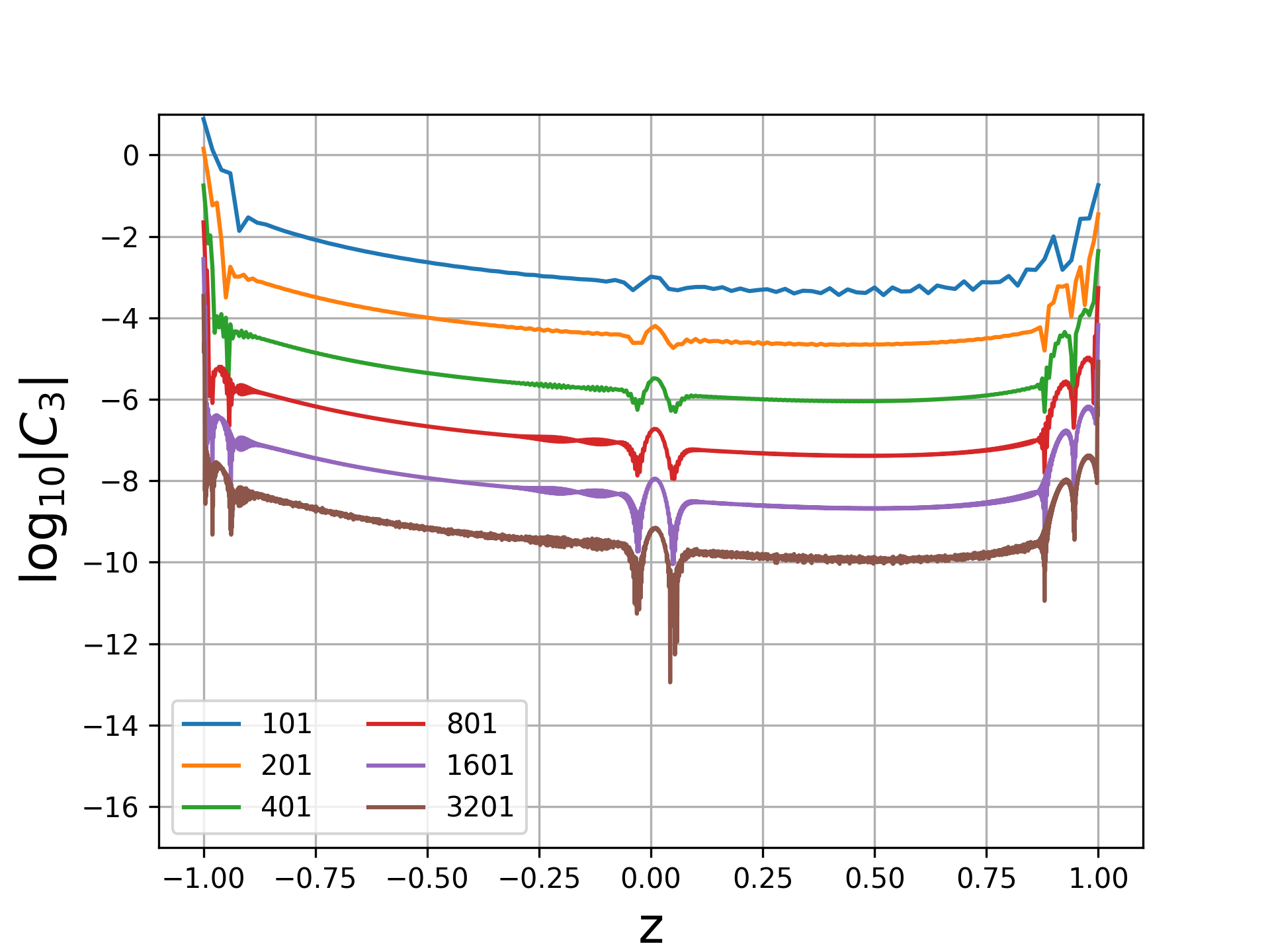}
    \caption{$\log_{10}|C_3|$ evaluated over the spatial grid at $t=1.4$ of the constraint $C_3$ for $\alpha=0.1$ and $\beta=0.7$ with $a=32$ and for an increasing number of spatial points.}\label{fig:NHConstraintConvergence}
\end{figure}

Further, we show convergence of the constraints at the correct order for a fixed amplitude in Fig. \ref{fig:NHConstraintConvergence}. The other constraints converge in an analogous manner.

\subsubsection{Griffith's 1975 solution}
\begin{figure}[H]
    \centering
    \begin{subfigure}[t]{0.45\textwidth}
        \centering
        \includegraphics[height=5.5cm]{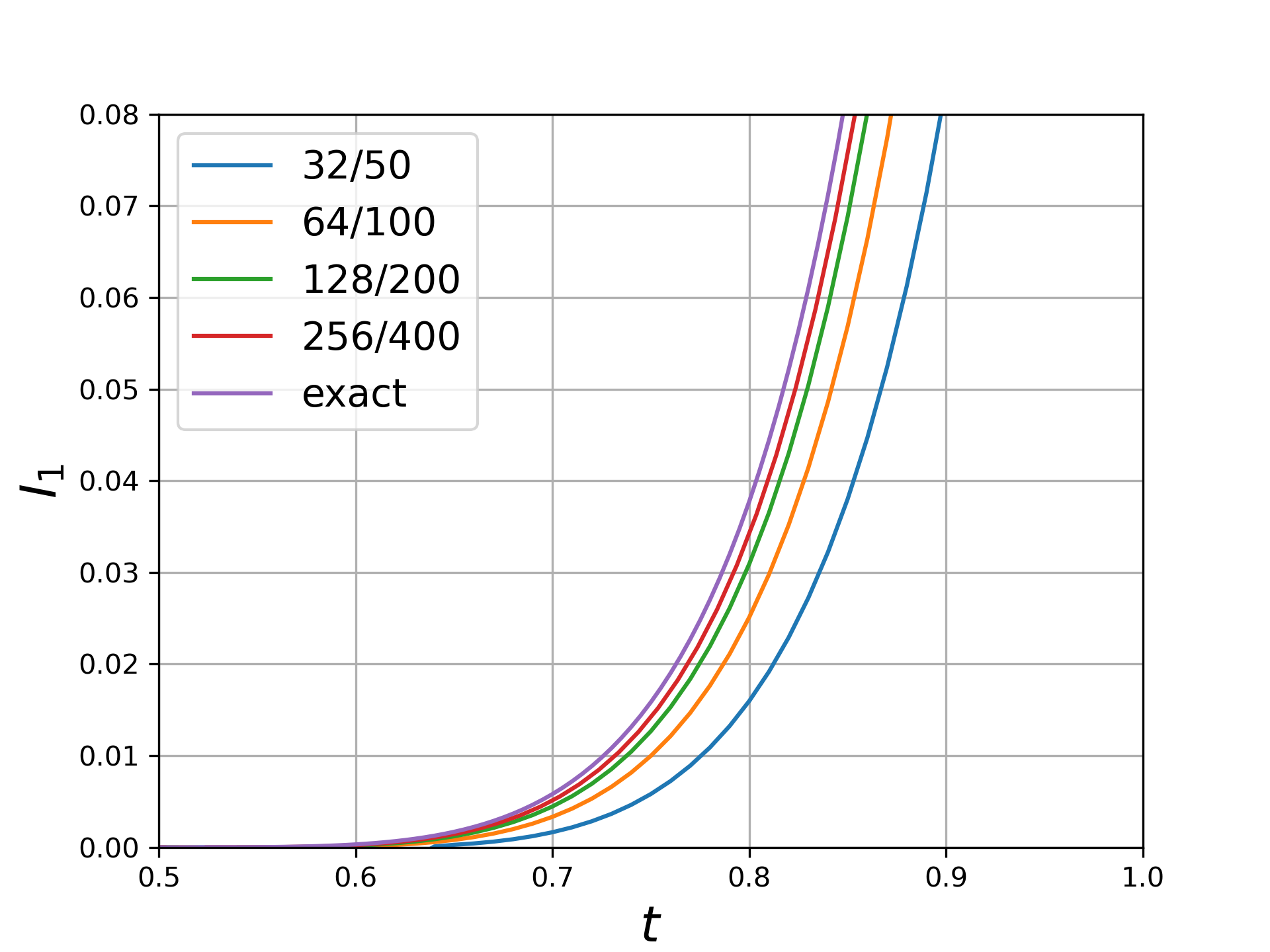}
        \caption{}
    \end{subfigure}%
    ~ 
    \begin{subfigure}[t]{0.45\textwidth}
        \centering
        \includegraphics[height=5.5cm]{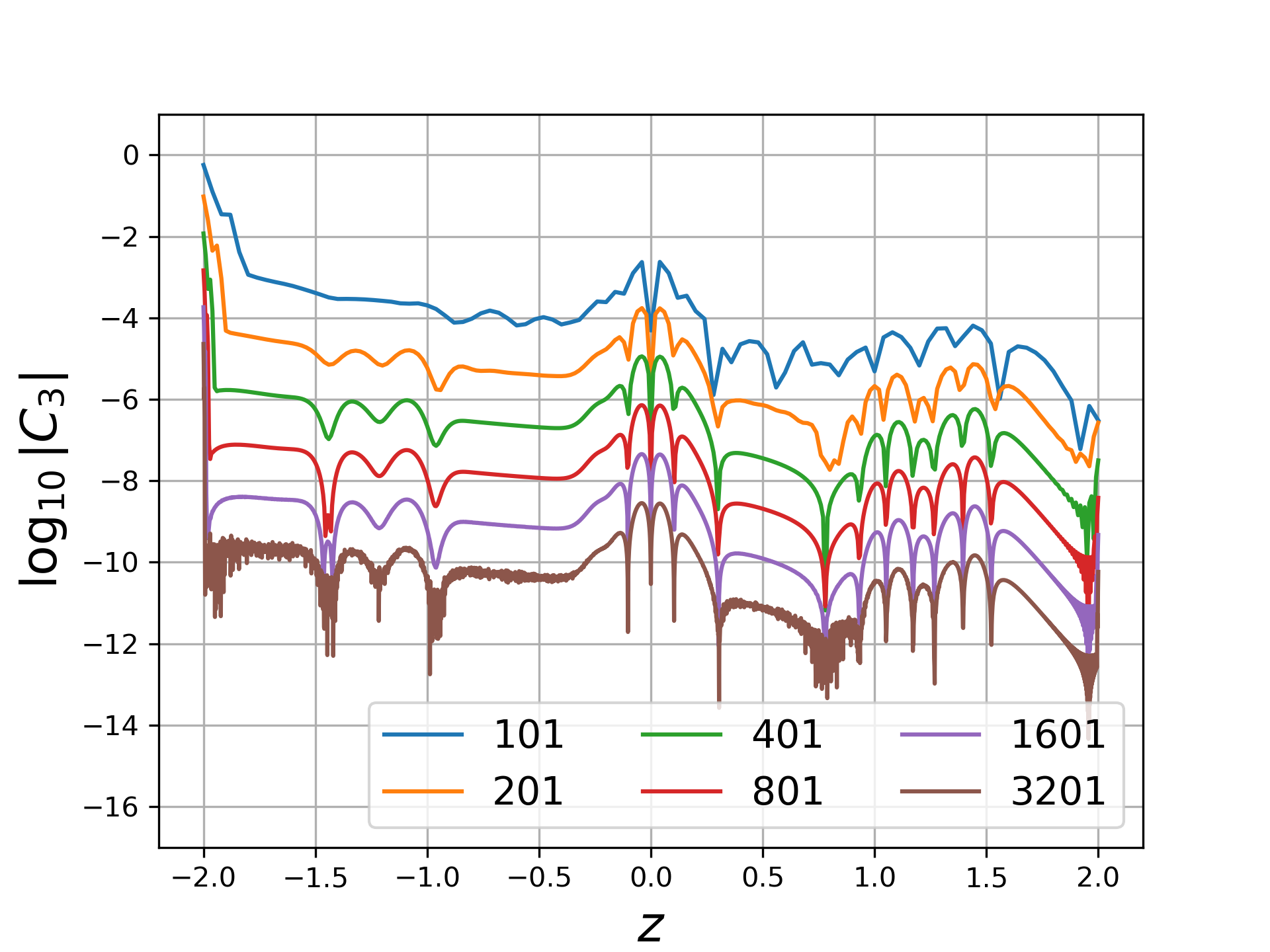}
        \caption{}
    \end{subfigure}
    \caption{(a) Convergence of the Weyl scalar invariant $I_1$ to Griffith's exact solution along $u=v$ for increasing $a,h$ and computational resolution. The curves from right to left represent simulations using $a=32,64,128,256$ and $h=50,100,200,400$ respectively, denoted as "a/h" in the legend, with the left-most curve being the exact solution. (b) $\log_{10}|C_3|$ evaluated over the spatial grid at $t=0.85$ for $a=64,\;h=100$ and for an increasing number of spatial points.}\label{fig:ConstraintConvergenceGriffiths}
\end{figure}

We now test the generalisation of our system to include coupling to the electromagnetic field by reproducing Griffith's exact solution of an impulsive gravitational wave and an electromagnetic shock wave \cite{griffiths1975colliding}. The boundary conditions are
\begin{gather} 
	\Psi_4(u, z_l) = \delta_N(a,u), \qquad
    \phi_0(v, z_r) = \Theta_N(h,v), \nonumber \\
    \Psi_0(v, z_r) = \phi_2(u, z_l) = 0, \qquad
    u(t, z_l) = v(t, z_r) = \frac{t}{\sqrt{2}}, \label{eq:GriffithsBCs}
\end{gather}
which together with the initial data Eq.~\eqref{eq:standardID} will reproduce the exact solution in the limits $a,h\rightarrow\infty$ with increasing resolution.
Numerical integration of our initial boundary value problem given by the evolution system Eqs.~\eqref{eq:EvEqs} with initial conditions Eq.~\eqref{eq:standardID} and boundary conditions Eqs.~\eqref{eq:GriffithsBCs} are performed for a selection of increasing $a,h$ and resolutions of the computational domain. The curve $u=v$ is no longer given by $z=0$ due to the symmetry $l^a \leftrightarrow n^a$ not holding. A simple root finding method applied to $u-v$ yields the $z$-value of $u=v$ on each timeslice onto which the relevant quantities can be interpolated. Along this curve the Weyl scalar invariant $I_1$ ($I_2$ is zero due to the realty of the setup) is seen to converge to the exact solution as $a\rightarrow\infty$, $h\rightarrow\infty$ and with increasing resolution. This is seen together with convergence of the constraints at the correct order for a fixed amplitude in Fig. \ref{fig:ConstraintConvergenceGriffiths}. The other constraints converge in an analogous manner.

\section{Non-linear scattering}\label{sec:nonlinearbehaviour}
Now that checks of correctness are completed and we are confident in the numerical implementation of the system, we begin to explore the non-linear scattering of electromagnetic and gravitational waves in situations not described by exact solutions. That is, in scenarios where the waves exhibit gradual rather than abrupt changes in profile, and where there could be variations in period, polarization, and amplitude. We focus largely on the waves post-scattering as well as the Weyl scalar curvature invariants $I_1$ and $I_2$, which represent in a physically meaningful way the strength of the local Weyl curvature as they are frame-independent quantities.

\subsection{Collisions of electromagnetic waves}\label{sec:collidingEMW}
The effect of the electromagnetic waves' polarisations plays a very minor role in the scattering process, and affect the gravitational waves only indirectly. This is clear by noticing that the evolution system does not have components of $\phi_{AB}$ appearing explicitly. Rather, they appear only indirectly through components of $\Phi_{AA'BB'}$ that are related to $\phi_{AB}$ via Eq.~\eqref{eq:PhiDef}. The only place where the polarisation will have an effect is in the evolution equations for the shears $\sigma$, $\sigma'$ and the gravitational waves $\Psi_0,\Psi_4$. However, as the magnitude of the polarisation takes values in $[-1,1]$, and the effect on the gravitational waves is non-linear, the effect is small. In fact, in the symmetric case where the system is unchanged under the 'prime' operation of \cite{penrose1984spinors} one sees that the polarisations do not play a role at all. Hence, we do not explore varying the polarisation of the electromagnetic wave profile.

\subsubsection{Smoothed shock profiles}
\begin{figure}[H]
    \centering
    \includegraphics[height=5.5cm]{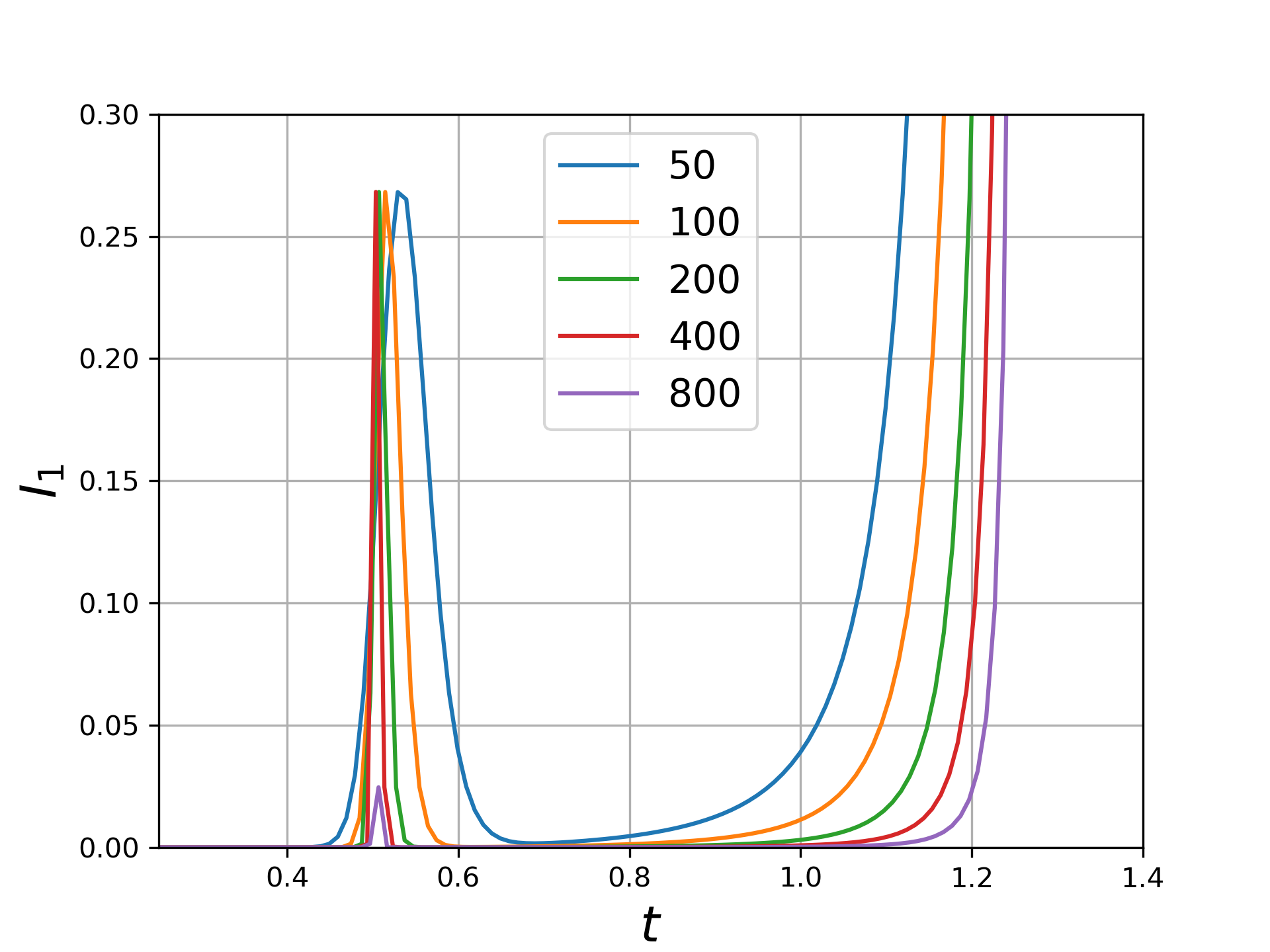}
    \caption{Convergence of the Weyl scalar invariant $I_1$ to the Bell-Szekeres solution along $u=v$ using boundary conditions Eq.~\eqref{eq:SmoothedShockBCs} with increasing $h$ and computational resolution. The curves from left to right represent simulations with $h$ as $50,100,200,400$ and $800$ respectively and $\alpha=0$.}\label{fig:BSI1Convergence}
\end{figure}

The Bell-Szekeres solution \cite{bell1974interacting} describes the scattering of two plane electromagnetic shock waves that generate plane impulsive gravitational waves. Of interest is that the scattering region does not contain a space-like singularity in the future, as is the usual case for plane wave collisions, but instead has a Cauchy horizon where the curvature remains bounded. Considering smoothed step-functions as boundary conditions, namely
\begin{gather} 
    \phi_0(v, z_r) = \Theta_N(h,v), \qquad
    \phi_2(u, z_l) = e^{2\pi i \alpha}\Theta_N(h,u), \nonumber \\
    \Psi_0(v, z_r) = \Psi_4(u, z_l) = 0, \qquad
    u(t, z_l) = v(t, z_r) = \frac{t}{\sqrt{2}}\label{eq:SmoothedShockBCs}
\end{gather}
one can comment on whether the curvature tensor remains bounded for deviations from the exact solution. It is clear from Fig.~\ref{fig:BSI1Convergence} that the Weyl invariant $I_1$ is approaching zero as one approaches step-function wave profiles, but is non-zero otherwise. Hence, conformal flatness and a future Cauchy horizon seem to be properties special to the shock wave profiles, and small deviations from them seem likely to lead to a future curvature singularity instead of a Cauchy horizon, due to the divergence of $I_1$.
 
\begin{figure}[H]
    \centering
    \begin{subfigure}[t]{0.45\textwidth}
        \centering
        \includegraphics[height=5.5cm]{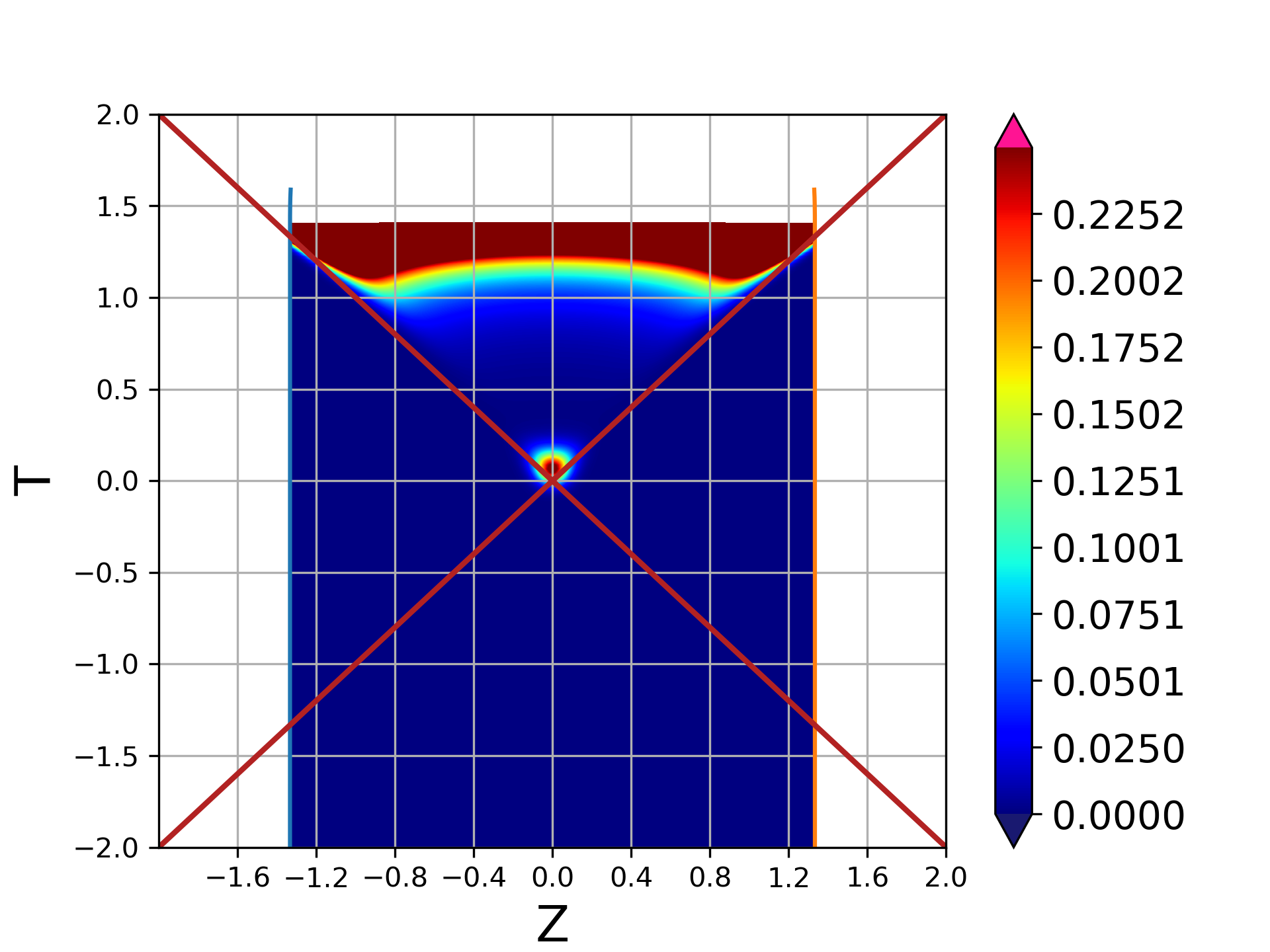}
    \end{subfigure}%
    ~ 
    \begin{subfigure}[t]{0.45\textwidth}
        \centering
        \includegraphics[height=5.5cm]{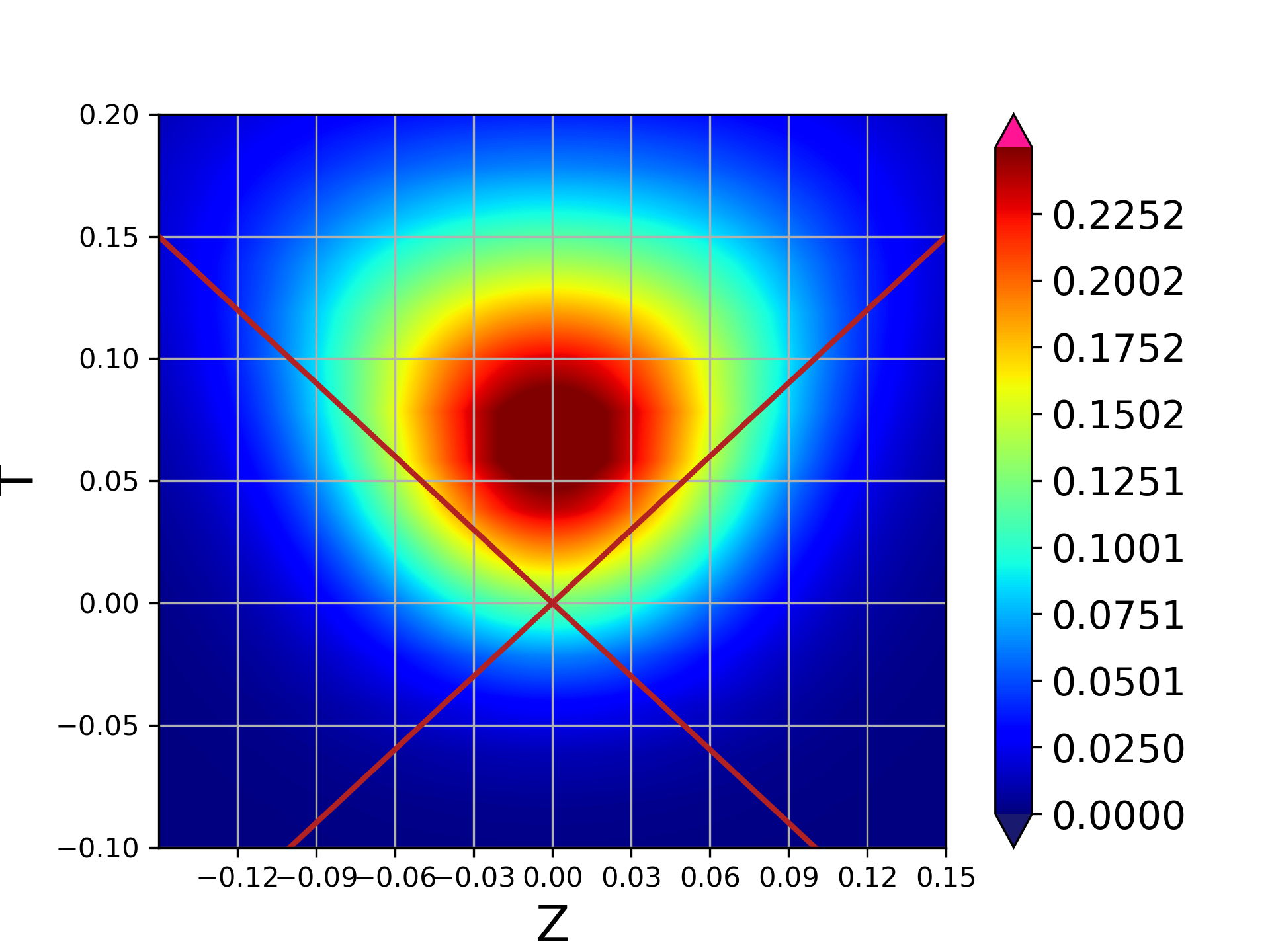}
    \end{subfigure}
    \caption{Conformal diagram contour plots showing the Weyl invariant $I_1$ over the computational domain using boundary conditions Eq.~\eqref{eq:SmoothedShockBCs} with $h=50$ and $\alpha=0$.}\label{fig:EMWEMWI1Contours2}
\end{figure}

Fig.~\ref{fig:EMWEMWI1Contours2} shows contour plots of the Weyl invariant $I_1$, indicating a behaviour very similar to that of two colliding gravitational waves with smoothed impulsive wave profiles as seen in \cite{frauendiener2014numerical}.

\begin{figure}[H]
    \centering
    \begin{subfigure}[t]{0.45\textwidth}
        \centering
        \includegraphics[height=5.5cm]{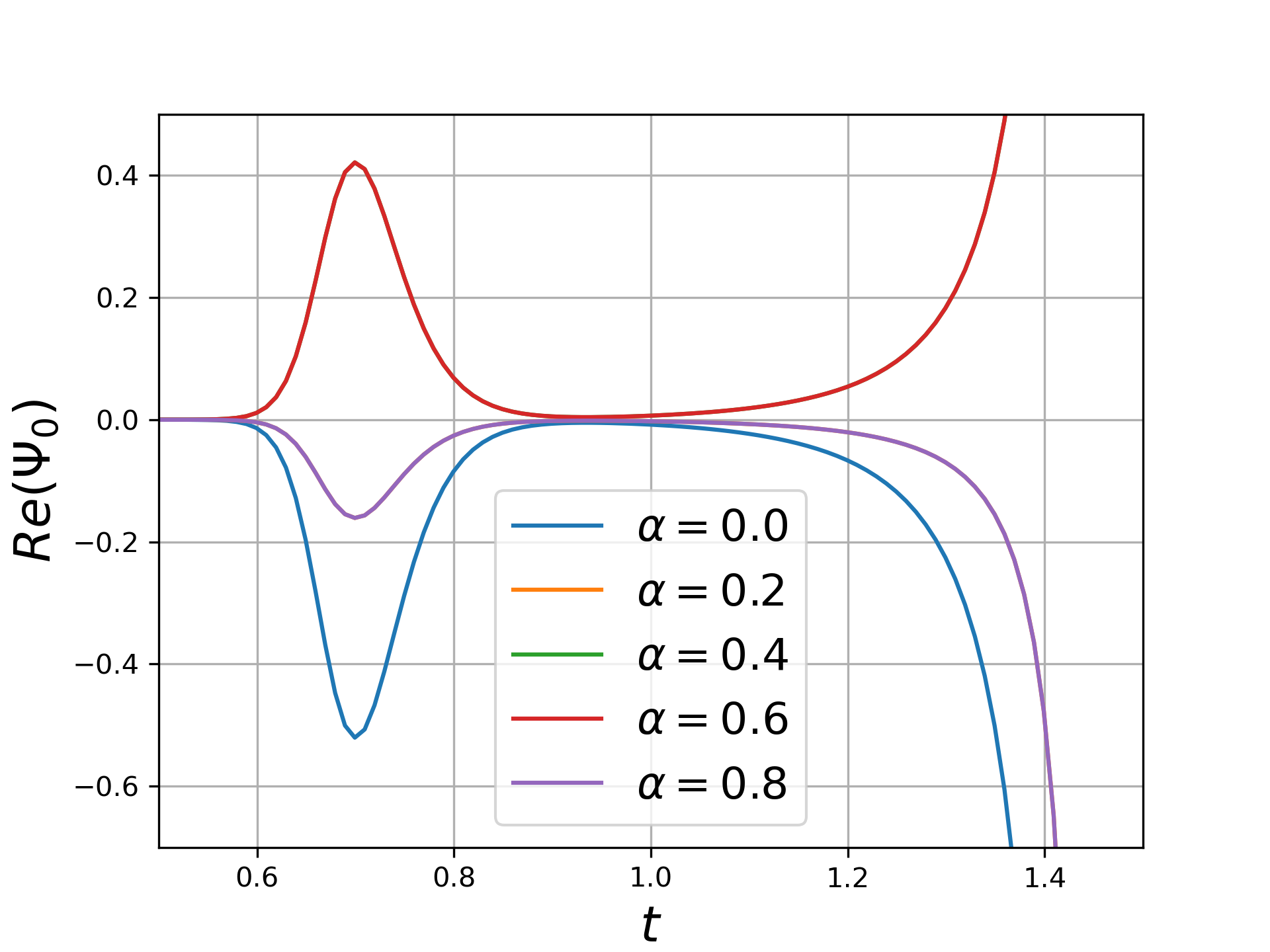}
    \end{subfigure}%
    ~ 
    \begin{subfigure}[t]{0.45\textwidth}
        \centering
        \includegraphics[height=5.5cm]{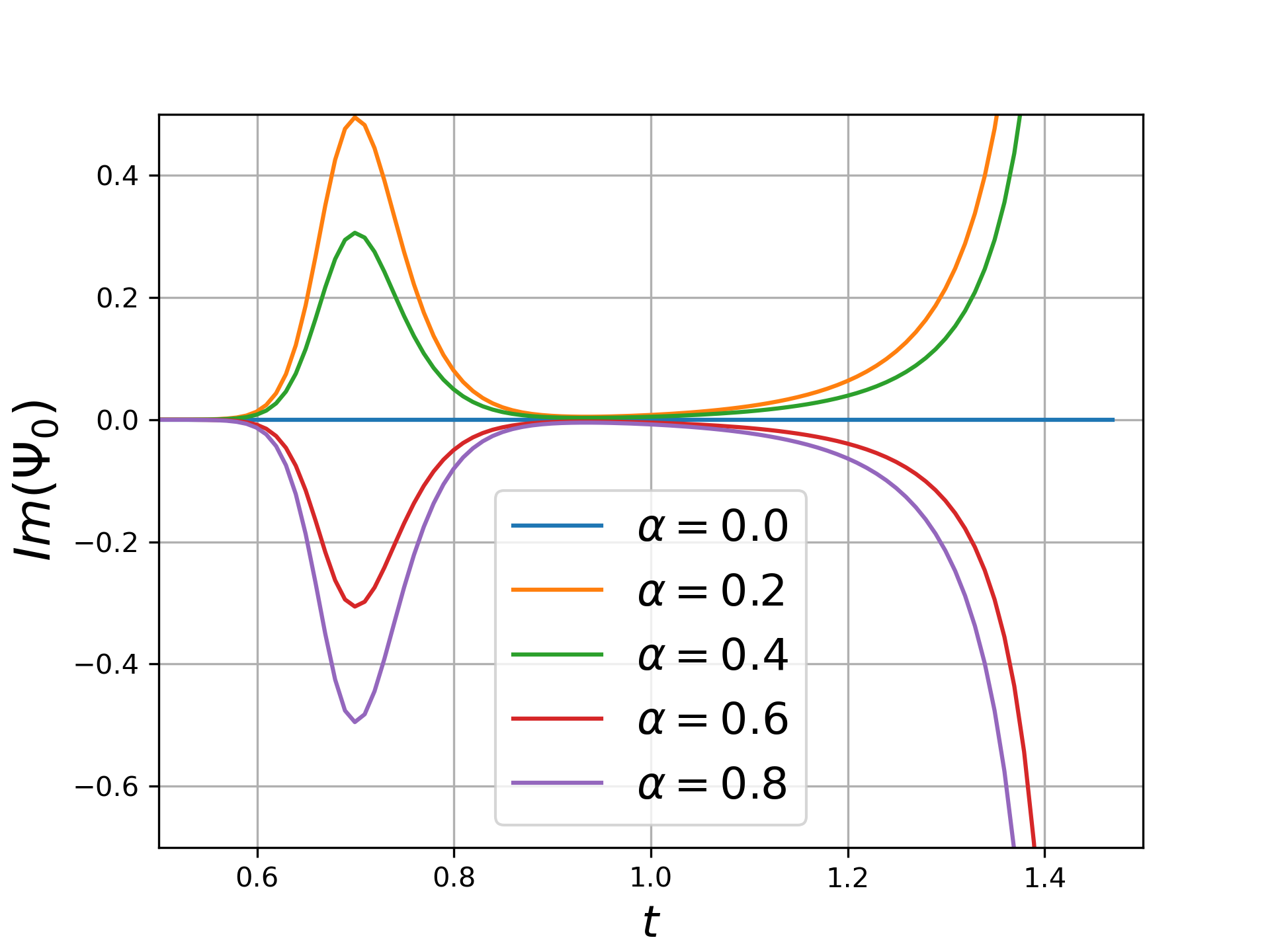}
    \end{subfigure}
    \caption{The generated real and imaginary parts of $\Psi_0$ along $u=v$ using boundary conditions Eq.~\eqref{eq:SmoothedShockBCs} with $h=50$. Note that in the plot of $Re(\Psi_0)$ the curves $\alpha=0.2,0.4$ are the same as $\alpha=0.8,0.6$ respectively.}\label{fig:psi0_ueqv_EMW_EMW_SmoothedShock}
\end{figure}

\begin{figure}[H]
    \centering
    \begin{subfigure}[t]{0.45\textwidth}
        \centering
        \includegraphics[height=5.5cm]{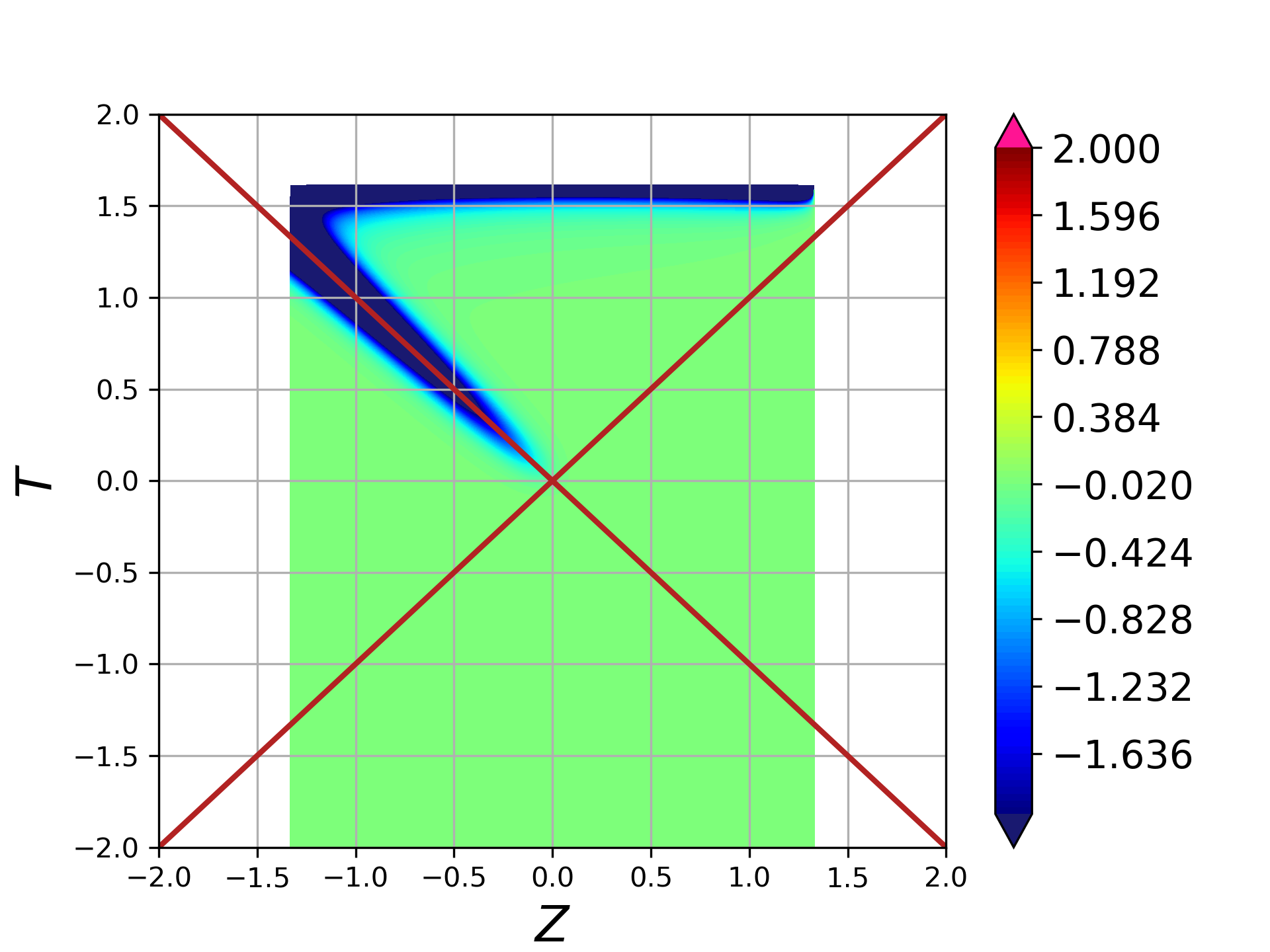}
    \end{subfigure}%
    ~ 
    \begin{subfigure}[t]{0.45\textwidth}
        \centering
        \includegraphics[height=5.5cm]{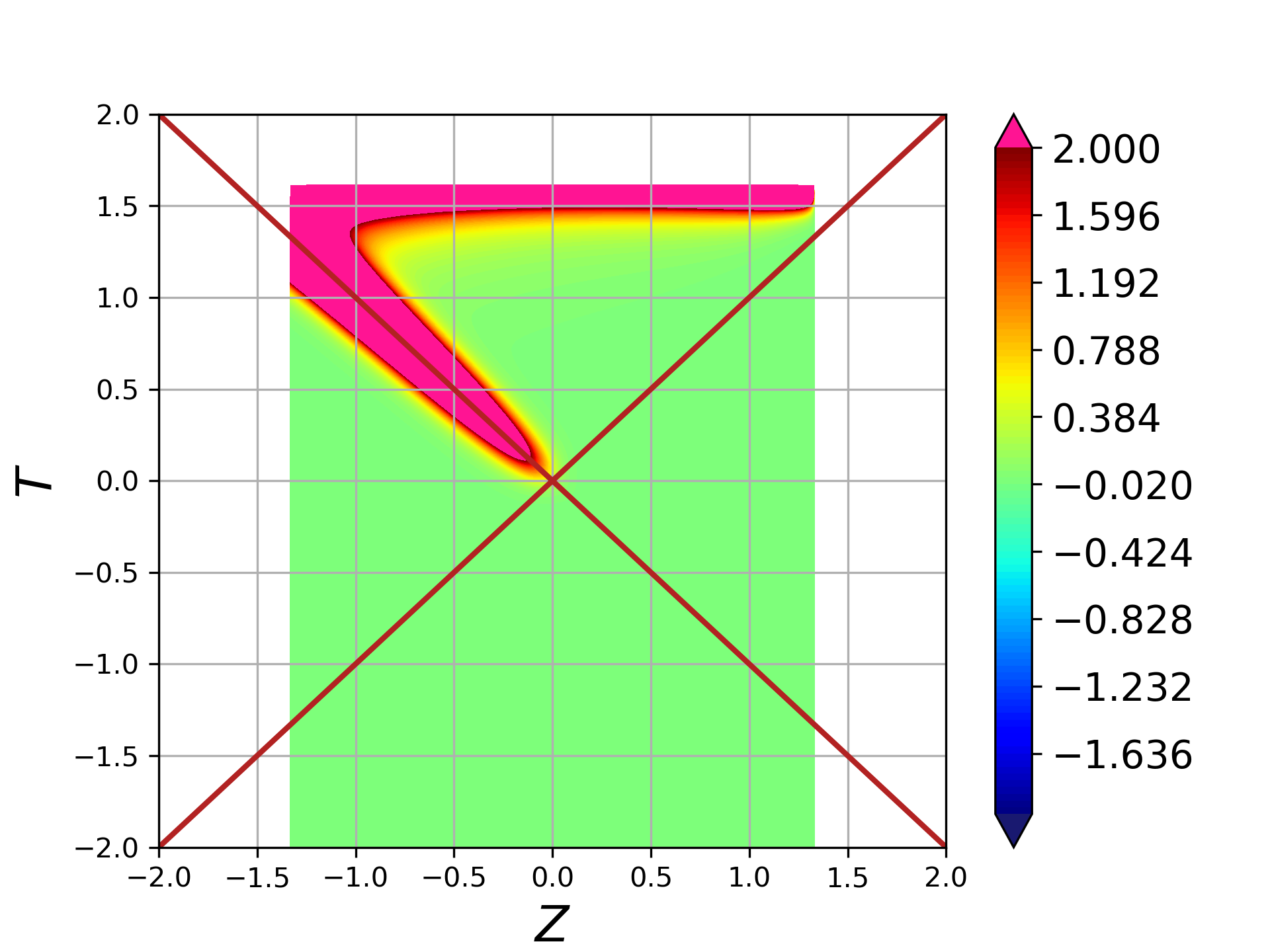}
    \end{subfigure}
    \caption{Conformal diagram contour plot showing the real and imaginary parts of $\Psi_0$ over the computational domain using boundary conditions Eq.~\eqref{eq:SmoothedShockBCs} with $h=50$ and $\alpha=0.2$.}\label{fig:Psi0_contour_EMW_EMW_SmoothedShock}
\end{figure}

It is found that the polarisations of the electromagnetic wave profiles have no effect on the Weyl invariants, with $I_2$ remaining zero in the scattering region. The main role they play is that they govern the polarisations of the generated gravitational waves due to the scattering. Further, only the relative polarisation between the electromagnetic waves influences the solution in the scattering region, a property shared with the Nutku-Halil solution. Fig.~\ref{fig:psi0_ueqv_EMW_EMW_SmoothedShock} showcases the generated $\Psi_0$ in the scattering region along $u=v$ for a range of polarisations $\alpha$ while Fig.~\ref{fig:Psi0_contour_EMW_EMW_SmoothedShock} presents a conformal diagram contour plot for the case $\alpha=0.2$. The relationships between the polarisation of the gravitational and electromagnetic radiation in the scattering region to the polarisation $\alpha$ of the $\phi_2$ is given by Tab.~\ref{tab:EMW_EMW_SmoothedShock_pols}.

\begin{table}[!h]
    \begin{center}
        \begin{tabular}{|c|c|c|c|c|} 
             \hline
             {} & $\Psi_0$ & $\Psi_4$ & $\phi_0$ & $\phi_2$ \\ [0.5ex] 
             \hline\hline
             $Pol$ & $\frac{1}{10}[(5-10\alpha)\mod 10]$ & $\frac{1}{10}[(5+10\alpha)\mod 10]$ & 0 & $\alpha$ \\ 
             \hline
        \end{tabular}
    \end{center}
    \caption{The polarisation $Pol\in[0,1)$ of $\Psi_0$, $\Psi_4$, $\phi_0$ and $\phi_2$ in the scattering region, written in terms of the initial polarisation $\alpha$ of $\phi_2$.}
    \label{tab:EMW_EMW_SmoothedShock_pols}
\end{table}

\subsubsection{Smoothed impulsive profiles}
A scenario that is completely missing from the exact solution literature is the collision of two electromagnetic plane waves that do not have a shock profile. For example, pulsars emit bursts of electromagnetic radiation rather than a constant intensity beam. As such, we explore what happens when we choose wave profiles that are smooth bumps described by the $\delta_N(a,t)$ profiles. The boundary conditions are chosen as

\begin{gather} 
	\phi_0(v, z_r) = \frac{1}{10}\delta_N(a,v), \qquad
    \phi_2(u, z_l) = \frac{1}{10}e^{2\pi i \alpha}\delta_N(a,u) \nonumber \\
    \Psi_0(v, z_r) = \Psi_4(u, z_l) = 0, \qquad
    u(t, z_l) = v(t, z_r) = \frac{t}{\sqrt{2}}. \label{eq:EMW_EMW_Bump_BCs}
\end{gather}

\begin{figure}[H]
    \centering
    \begin{subfigure}[t]{0.45\textwidth}
        \centering
        \includegraphics[height=5.5cm]{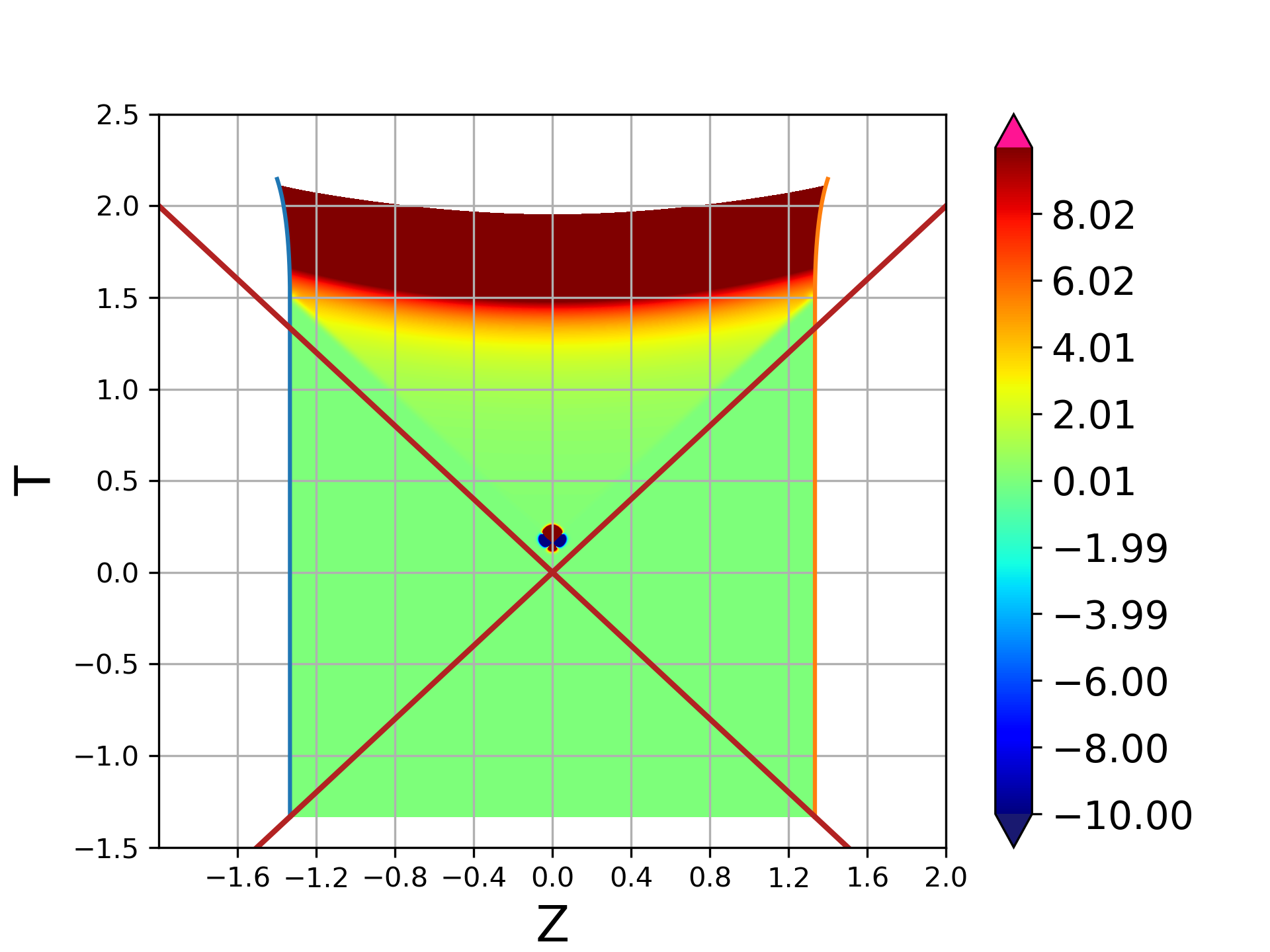}
    \end{subfigure}%
    ~ 
    \begin{subfigure}[t]{0.45\textwidth}
        \centering
        \includegraphics[height=5.5cm]{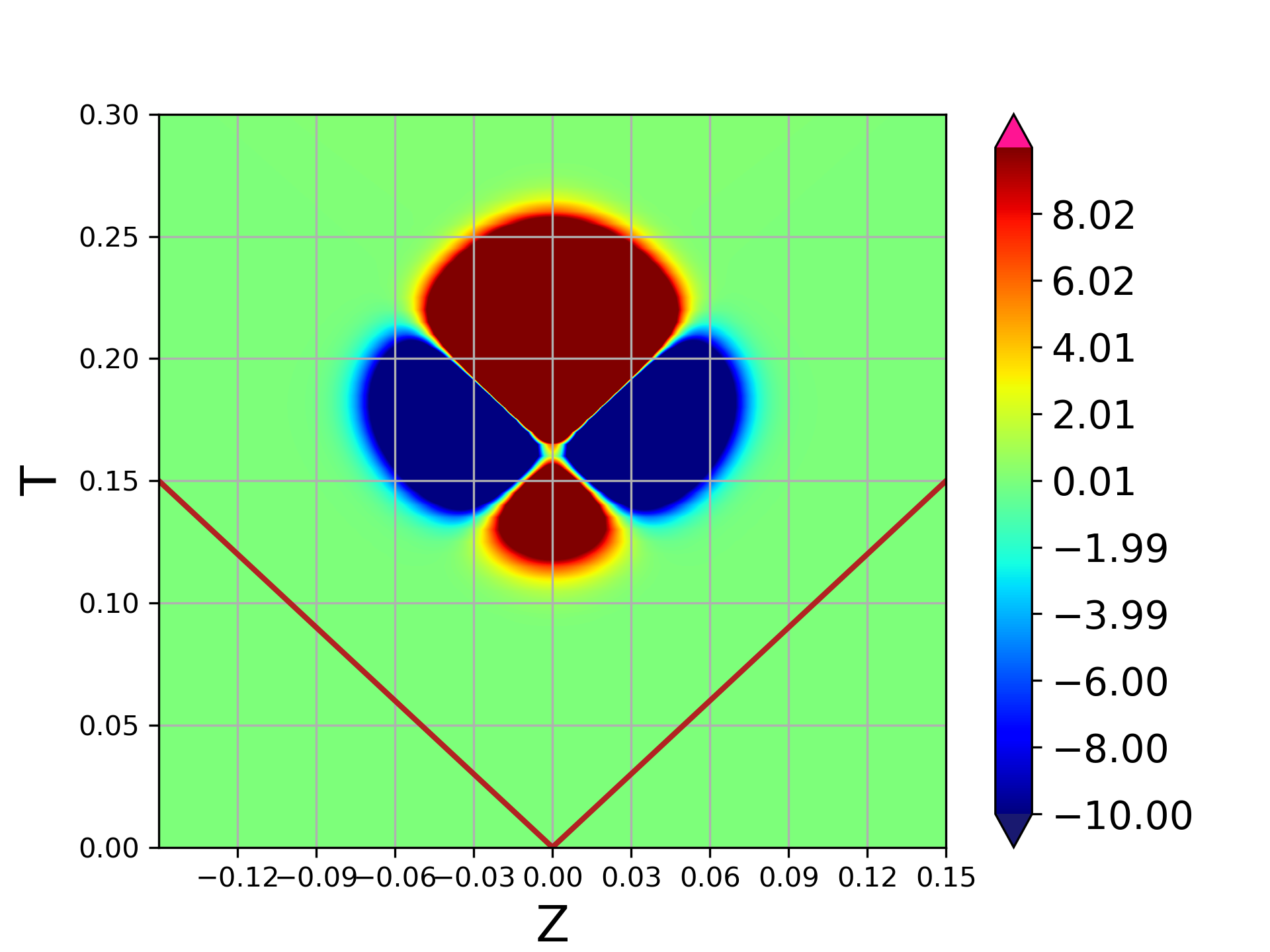}
    \end{subfigure}
    \caption{Conformal diagram contour plots showing the Weyl invariant $I_1$ over the computational domain using boundary conditions given by Eq.~\eqref{eq:EMW_EMW_Bump_BCs} with $a=32$ and $\alpha=0$.}\label{fig:EMWEMWI1Contours}
\end{figure}

Fig.~\ref{fig:EMWEMWI1Contours} shows contour plots of the Weyl invarant $I_1$ and there is a significant difference during the collision compared to the smoothed shock wave profile, where it now takes negative values and exhibits a quadrupolar-like appearance. 

It is found again that the scattering region depends only on the relative polarisation between the electromagnetic waves and the Weyl invariant $I_2$ remains zero for any choice of $\alpha$.

\begin{figure}[H]
    \centering
    \begin{subfigure}[t]{0.45\textwidth}
        \centering
        \includegraphics[height=5.5cm]{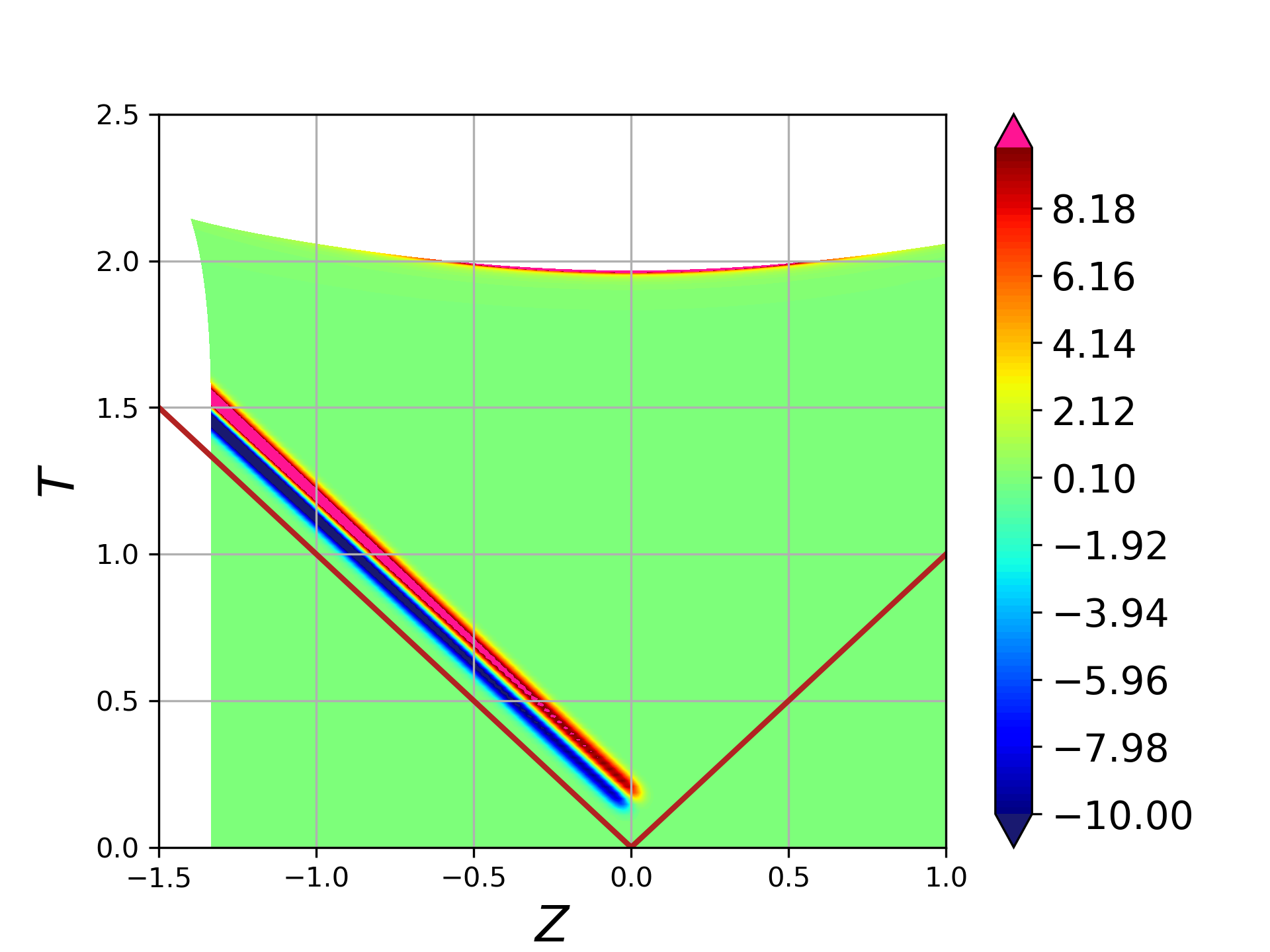}
        \caption{$Re(\Psi_0)$}
    \end{subfigure}%
    ~ 
    \begin{subfigure}[t]{0.45\textwidth}
        \centering
        \includegraphics[height=5.5cm]{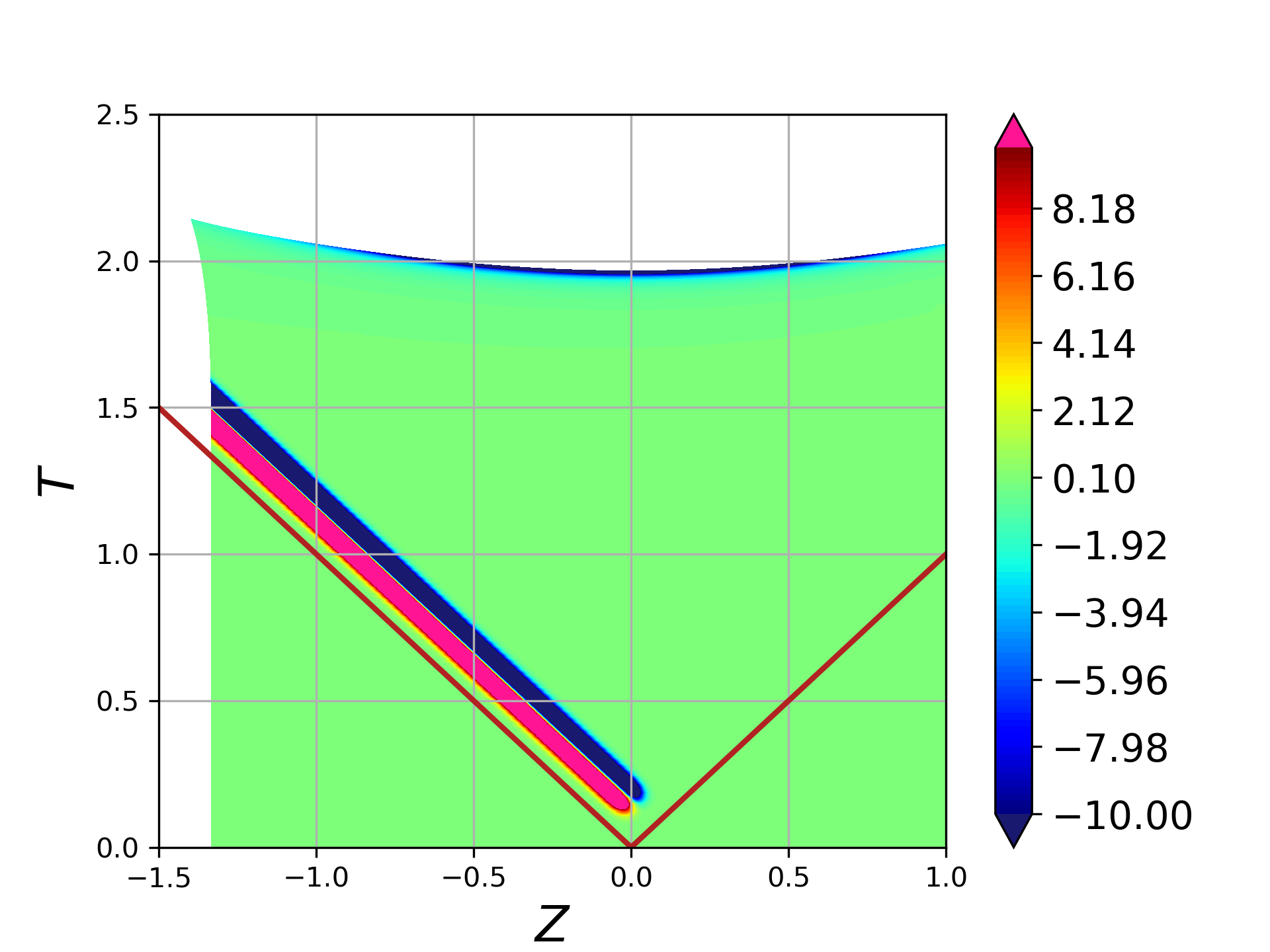}
        \caption{$Im(\Psi_0)$}
    \end{subfigure}
    \caption{Conformal diagram contour plot showing the real and imaginary parts of $\Psi_0$ over the computational domain using boundary conditions Eq.~\eqref{eq:EMW_EMW_Bump_BCs} with $a=32$ and $\alpha=0.2$.}\label{fig:Psi0_contour_EMW_EMW_SmoothedDelta}
\end{figure}

Fig.~\ref{fig:Psi0_contour_EMW_EMW_SmoothedDelta} showcases one of the generated gravitational waves, which has a dramatically different appearance to that generated by colliding electromagnetic waves with a shock wave profile. In particular, the gravitational radiation that is generated from the scattering process has a particular polarisation for a short duration, followed by another polarisation within the rest of the scattering region. Whilst the polarisation for $\phi_0$ remains constant before and after scattering, $\phi_2$ also changes polarisation. The results are tabulated in Tab.~\ref{tab:EMW_EMW_SmoothedDelta_pols}. 

\begin{table}[!h]
    \begin{center}
        \begin{tabular}{|c|c|c|c|c|} 
             \hline
             {} & $\Psi_0$ & $\Psi_4$ & $\phi_0$ & $\phi_2$ \\ [0.5ex] 
             \hline\hline
             $Pol_1$& $\frac{1}{10}[(5-10\alpha)\mod 10]$ & $\frac{1}{10}[(5+10\alpha)\mod 10]$ & 0 & $\alpha$ \\ 
             \hline
             $Pol_2$ & $\frac{1}{10}[-10\alpha \mod 10]$ & $\alpha$ & 0.5 & $\frac{1}{10}[(5+10\alpha)\mod 10]$ \\ 
             \hline
        \end{tabular}
    \end{center}
    \caption{The polarisations $Pol_1,Pol_2\in[0,1)$ of $\Psi_0$, $\Psi_4$, $\phi_0$ and $\phi_2$ in the scattering region, written in terms of the initial polarisation $\alpha$ of $\phi_2$. The first row is immediately after collision while the second is when this changes and in the remainder of the scattering region.}
    \label{tab:EMW_EMW_SmoothedDelta_pols}
\end{table}

\subsection{Collisions of gravitational waves}\label{sec:collisions_GWs}
In \cite{frauendiener2014numerical, frauendiener2021can}, Frauediener et.~al analyse the non-linear scattering of gravitational plane waves with smoothed impulsive profiles. Of further interest is to understand what effect the choice of polarisation has on this process, as well as whether having multiple `bumps' in the wave profile yields new effects. Hence we extend these results and that of the Nutku-Halil solution \cite{nutku1977colliding} by considering the collision of two gravitational waves within the scattered region of this solution. Namely, we consider boundary conditions

\begin{gather} 
	\Psi_0(v, z_r) = \delta_N(a,v) + e^{2\pi i \gamma}\delta_N(a,v - 2\pi), \qquad
    \Psi_4(u, z_l) = e^{2\pi i \alpha}\delta_N(a,u) + e^{2\pi i \beta}\delta_N(a,u - 2\pi), \nonumber \\
    \phi_0(v, z_r) = \phi_2(u, z_l) = 0, \qquad
    u(t, z_l) = v(t, z_r) = \frac{t}{\sqrt{2}}.  \label{eq:DoubleGWs}
\end{gather}

We find the polarisations of $\Psi_0$ and $\Psi_4$ retain their original values of zero and $\alpha$ respectively for a short duration in the scattering region, but exhibit a phase change by $\alpha$ in the rest of the scattering region.  The results are tabulated in Tab.~\ref{tab:GG_pols}.

\begin{table}[!h]
    \begin{center}
        \begin{tabular}{|c|c|c|c|c|} 
             \hline
             {} & $\Psi_0$ & $\Psi_4$ \\ [0.5ex] 
             \hline\hline
             $Pol_1$ & $0$ & $\alpha$ \\ 
             \hline
             $Pol_2$ & $\alpha$ & $\frac{1}{10}[20\alpha\mod10]$ \\
             \hline
        \end{tabular}
    \end{center}
    \caption{The polarisations $Pol_1,Pol_2\in[0,1)$ of $\Psi_0$ and $\Psi_4$ in the scattering region, written in terms of the initial polarisation $\alpha$ of $\psi_4$. The first row is immediately after the first collision while the second is after all collisions.}
    \label{tab:GG_pols}
\end{table}

\begin{figure}[H]
    \centering
    \begin{subfigure}[t]{0.45\textwidth}
        \centering
        \includegraphics[height=5.5cm]{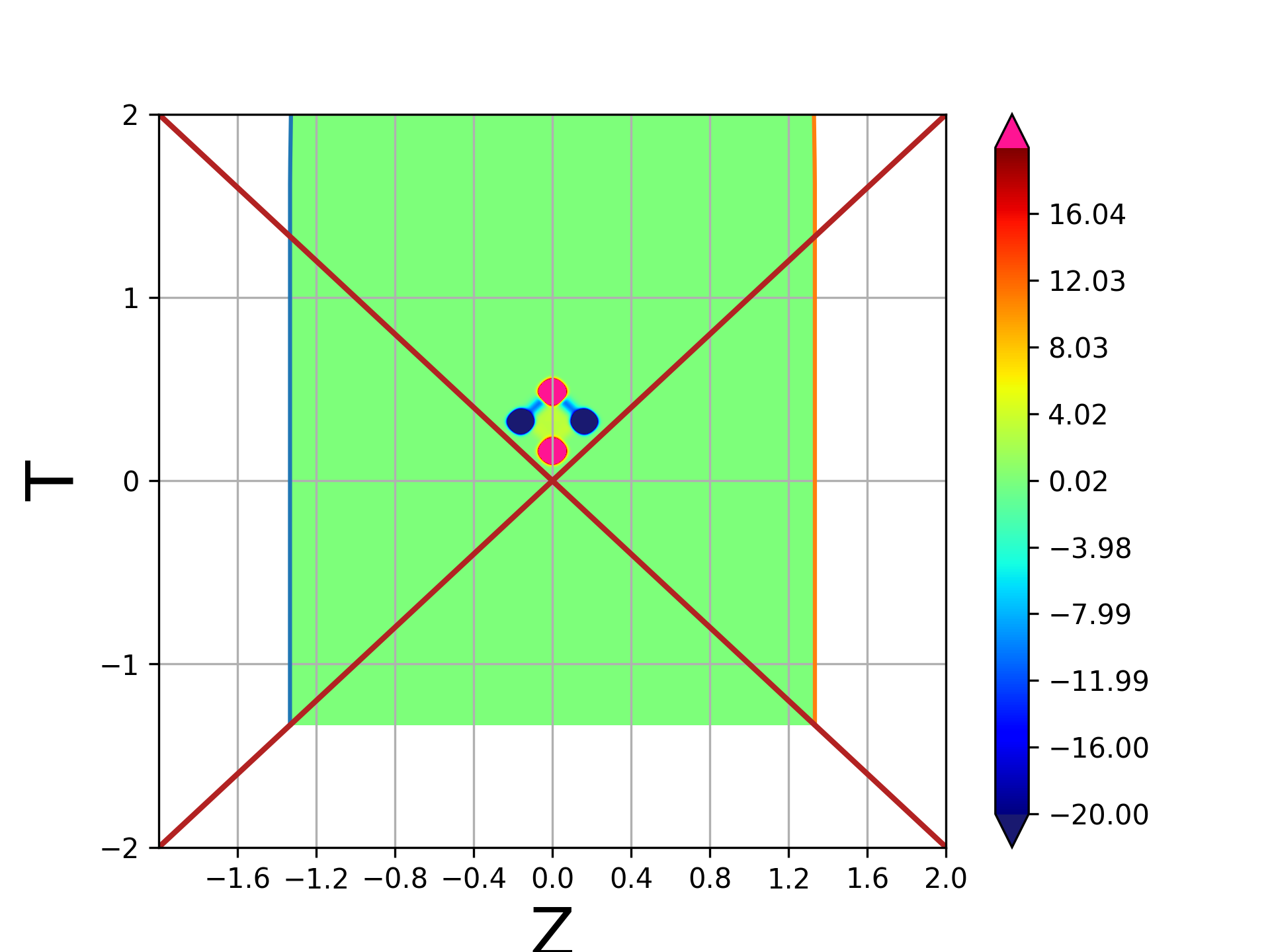}
    \end{subfigure}%
    ~ 
    \begin{subfigure}[t]{0.45\textwidth}
        \centering
        \includegraphics[height=5.5cm]{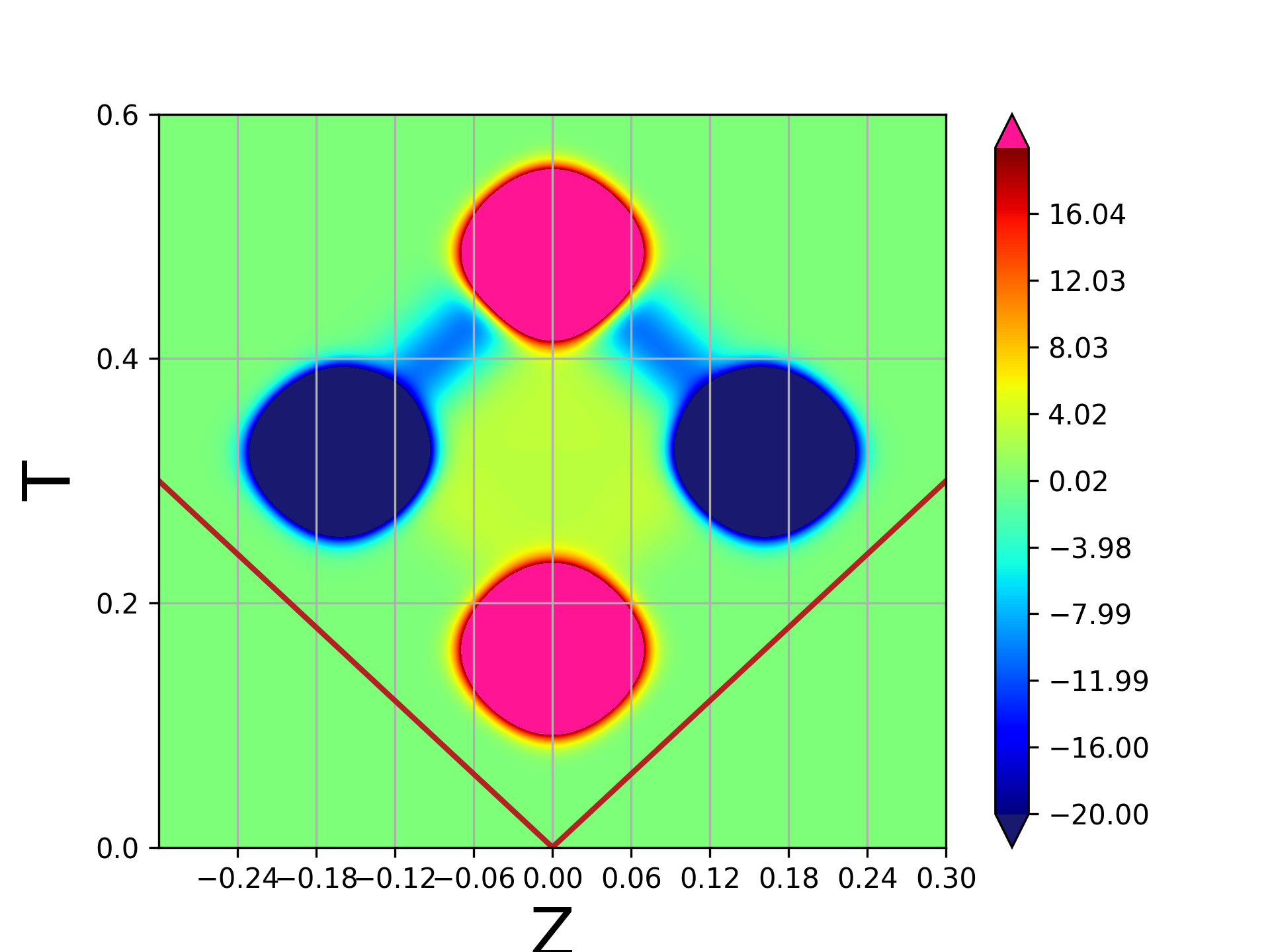}
    \end{subfigure}
    \caption{Conformal diagram contour plots showing the Weyl invariant $I_1$ over the computational domain using boundary conditions Eq.~\eqref{eq:DoubleGWs} with $a=32$, $\gamma = \beta = 0.5$ and $\alpha = 0$.}\label{fig:GWGW005005Contours}
\end{figure}

The contour plot of the Weyl invariant $I_1$ for $\gamma = \beta = 0.5$ and $\alpha = 0$ is shown in Fig.~\ref{fig:GWGW005005Contours}.  Positive peaks in $I_1$ are generated during the first and last collisions, where each pair of colliding waves have co-aligned polarisations.  Negative troughs in $I_1$ are generated during the second pair of collisions, where each pair of colliding waves have polarities exactly out of phase.  Of note is that the long-term behaviour for the spacetime generated by this particular set of boundary conditions exhibits a near-zero scalar curvature invariant $I_1$: the incident waves essentially `cancel out' in the scattering region after all collisions are completed and convergence tests show that this value does not go to zero. This small, remaining value of $I_1$ can be thought of as a measure of the non-linearity of the system because in the linear case one would expect an exact cancellation. It is for this reason that the simulation takes a much longer proper time to reach the curvature singularity compared to the Nutku-Halil solution, but does eventually reach it at $t\approx3.7195$. Note that Fig.~\ref{fig:GWGW005005Contours} has been truncated here as there is little to see before the curvature singularity is reached.

\begin{figure}[H]
    \centering
    \begin{subfigure}[t]{0.45\textwidth}
        \centering
        \includegraphics[height=5.5cm]{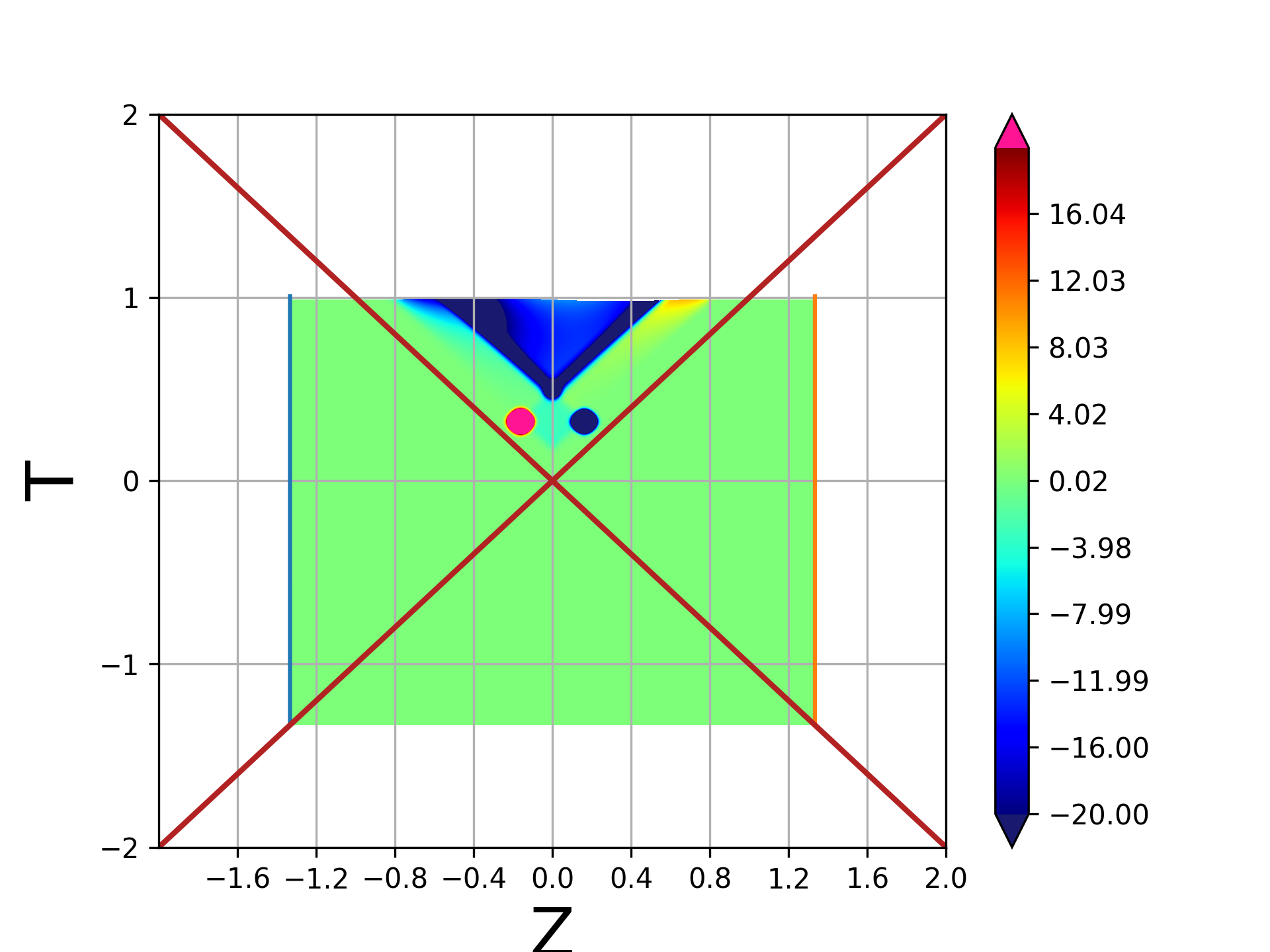}
        \caption{$I_1$}
    \end{subfigure}%
    ~ 
    \begin{subfigure}[t]{0.45\textwidth}
        \centering
        \includegraphics[height=5.5cm]{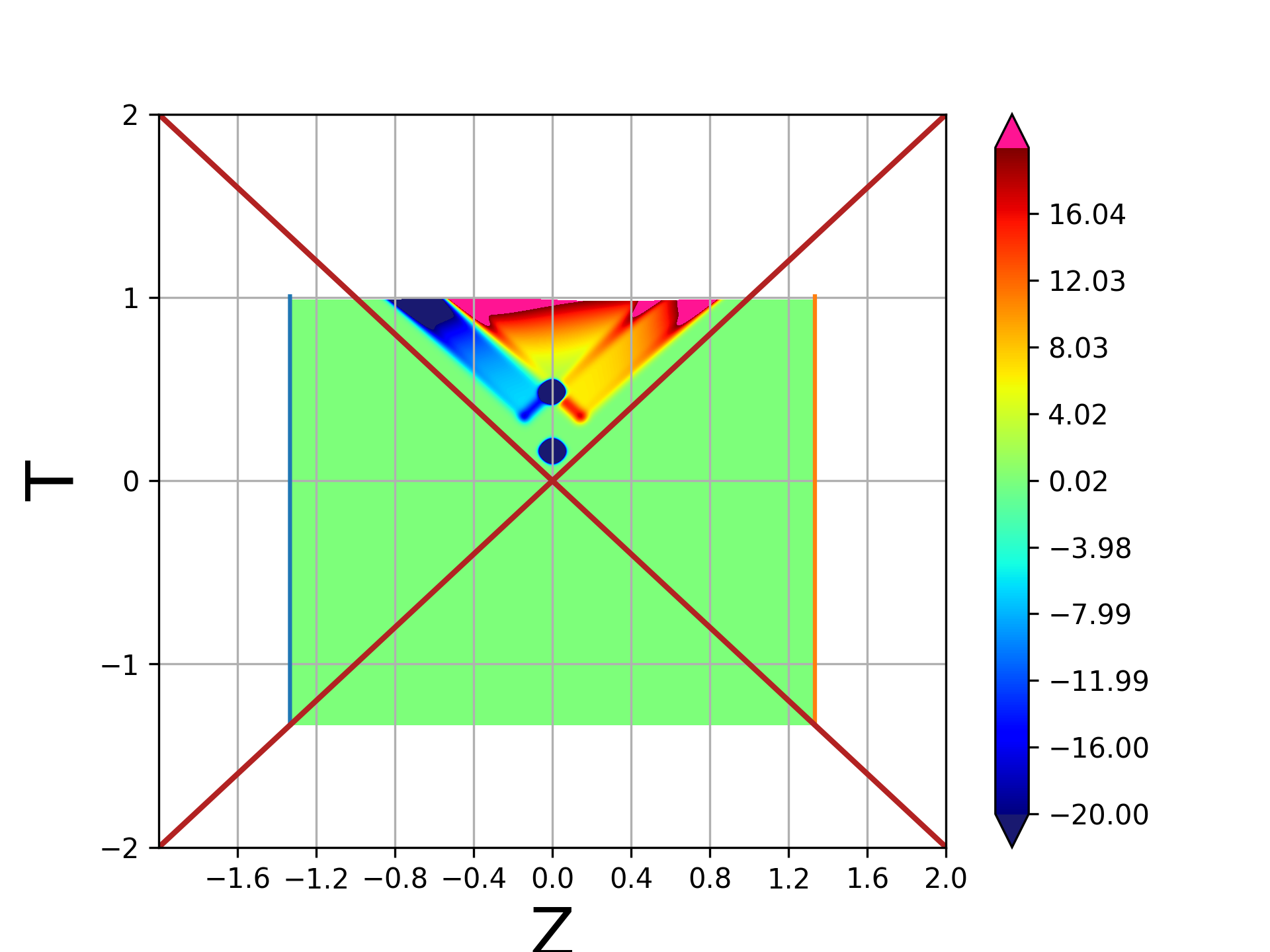}
        \caption{$I_2$}
        \label{fig:007502505_I2}
    \end{subfigure}
    \caption{Conformal diagram contour plots showing the Weyl invariants $I_1$ and $I_2$ over the computational domain using boundary conditions Eq.~\eqref{eq:DoubleGWs} with $a=32$, $\gamma = 0.75$, $\alpha = 0.25$ and $\beta = 0.5$.}\label{fig:GWGW007502505Contours}
\end{figure}

Varying the polarisation to yield a non-zero $I_2$ shows a complicated variety of different scenarios involving $I_1$ and $I_2$ diverging to either positive or negative infinity, even in the case when $\beta = \gamma$. For example, Fig.~\ref{fig:GWGW007502505Contours} shows contour plots of the Weyl invariants $I_1$ and $I_2$ for $\gamma = 0.25$, $\alpha = 0.75$ and $\beta = 0.5$.  Considering the long-term behaviour, $I_1$ is negative in a greater portion of the scattering region, and initially peaks on the left side of the domain and troughs on the right. Meanwhile, $I_2$ initially troughs, and the collision yields a negative $I_2$ which propagates with $\Psi_0$ and a positive $I_2$ which propagates with $\Psi_4$.

The values of $I_1$ and $I_2$ take on four distinct values at each of the four unique points of collisions between incident wave pulses in the scattering region.  That is, when the first two pulses collide, when the first $\Psi_0$ pulse collides with the second $\Psi_4$ pulse and vice versa, and finally when the second pulses collide.  A positive peak in $I_1$ is produced when a pair of colliding pulses are in phase, and a negative peak in $I_1$ is produced when this pair is out of phase.  The value of $I_1$ here is an extremum when the pulses are exactly in phase or exactly out of phase.  The real part of the curvature generated by each pulse cancels out when the colliding pulses have a phase difference of 0.25. 
 Meanwhile, the imaginary part of the curvature, $I_2$, of a colliding pair of pulses cancels out when the pulses have a phase difference of 0.5. The value of $I_2$ takes on a maximum value when the polarisation of the $\Psi_0$ pulse is 0.25 less than the polarisation of the $\Psi_4$ pulse, and takes on a minimum value when the polarisations are vice versa.  Extremum values of $I_1$ and $I_2$ which propagate with the incident waves are generated when the second set of pulse collisions yield zero real or imaginary curvature respectively, as shown in Fig.~\ref{fig:007502505_I2}. 

The of the new and interesting features detailed above, two in particular stand out. Firstly, there exists polarisations whereby the scattering process yields very small (relative to other choices of the polarisations) Weyl scalar curvature invariants, a feature that is not found until more than one wave is generated from each boundary. Secondly, there are choices of polarisation that generate large (relative to other choices of the polarisations) Weyl scalar curvature invariants that propagate at the same speed and with the gravitational waves. This indicates that after scattering, the waves are causing substantially larger curvature deformations.

\subsection{Collisions of electromagnetic and gravitational waves}

\subsubsection{Smoothed shock electromagnetic profile}
A solution of Griffiths \cite{Griffiths1975} describes the scattering of a plane gravitational impulsive wave and a plane electromagnetic step function wave. The electromagnetic wave is partially reflected during scattering but the gravitational wave is not, while the scattering region contains a curvature singularity. We replicate this system but with smoothed impulsive and shock boundary conditions

\begin{gather} 
	\phi_0(v, z_r) = \Theta_N(h,v), \qquad
    \Psi_4(u, z_l) = \delta_N(a,u), \nonumber \\
    \Psi_0(v, z_r) = \phi_2(u, z_l) = 0, \qquad
    u(t, z_l) = v(t, z_r) = \frac{t}{\sqrt{2}}.  \label{eq:GWEMWSmoothedShock}
\end{gather}

\begin{figure}[H]
    \centering
    \begin{subfigure}[t]{0.75\textwidth}
        \centering
        \includegraphics[height=7.5cm]{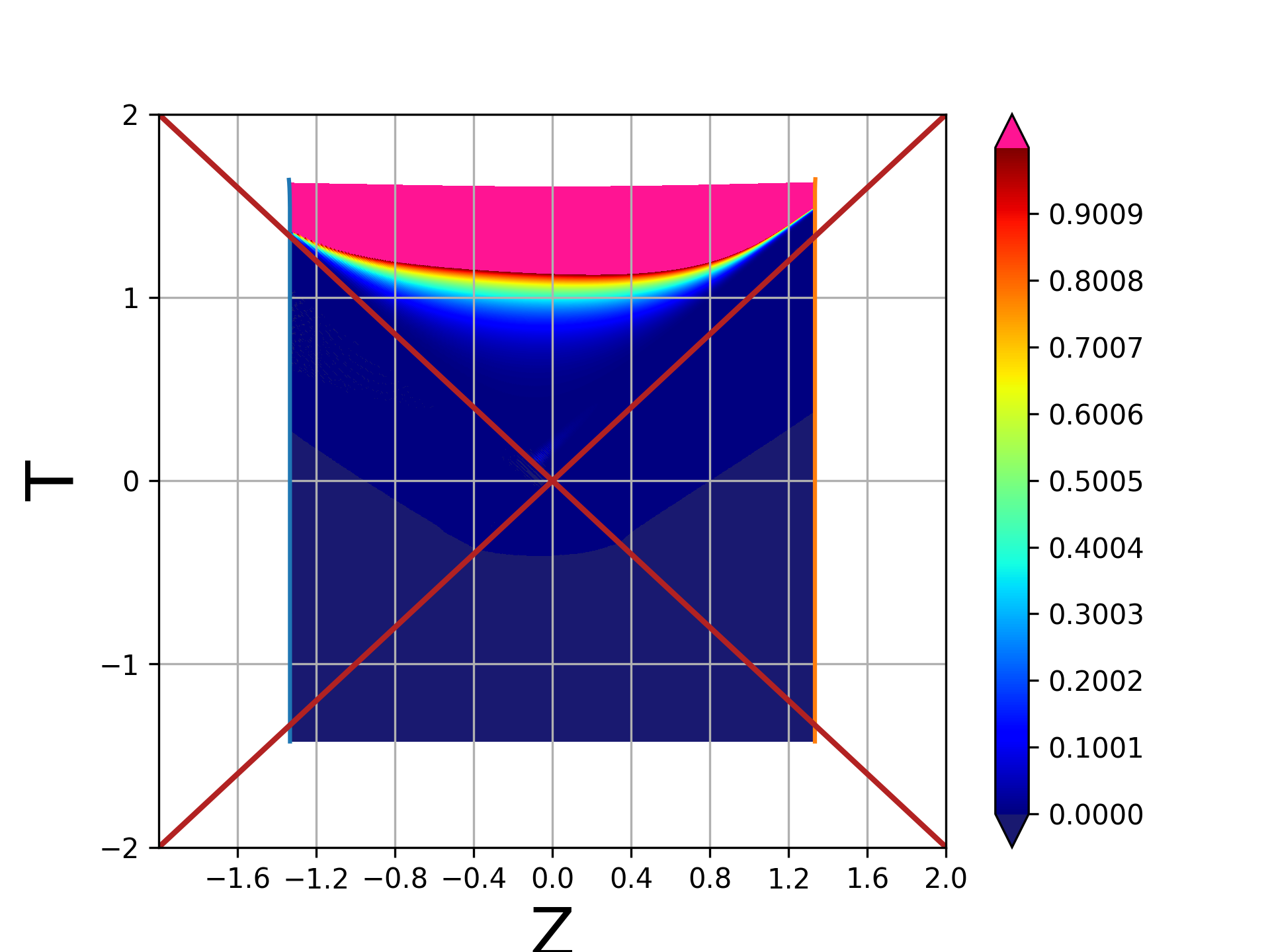}
    \end{subfigure}%
    \caption{Conformal diagram contour plots showing the Weyl invariant $I_1$ over the computational domain using boundary conditions Eq.~\eqref{eq:GWEMWSmoothedShock} with $a=32$ and $h = 800$.}\label{fig:GWEMWShock00Contours}
\end{figure}

Fig.~\ref{fig:GWEMWShock00Contours} exhibits the resulting curvature singularity generated by these boundary conditions.  However,it is found that generation of the electromagnetic wave $\phi_2$ does not occur. This indicates that this feature is a special case given by the idealized boundary conditions of the exact solution, and would not occur in more realistic circumstances.

\subsubsection{Smoothed impulsive electromagnetic profile}
Here we go beyond the literature of exact solutions and consider an electromagnetic wave profile that is a smoothed impulse.  We consider the following boundary conditions,

\begin{gather} 
	\phi_0(v, z_r) = \frac{1}{10}\delta_N(a,v), \qquad 
    \Psi_4(u, z_l) = \delta_N(a,u) \nonumber \\
    \phi_2(u, z_l) = \Psi_0(v, z_r) = 0, \qquad
    u(t, z_l) = v(t, z_r) = \frac{t}{\sqrt{2}}.  \label{eq:GWEMWSmoothedImpulse}
\end{gather}

\begin{figure}[H]
    \centering
    \begin{subfigure}[t]{0.45\textwidth}
        \centering
        \includegraphics[height=5.5cm]{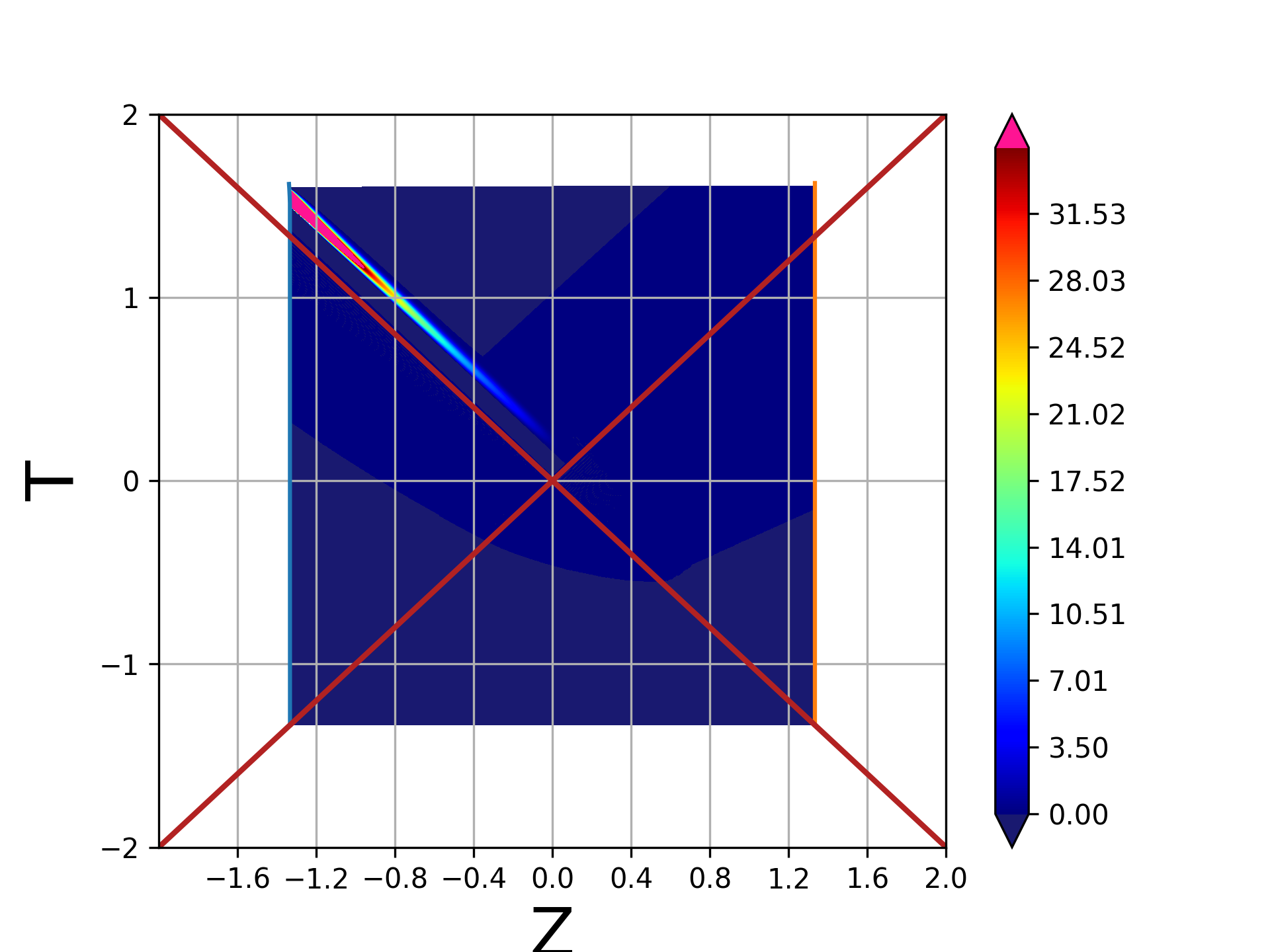}
        \caption{$\Psi_0$}
        \label{fig:psi0_GE}
    \end{subfigure}%
    ~ 
    \begin{subfigure}[t]{0.45\textwidth}
        \centering
        \includegraphics[height=5.5cm]{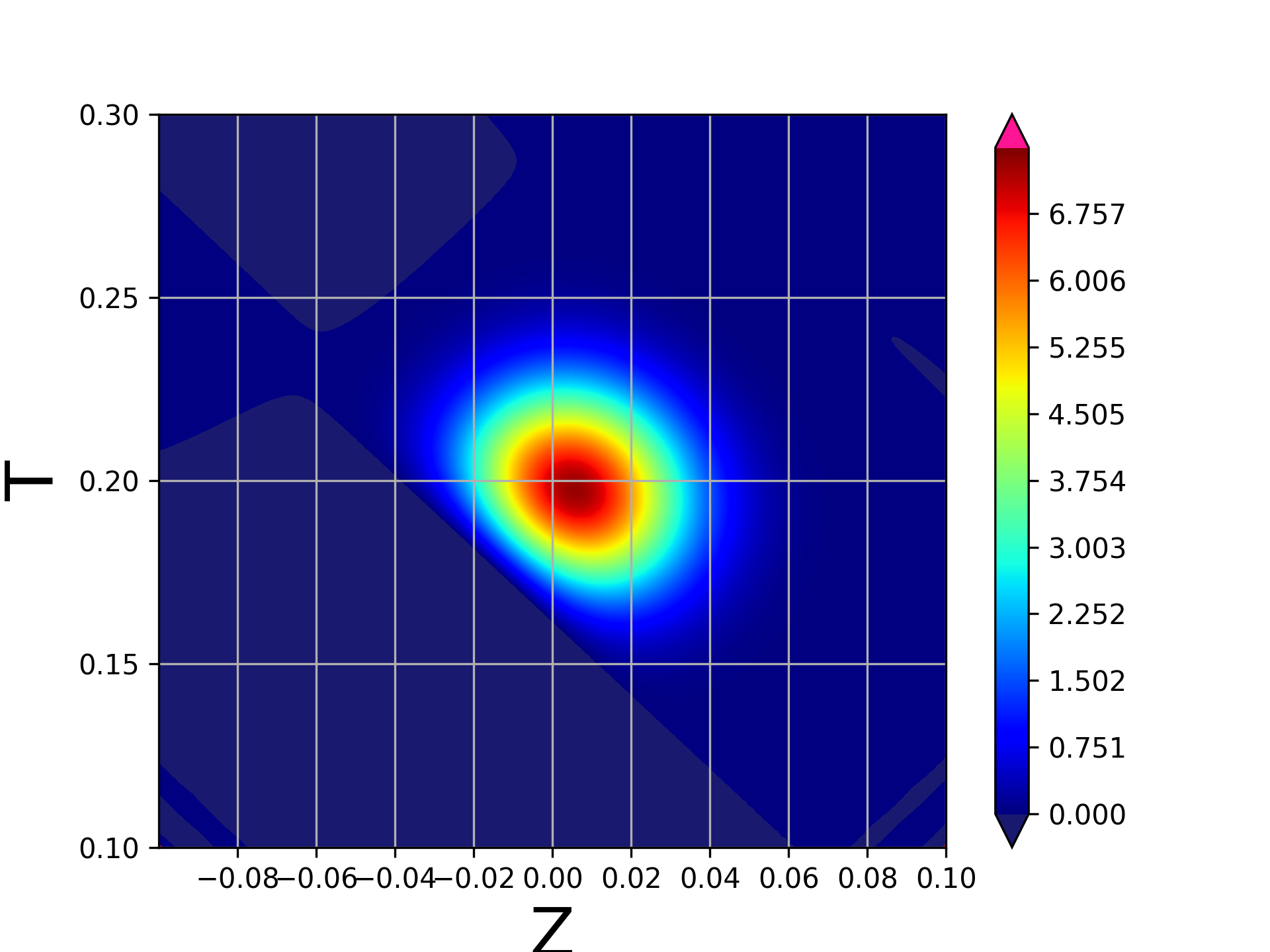}
        \caption{$I_1$}
        \label{fig:GWEMWImpulse00Contours}
    \end{subfigure}
    \caption{Conformal diagram contour plots showing $\Psi_0$ and $I_1$ over the computational domain using boundary conditions Eq.~\eqref{eq:GWEMWSmoothedImpulse} with $a=32$.}
\end{figure}

Unlike the previous case, we observe that the current case generates a localized positive $I_1$ in the scattering region; compare Figs~\ref{fig:GWEMWShock00Contours} and \ref{fig:GWEMWImpulse00Contours}. Further, Fig.~\ref{fig:psi0_GE} shows that a gravitational wave, rather than an electromagnetic wave, is generated as a result of the collision.  

It is found that the relative polarisation of the gravitational and electromagnetic waves does not affect the Weyl invariants, with $I_2$ once again remaining zero in the scattering region. However, for both the current and previous cases, the relative polarisations of the gravitational and electromagnetic waves does affect the polarisations of the original and generated gravitational waves in the scattering region.  The polarisation of the active electromagnetic component $\phi_2$ remains unchanged.

\begin{table}[!h]
    \begin{center}
        \begin{tabular}{|c|c|c|c|c|} 
             \hline
             {} & $\Psi_0$ & $\Psi_4$ & $\phi_0$ & $\phi_2$ \\ [0.5ex] 
             \hline\hline
             $Pol_1$ & $\frac{1}{10}[(5-10\alpha)\mod 10]$ & $\alpha$ & 0 & $\alpha$ \\ 
             \hline
             $Pol_2$ & $\frac{1}{10}[10(1-\alpha)\mod 10]$ & $\frac{1}{10}[(5+10\alpha)\mod 10]$ & 0 & $\alpha$ \\ 
             \hline
        \end{tabular}
    \end{center}
    \caption{The polarisations $Pol_1,Pol_2\in[0,1)$ of $\Psi_0$, $\Psi_4$, $\phi_0$ and $\phi_2$ in the scattering region, written in terms of the initial polarisation $\alpha$ of $\phi_2$. The first row is immediately after collision while the second is when this changes and in the remainder of the scattering region.}
    \label{tab:GE_pols}
\end{table}

\subsubsection{Gravitational spotlights}
In this section we look at the behaviour of colliding electromagnetic and gravitational waves where the waves propagate together as pairs. Namely, we look at the following boundary conditions 

\begin{gather} 
	\phi_0(v, z_r) = \frac{1}{10}e^{2\pi i \beta}\delta_N(a,v), \qquad
    \phi_2(u, z_l) = \frac{1}{10}e^{2\pi i \alpha}\delta_N(a,u), \nonumber \\
    \Psi_0(v, z_r) = e^{2\pi i \beta}\delta_N(a,v), \qquad
    \Psi_4(u, z_l) = e^{2\pi i \alpha}\delta_N(a,u), \nonumber \\
    u(t, z_l) = v(t, z_r) = \frac{t}{\sqrt{2}},  \label{eq:GWEMW_GWEMW_Bump_BCs}
\end{gather}

where we keep the polarisation of $\phi_0, \Psi_0$ and $\phi_2,\Psi_4$ aligned to reduce the parameter space of our investigation.

We find that in this case the solution after scattering is a function of both the polarisations $\alpha$ and $\beta$, not just the relative polarisation. Further, the change of polarisation during scattering does not have a simple relationship as we have found previously. As in the case of colliding smoothed impulsive electromagnetic waves, the scattered gravitational waves pick up two distinct polarisations. Further, a new feature is found that is absent in the collision of two gravitational or two electromagnetic waves. Namely, the Weyl invariants obtains a profile that analogises the scattered gravitational wave profile, albeit with a different polarisation, and propagates with the gravitational wave. This was also found in Sec.~\ref{sec:collisions_GWs} with collisions of pairs of gravitational waves. Previously, the Weyl invariants changed via a smoothed step function in the scattering region. It is found that, as expected, the Weyl invariants diverge. However, it is noted that whether they diverge to positive or negative infinity is spatially dependent. There exists a lightcone which bifurcates the positive and negative divergence.

\begin{figure}[H]
    \centering
    \begin{subfigure}[t]{0.45\textwidth}
        \centering
        \includegraphics[height=5.5cm]{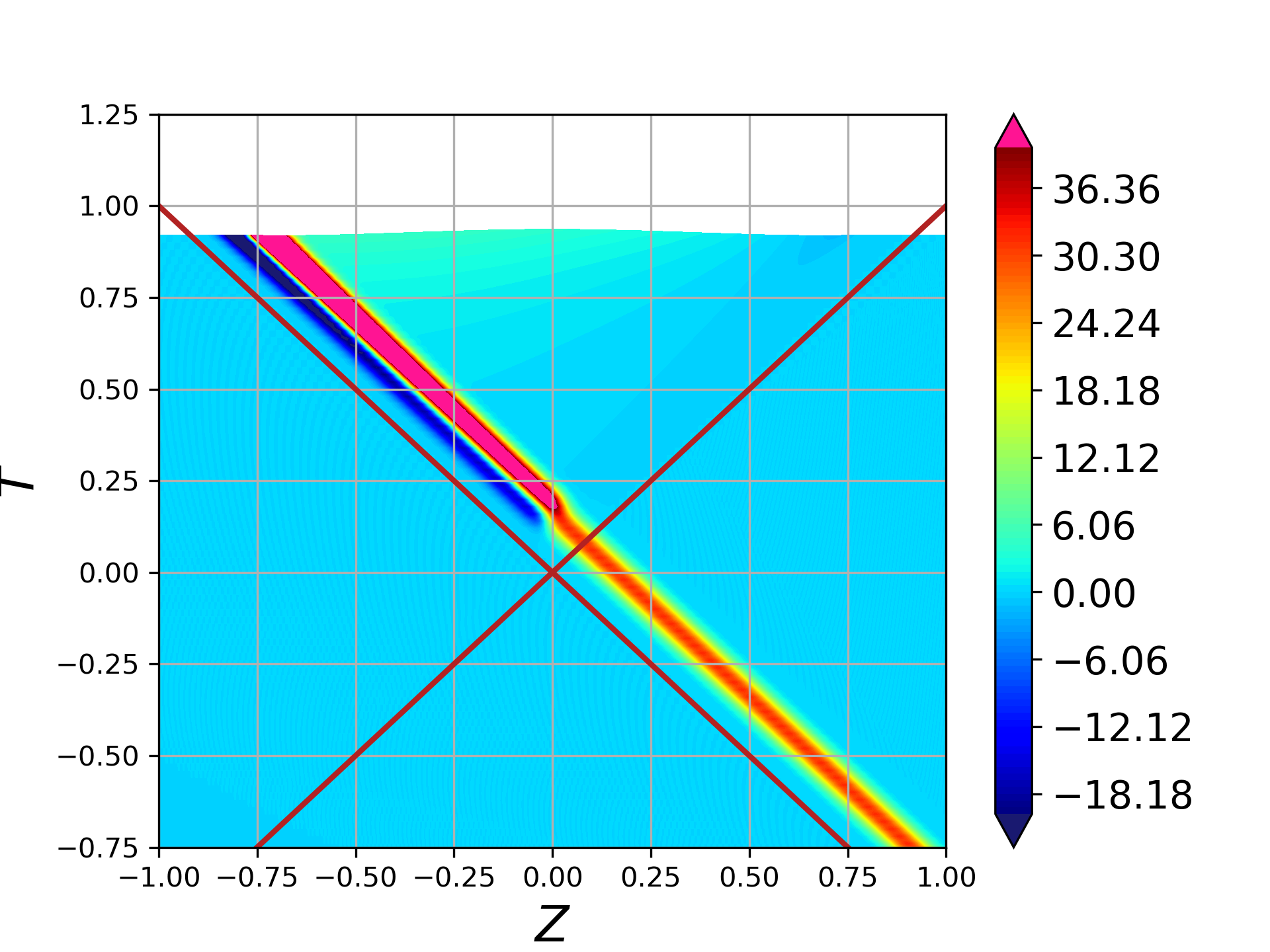}
        \caption{$\Psi_0$}
    \end{subfigure}%
    ~ 
    \begin{subfigure}[t]{0.45\textwidth}
        \centering
        \includegraphics[height=5.5cm]{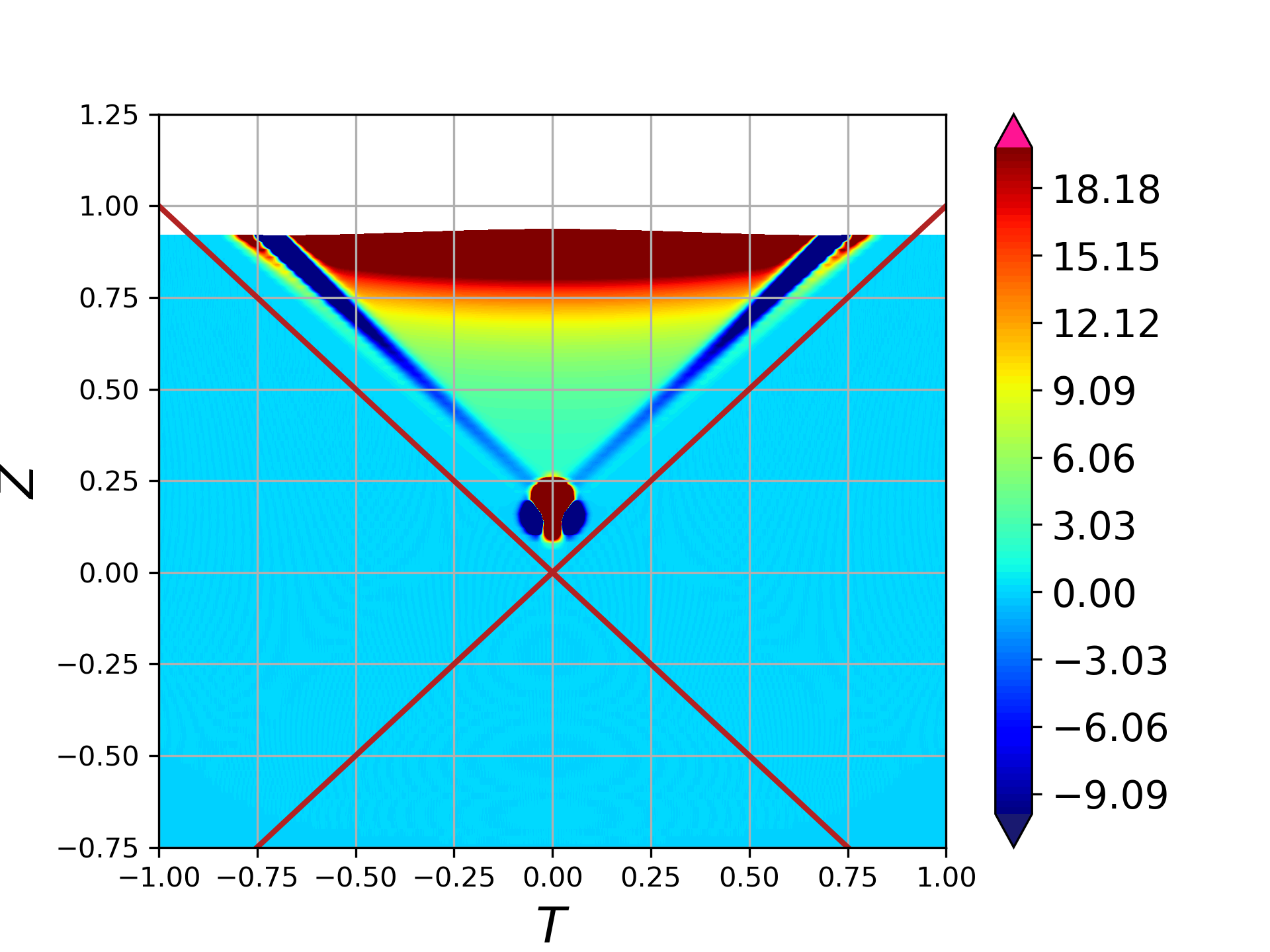}
        \caption{$I_1$}
    \end{subfigure}
    \caption{Conformal diagram contour plots showing $\Psi_0$ and $I_1$ over the computational domain using Eq.~\eqref{eq:GWEMW_GWEMW_Bump_BCs} with $a=32$ and $\alpha=\beta=0$.}\label{fig:I1_psi0re_spotlight_contours}
\end{figure}

\section{Toward an application to astrophysics}\label{sec:astro}
Plane waves are often used as an approximation to a spherical wave far from the source. Close to the source however, they can be used to approximate a small piece of the spherical wave. With the choice $z_l=-2,\;z_r=2$ and $A=4$ we find that in Minkowski space-time, the proper length of our computational domain is $1m$. Non-linear scattering close to the source may affect and change the asymptotic gravitational or electromagnetic wave even in the linear regime. In this section, we take the waves' profiles to be sinusoidal functions without compact support, rather than having a single `bump' as we did in preceding sections. We study the non-linear scattering between gravitational and electromagnetic waves and highlight what potentially observable effects this process has on the electromagnetic wave.

The astrophysical sites of high-amplitude gravitational waves are the mergers of compact objects. While gravitational waves may be observable long before the merger itself, the highest strains will occur at the time of the merger. Mergers occur at a rate of order $30\ \mathrm{Gpc}^{-3}\ \mathrm{yr}^{-1}$ \cite{abbott2023population,callister2024} in the LIGO band, corresponding to masses from a few to a few 10s of solar masses. Mergers of more massive black holes detectable in LISA or with pulsar timing arrays will be even rarer per unit volume in the Universe. We therefore expect that nearly all mergers we could use to observe the interactions between electromagnetic and gravitational waves will occur at cosmological distances. The electromagnetic source that illuminates the merger needs to be bright enough to be observed, compact enough that a non-neglible fraction of the  light we observe from it passes through the merger region, and if it is not local to the merger, common enough on the sky to have a chance of illuminating a merger. We discuss plausible candidates below following an enumeration of some of the effects on the electromagnetic wave. 

\subsection{Strain, time delay and frequency shift}\label{sec:physicalparams}
In order to make links to physically realistic processes, we need to compute quantities that have a direct relevance. 

In \cite{frauendiener2021can} it was shown that with the gauge choice $B=0$, which is what is used throughout this paper, our gauge is in fact the Gau\ss\;gauge. This means that curves represented by fixed spatial coordinates are timelike metric geodesics, and our time parameter is the proper time along them. This is ideal for computing the strain induced by a gravitational wave on free-falling observers. We do this by picking two antipodal points on the unit circle within the planes of symmetry and measure their strain through
\begin{equation}
    h = \frac{\delta L}{L}, \qquad
    L = \int_\gamma\sqrt{-g(\dot\gamma,\dot\gamma})\,\text{d}s,
\end{equation}
where $L$ is the proper distance between two points on the plane at a given time, $\delta L = L - L_0$ is the difference between $L$ at a given time and at the initial time, $\gamma=(\lambda,m\lambda)$ is the line connecting the two points and $\dot\gamma$ is the tangent vector with respect to some parameter along the curve. Parametrizing the curve using the radius and writing the induced metric on the planes of symmetry in polar coordinates, one finds
\begin{equation}
    L = 2\sqrt{2}\left(
    \frac{\Big{(}\sin(\theta_0)\xi - \cos(\theta_0)\eta\Big{)}\Big{(}\cos(\theta_0)\bar{\eta} - \sin(\theta_0)\bar{\xi}\Big{)}}
    {(\xi\bar{\eta} - \bar{\xi}\eta)^2}
    \right)^{1/2},
\end{equation}
for a given angle $\theta_0$. The strain induced by the gravitational waves gives us physical, unitless information regarding the curvature induced by scattering (through its affect on neighbouring free-falling observers) and hence can be used to compare to other physical processes including what processes emit gravitational waves that induce such a strain.

A number of physical systems exhibit interactions between gravitational and electromagnetic waves.  A \textit{pulsar} is a rotating neutron star which emits electromagnetic radiation, typically in the radio spectrum.  Due to the extremely regular rotational period of the pulsar, this radiation is detected as regular \textit{pulses} by an Earth-based detector. Many pulsars together constitute a \textit{pulsar timing array} (PTA) \cite{Background}.  Variables such as the intensity, shape and time of arrival (TOA) of the pulses when measured on Earth are strongly correlated to the physical properties of the pulsar.  One can model these physical properties, accounting for effects caused by the detector itself and the propagation of the wave through the interstellar medium.  \textit{Timing residuals} between the predicted and observed TOAs then indicate the presence of other effects which have not yet been accounted for, such as the propagation of a gravitational wave through the path of the radio pulse \cite{Pulsars}.  When a pulsar photon interacts with a gravitational wave during its propagation towards Earth, it will experience a change in its proper length \cite{Background}.  Studies of large PTA samples have also yielded evidence for a gravitational wave background with a frequency range of 2-30 nHz \cite{Background}.

We can measure time delay of an electromagnetic wave due to scattering in the following way: We generate an electromagnetic wave from the right boundary at $t=0$ and measure the proper time of the left boundary between $t=0$ and the time when the maximum/minimum of the first peak/trough reaches it. This is simple in our gauge, as the proper time along constant spatial curves is just the coordinate time. We can then perform the same calculation, but now for the case of a gravitational wave being generated from the left boundary as well. The difference between the time taken for the electromagnetic wave to reach the left boundary in this case and the previous case gives us our time delay. It is important to note however, that if one considers the space-time produced with exact Minkowski initial and boundary data, but modified to have a single gravitational wave generated from a boundary, the proper length of the spatial domain will increase over time, since $A$ decreases. Then one must be careful to separate whether changes in the electromagnetic wave's TOA are due to the boundaries bending or the scattering process in between. It is found that for the situations we consider in the subsequent sections, the scattering process dominates changes to the TOA. To exemplify this, consider the boundary conditions Eq.~\eqref{eq:GWEMW_Train_BCs} with frequencies $f_\phi = f_\Psi = 10$ and gravitational wave amplitude $a=32$. If we use these boundary conditions but with the alteration that $\phi_0(v,z_r)=0$, the increase in proper distance between the boundaries once the gravitational wave reaches the opposite boundary is of the order $O(10^{-11})$. However, with the addition of an electromagnetic wave as in Eq.~\eqref{eq:GWEMW_Train_BCs}, the proper distance between boundaries at the same time has now increased by an order of $O(10^{-5})$. Thus, it is clear that in the situations outlined below, the change to the TOA is dominated by the scattering process rather than the bending of the boundaries before scattering. 


In the non-linear scattering between gravitational and electromagnetic waves, a time-dependent frequency shift in the electromagnetic wave is also expected due to the metric dynamically changing a non-trivial amount with the presence of strong Weyl curvature. This will impinge on the time taken for the electromagnetic wave to pass by a free-falling observer given by $z=$ constant, hence a shift in frequency should be observed. Note that there is no reason to expect, especially when continuously scattered by gravitational waves, that this frequency shift should be constant during scattering, and will no doubt be temporally and spatially dependent during the scattering process.

In the proceeding sections, we present results showcasing the strain, time delay and frequency shift of the electromagnetic wave during scattering with a gravitational wave and the prominent features. A manageable frequency of $10$ m$^{-1}$ is taken for both the electromagnetic and gravitational waves to ascertain the behaviour of strain, time delay and frequency shift. Although this is a good choice for the electromagnetic wave, corresponding to frequencies routinely observed in the radio, this would be a higher frequency for the gravitational wave than what a compact binary merger would produce for example. In Sec.~\ref{sec:morerealistic}, we investigate the trend as we take lower and lower gravitational wave frequencies, heading toward those of physical relevance. We note that constraint satisfaction always remains smaller than the presented results.

\subsection{Collision of a gravitational wave with an electromagnetic wave}
\begin{figure}[H]
    \centering
    \begin{subfigure}[t]{0.45\textwidth}
        \centering
        \includegraphics[height=5.5cm]{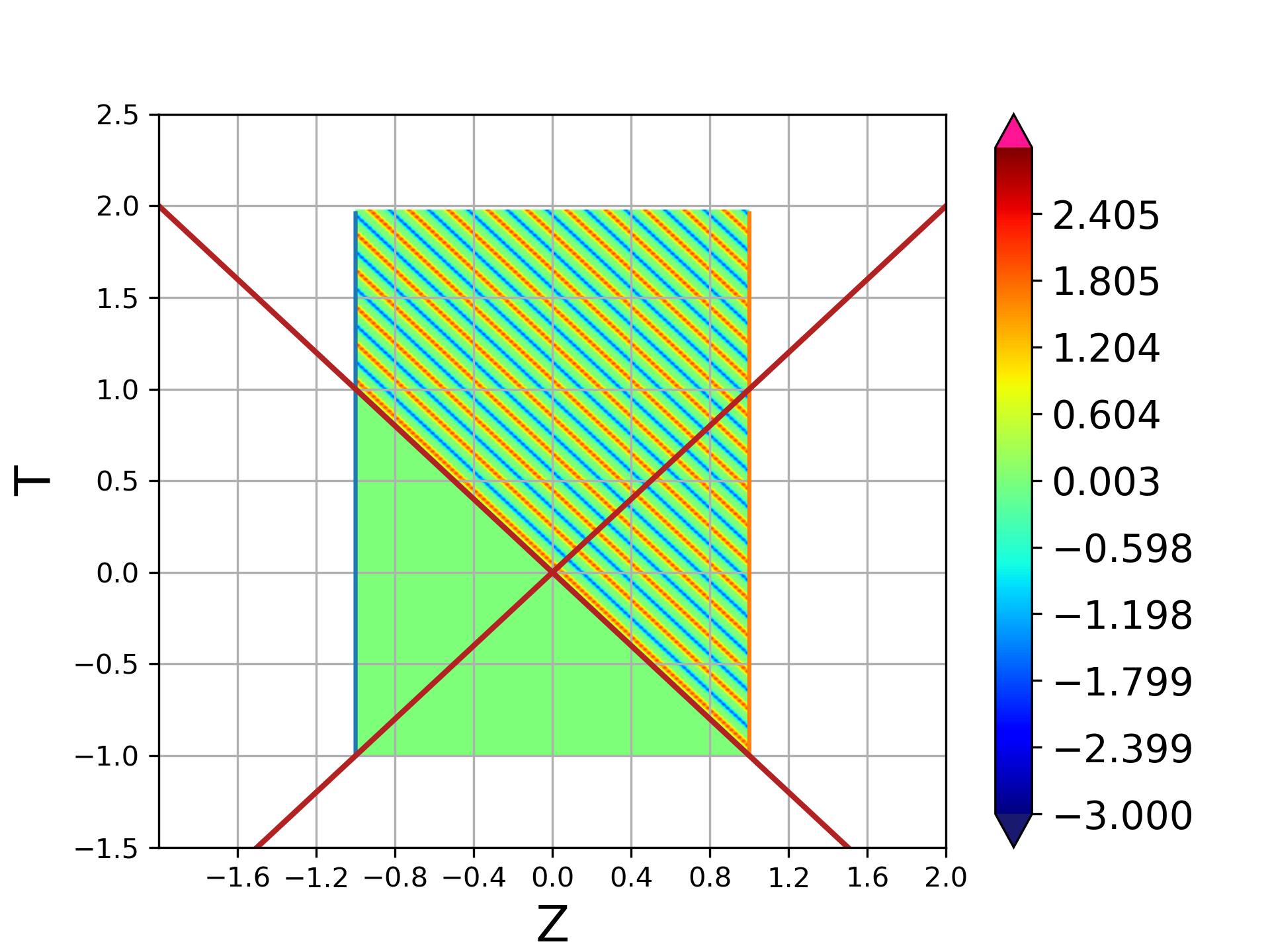}
        \caption{$\phi_0$}
    \end{subfigure}%
    ~ 
    \begin{subfigure}[t]{0.45\textwidth}
        \centering
        \includegraphics[height=5.5cm]{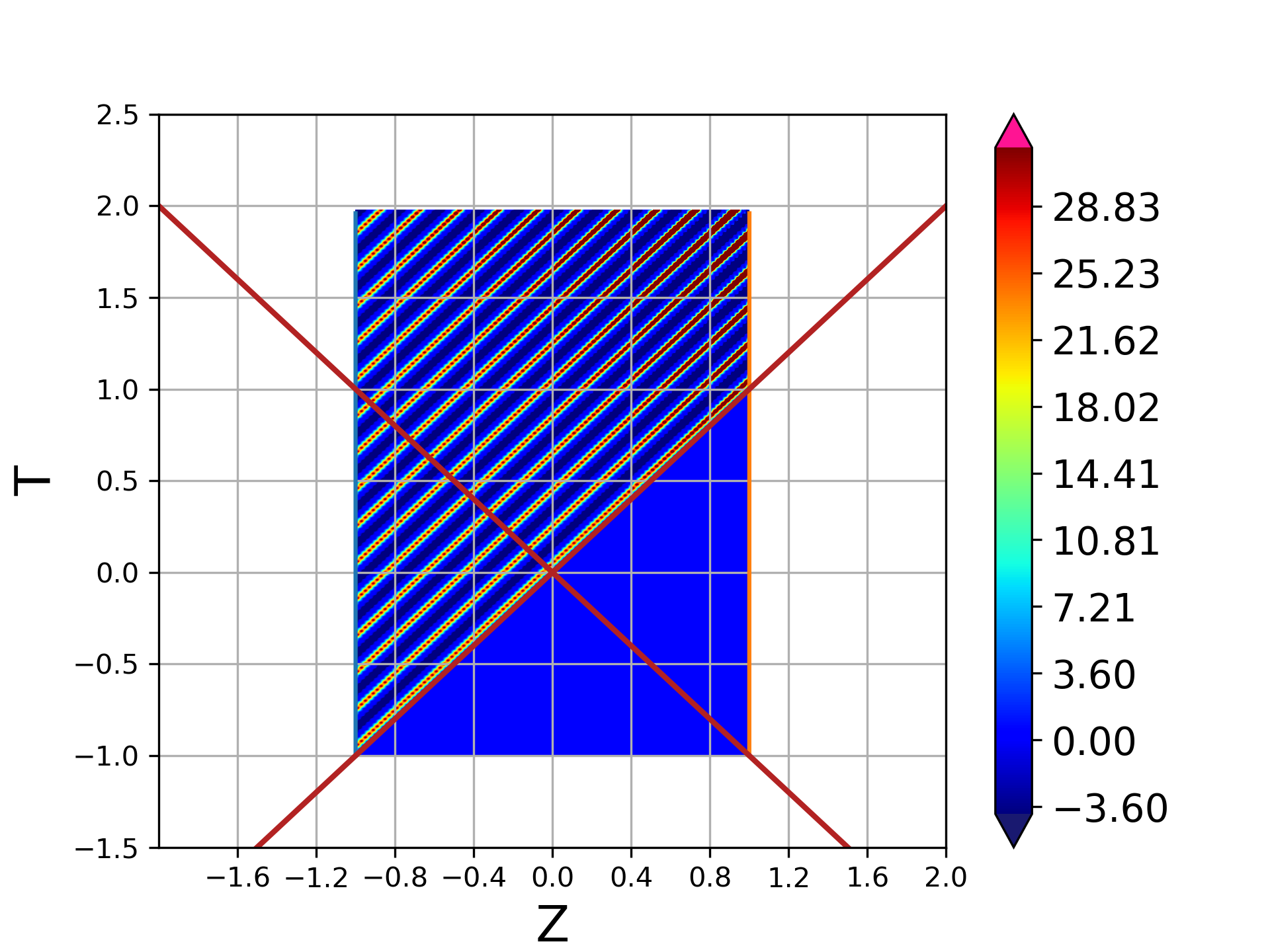}
        \caption{$\Psi_4$}
    \end{subfigure}
    \caption{Conformal diagram contour plots showing the electromagnetic wave $\phi_0$ and gravitational wave $\Psi_4$ over the computational domain using Eq.~\eqref{eq:GWEMW_Train_BCs} with $a=32$ and $f_{\phi}, f_{\Psi}=10$.}\label{fig:ContourPlots_TrainEMWGW}
\end{figure}

Introducing a new wave profile

\begin{equation}
    \tilde{\delta}_N(a,f,t) := 
    \begin{cases} 
	    a\sin^7(2\pi f t) & 0\leq t\leq\frac{1}{f} \\
	    0 & \text{otherwise}
	\end{cases},
\end{equation}
where $f$ is the frequency in m$^{-1}$, the boundary conditions are taken to be
\begin{gather} 
    \phi_0(t, z_r) = \tilde{\delta}_N(2,f_{\phi},t), \qquad
    \phi_2(t, z_l) = 0, \nonumber \\
    \Psi_0(t, z_r) = 0, \qquad
    \Psi_4(t, z_l) = \tilde{\delta}_N(a,f_{\Psi},t), \nonumber \\
    u(t, z_l) = v(t, z_r) = \frac{t}{\sqrt{2}}, \label{eq:GWEMW_Train_BCs}
\end{gather}
where $f_{\phi}, f_{\Psi}=10$ m$^{-1}$ is chosen to allow multiple periods to be obtained within the simulation. Note that now we are now writing the waves with respect to the proper time on the boundary $t$ rather than the null coordinates $u$ or $v$ as have been done previously. The situation is shown in Fig.~\ref{fig:ContourPlots_TrainEMWGW}.

Along the timelike geodesic $z=-1/2$, which only has a non-zero $\phi_0$ during the scattering with $\Psi_4$, we compute the dominant frequency of the electromagnetic wave $\phi_2$ by trimming the data between first and last local maximum to ensure periodicity of the data, performing a discrete Fourier transform and finding the dominant frequency in m$^{-1}$. We do this for spatial resolutions of $401,801,1601,3201,6401$ to confirm convergence and take the highest resolution result. We find that there is a frequency shift in the negative direction, which increases with the amplitude of the gravitational wave, as one would expect. This is what one would expect given the discussion in the above section. Tab.~\ref{tab:GW_EMW_Train_Freqshifts} showcases the frequency shift, which has a power law best fit of $|f - f_0| \approx 1.73\times10^{-6}a^{2.22}$, where $f_0=10$ is the initial frequency before scattering. 

\begin{table}[!h]
    \begin{center}
        \begin{tabular}{|c||c|c|c|c|c|} 
             \hline
             GW amplitude $a$ & 16 & 32 & 64 & 128 \\[0.5ex]
             \hline
             Freq. shift in m$^{-1}$. & -0.00179 & -0.00357 & -0.01783 & -0.08323 \\ 
             \hline
        \end{tabular}
    \end{center}
    \caption{The frequency shift of the electromagnetic wave $\phi_0$ from $10$ m$^{-1}$ along the timelike geodesic $z=-1/2$ when scattered by a gravitational wave $\Psi_4$ with an initial frequency of $10$ m$^{-1}$ for a variety of gravitational wave amplitudes $a$. The first six periods are taken.}
    \label{tab:GW_EMW_Train_Freqshifts}
\end{table}

Frequency shifts of this magnitude are directly observable for instance in ALMA \cite{cortes_ALMA_2023}, which can measure frequency channels as narrow as 3.8 kHz at observed frequencies of 110 GHz, yielding a detectable relative frequency shift of $\sim 3\times10^{-8}$. In optical spectroscopy the highest resolution spectrographs can detect more modest frequency shifts of order $10^{-5}$ \cite{vogt1994}. The key questions around whether frequency shifts will be detectable in realistic scenarios are how the frequency shift scales to the much longer wavelengths of most astrophysical gravitational waves, and whether the frequency shift induced can be distinguished from astrophysical effects, particularly those that cause variability in the shape of the spectral feature being observed. 

\begin{figure}[H]
    \centering
    \begin{subfigure}[t]{0.45\textwidth}
        \centering
        \includegraphics[height=5.5cm]{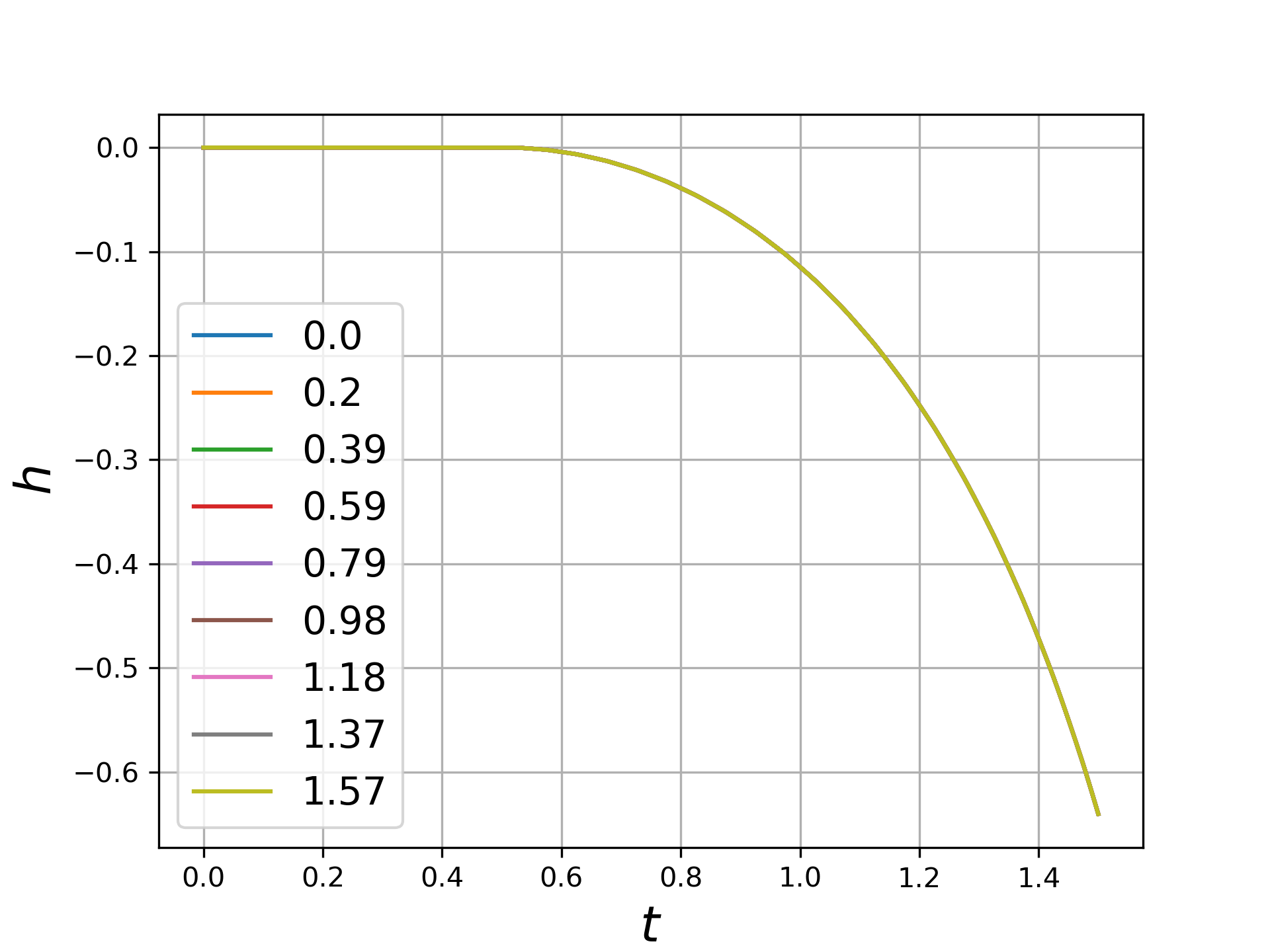}
        \caption{$a=10^{-5}$}
    \end{subfigure}%
    ~ 
    \begin{subfigure}[t]{0.45\textwidth}
        \centering
        \includegraphics[height=5.5cm]{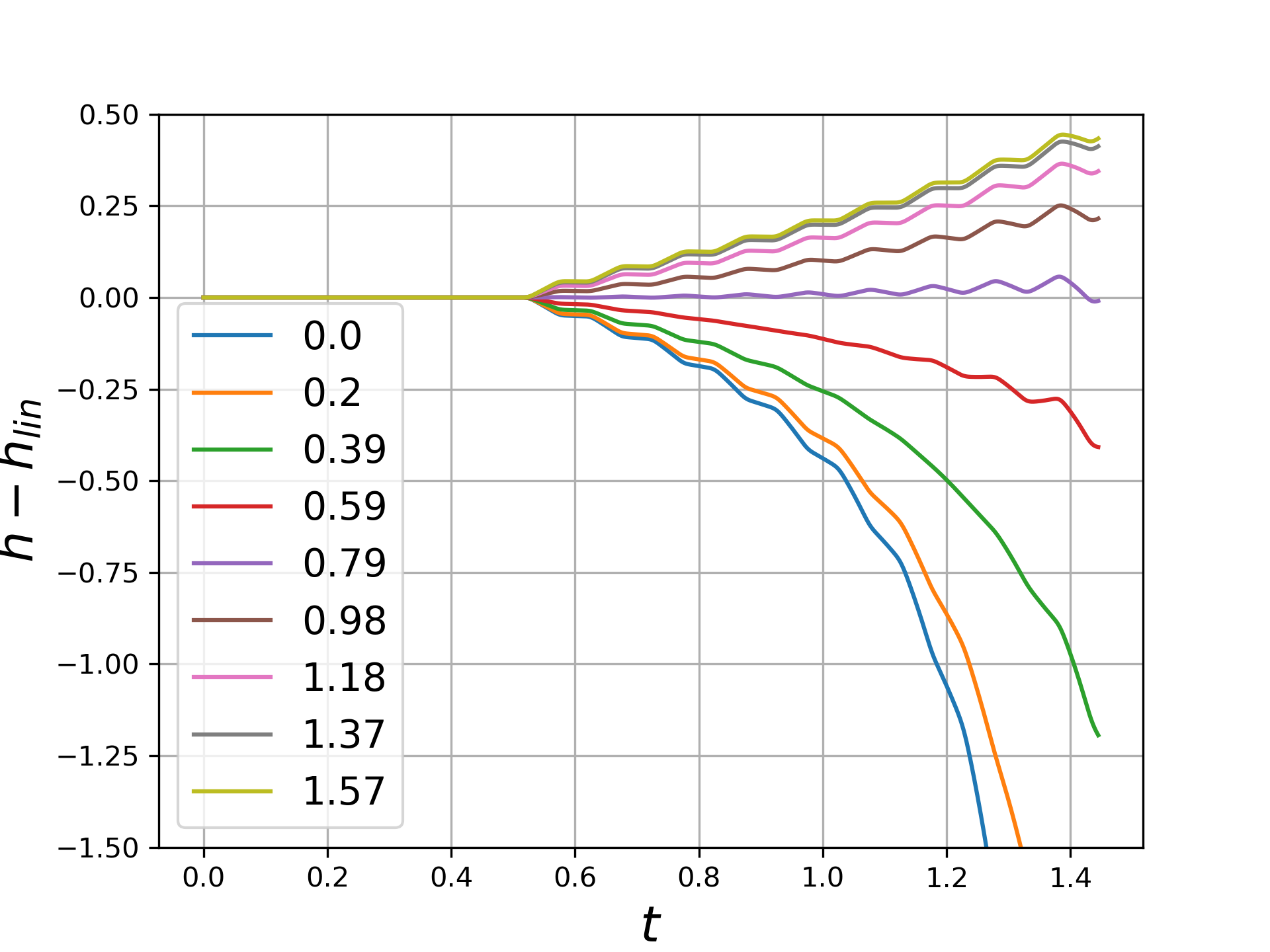}
        \caption{$a=128$ with linear result subtracted}
    \end{subfigure}
    \caption{The strain $h$ computed for various angles between 0 and $\pi/2$ for (a) the linear case of $a=10^{-5}$ and (b) the the non-linear case with $a=128$ and the result of (a) subtracted.}\label{fig:GW_EMW_Strain}
\end{figure}

The strain is shown in Fig.~\ref{fig:GW_EMW_Strain} for both the linear and non-linear case to emphasise the non-linear features and for a variety of angles within the plane of symmetry at $z=0$ over time. The choice $a=128$ is chosen to maximise the non-linear effect. As our gravitational wave polarisation corresponds to a `plus-wave', one can clearly see, especially initially when the space-time first deviates from Minkowski space-time, that the largest positive strain occurs along $\theta=\pi/2$, while the largest negative strain occurs along $\theta=0$, as expected. It is clear that with the non-linear gravitational wave profile $\tilde{\delta}_N(128,10,t)$ the strain becomes of order $O(10^{-1})$. This is certainly possible in the space-time region immediately surrounding a compact binary merger, which would induce extremely large strains.

Another interesting feature is time delay. This is the difference in time taken for the electromagnetic wave to travel from, in our case, from the right boundary to the left boundary, and is the main ingredient in the Pulsar Timing Array's recent results concerning the stochastic gravitational wave background \cite{antoniadis2023second,antoniadis2023second2,antoniadis2023second3}. Without any gravitational wave scattering, this takes a time of $t=1$. The results for the current situation with varying gravitational wave amplitudes $a$ are given in Tab.~\ref{tab:GW_EMW_Train_Freqshifts}.

\begin{table}[!h]
    \begin{center}
        \resizebox{\textwidth}{!}{
        \begin{tabular}{|c||c|c|c|c|c|c|c|} 
             \hline
             Peak number & 2 & 3 & 4 & 5 & 6 & 7 \\ [0.5ex]
             \hline\hline
             Time delay $a=32$ & $4.5531\times10^{-6}$ & $2.235\times10^{-5}$ & $7.189\times10^{-5}$ & $1.787\times10^{-4}$ & $3.808\times10^{-4}$ & $7.357\times10^{-4}$ \\ [0.5ex]
             \hline
             Time delay $a=64$ & $1.862\times10^{-5}$ & $9.490\times10^{-5}$ & $3.095\times10^{-4}$ & $7.802\times10^{-4}$ & $1.682\times10^{-3}$ & - \\ [0.5ex]
             \hline
             Time delay $a=128$ & $1.00\times10^{-4}$ & $5.31\times10^{-4}$ & $1.86\times10^{-3}$ & $5.3\times10^{-2}$ & - & - \\ 
             \hline
        \end{tabular}}
    \end{center}
    \caption{The proper time delay of $\phi_0$ in relativistic units compared against the case with no gravitational wave scattering. Simulations are done with a resolution of 6401 spatial points.}
    \label{tab:GW_EMW_Train_Freqshifts}
\end{table}

To obtain physical units, these time delays need to be divided by $c$, yielding a maximum of $10^{-11}$ seconds. Pulsar timing arrays can achieve typical errors on times of arrival of order $10^{-9}$ seconds \cite{Background}. While these numbers do not seem incredibly far apart, several large effects need to be included to compare them. First, as with the frequency shift, the size of the time delay needs to be scaled to the much longer wavelengths of astrophysical gravitational waves. Second, the pulsars being timed as part of pulsar timing arrays are all in the Milky Way, meaning that the amplitude of the gravitational waves that affect their arrival times is far lower than the non-linear case we are considering. The recent successful detection of a gravitational wave background by PTAs demonstrate that this effect is detectable and will only grow more so with longer time baselines and the monitoring of more pulsars. To probe the non-linear regime requires observing arrival time delays at cosmological distances where the high precision of pulsars are unavailable. This may nonetheless be possible if the delays are long enough.

\subsection{Collision of a gravitational wave with an electromagnetic and gravitational wave}
Now we consider the same situation as above, but with the addition of a gravitational wave propagating with the electromagnetic wave. The boundary conditions are taken to be

\begin{gather} 
    \phi_0(v, z_r) = \tilde{\delta}_N(2,f_{\phi},t), \qquad
    \phi_2(u, z_l) = 0, \nonumber \\
    \Psi_0(v, z_r) = \tilde{\delta}_N(a,f_{\Psi},t), \qquad
    \Psi_4(u, z_l) = \tilde{\delta}_N(a,f_{\Psi},t), \label{eq:GWEMW_Train_BCs2}
\end{gather}
where $f_{\phi}, f_{\Psi} =10$ m$^{-1}$ is chosen to allow multiple periods to be obtained straightforwardly. The situation looks the same as in Fig.~\ref{fig:ContourPlots_TrainEMWGW} but with the addition of the gravitational wave $\Psi_0$ accompanying the electromagnetic wave $\phi_0$.

Performing the same analysis as in the previous section along $z=-1/2$ we find the interesting feature that the frequency is shifted now in the positive direction, see Tab.~\ref{tab:GW_GWEMW_Train_Freqshifts}, and is roughly approximated with $|f - f_0| \approx 2\times10^{-5}a^{2.1238}$, where $f_0=10$ is the frequency before scattering. Notably, the frequency is shift is larger than the previous case by an order of magnitude, which must be due the electromagnetic wave having a gravitational wave counterpart. The geometrical interpretation of a shift to higher frequencies is that the proper length of the spatial domain has been decreasing.

This effect is unlikely to be observable directly because it is rare that a gravitational wave and electromagnetic wave will propagate together. Events like neutron star-neutron star mergers that produce both types of waves will generally produce light later than the gravitational waves owing to the optical depth of the ejecta. The frequency shift could conceivably affect the propagation of photons within the optically-thick ejecta.

\begin{table}[!h]
    \begin{center}
        \begin{tabular}{|c||c|c|c|c|} 
             \hline
             GW amplitude $a$ & 16 & 32 & 64 & 128 \\ [0.5ex]
             \hline
             Freq. shift in m$^{-1}$. & 0.00751 & 0.03261 & 0.13695 & 0.59883 \\
             \hline
        \end{tabular}
    \end{center}
    \caption{The frequency shift of the electromagnetic wave $\phi_0$, which is accompanied by a gravitational wave $\Psi_0$ with the same initial frequency, from $10$ m$^{-1}$ along the timelike geodesic $z=-1/2$ when scattered by a gravitational wave $\Psi_4$ with an initial frequency of $10$ m$^{-1}$ for a variety of gravitational wave amplitudes $a$. The first four periods are taken.}
    \label{tab:GW_GWEMW_Train_Freqshifts}
\end{table}

\begin{figure}[H]
    \centering
    \begin{subfigure}[t]{0.45\textwidth}
        \centering
        \includegraphics[height=5.5cm]{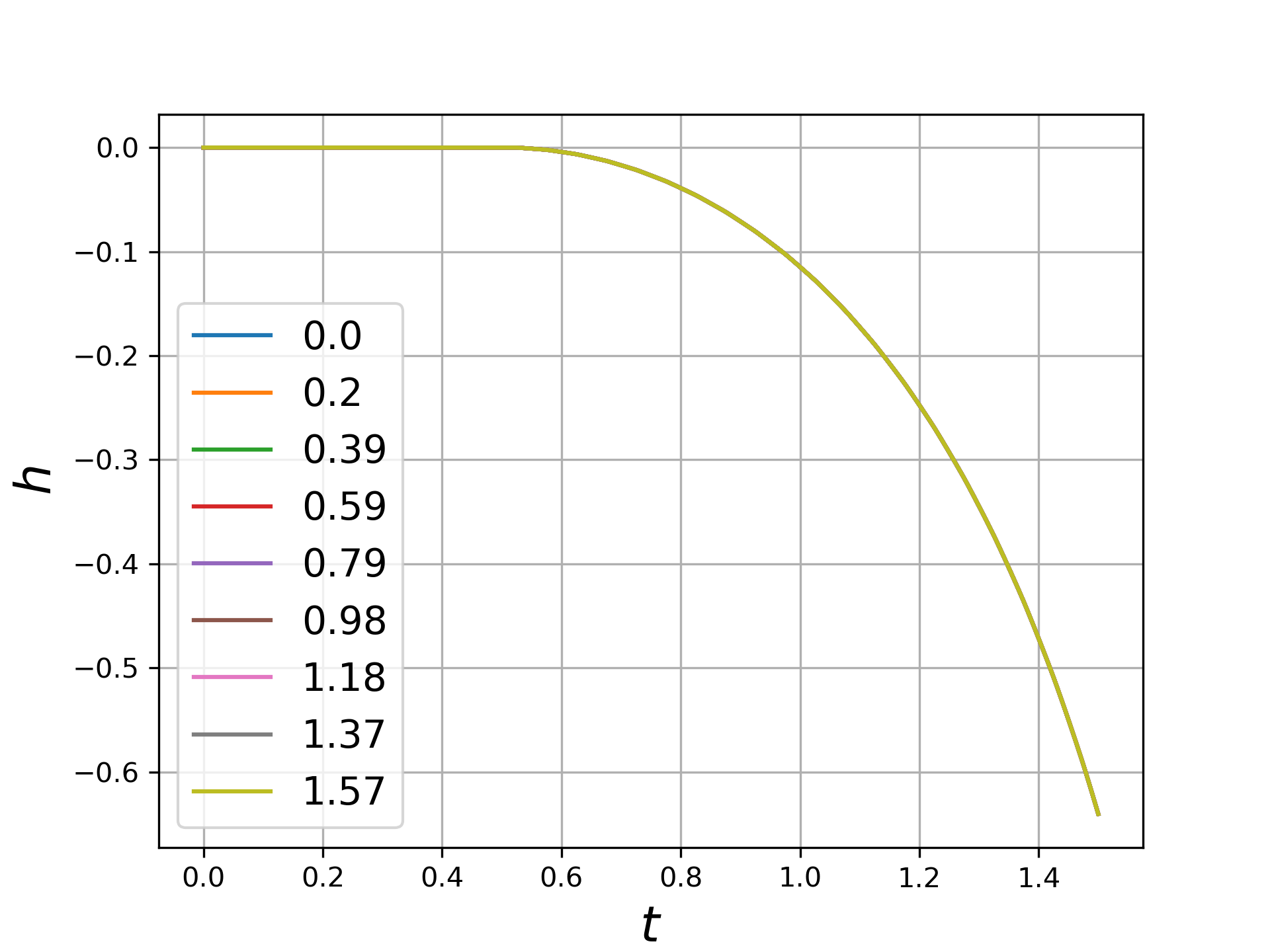}
        \caption{$a=10^{-5}$}
    \end{subfigure}%
    ~  
    \begin{subfigure}[t]{0.45\textwidth}
        \centering
        \includegraphics[height=5.5cm]{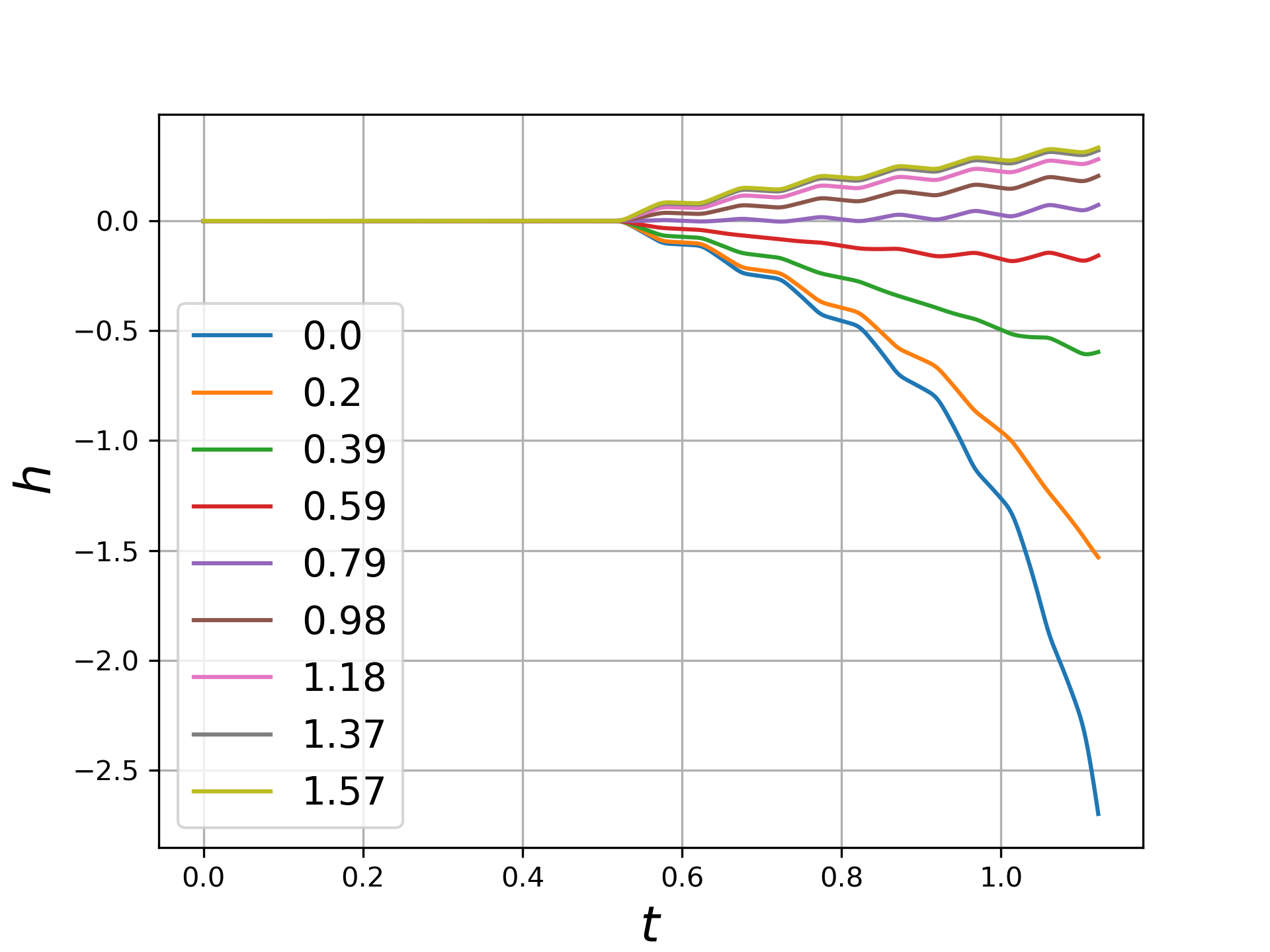}
        \caption{$a=128$ with linear result subtracted}
    \end{subfigure}
    \caption{The strain $h$ computed for various angles between 0 and $\pi/2$ for (a) the linear case of $a=10^{-5}$ and (b) the the non-linear case with $a=128$ and the result of (a) subtracted.}\label{fig:GW_GWEMW_Strain}
\end{figure}

The strain is shown in Fig.~\ref{fig:GW_GWEMW_Strain} for both the linear and non-linear case and for a variety of angles within the plane of symmetry at $z=0$ over time in the same way as the previous section. It is clear that with the non-linear gravitational wave profile $\tilde{\delta}_N(128,10,t)$ the strain becomes of order unity in this case.

The results for the time delay with varying gravitational wave amplitudes $a$ are given in Tab.~\ref{tab:GW_GWEMW_Train_Freqshifts2}. Again, this is roughly an order of magnitude larger than the previous case.

\begin{table}[!h]
    \begin{center}
        \resizebox{\textwidth}{!}{
        \begin{tabular}{|c||c|c|c|c|c|c|} 
             \hline
             Peak number & 2 & 3 & 4 & 5 & 6\\ [0.5ex]
             \hline\hline
             Time delay $a=32$ & $-2.095\times10^{-4}$ & $-7.231\times10^{-4}$ & $-1.566\times10^{-3}$ & $-2.772\times10^{-3}$ & $-4.387\times10^{-3}$  \\ [0.5ex]
             \hline
             Time delay $a=64$ & $-8.71\times10^{-4}$ & $-3.01\times10^{-3}$ & $-0.652\times10^{-3}$ & - & - \\ [0.5ex]
             \hline
             Time delay $a=128$ & $-4.182\times10^{-3}$ & - & - & - & - \\ 
             \hline
        \end{tabular}}
    \end{center}
    \caption{The proper time delay of $\phi_0$ in relativistic units compared against the case with no gravitational wave scattering. Simulations are done with a resolution of 6401 spatial points.}
    \label{tab:GW_GWEMW_Train_Freqshifts2}
\end{table}

It is not surprising that this case, which includes the scattering of two gravitational waves, creates larger curvature changes. This is manifestly seen in the expression for the Weyl scalar curvature invariants for example, which contain a $\Psi_0 \Psi_4$ term. Due to the larger curvature changes in this case, the simulation reaches the future curvature singularity much faster than the previous one.

\subsection{Collisions with more realistic frequencies}\label{sec:morerealistic}
A reasonable upper estimate for the frequency of the strain induced from gravitational waves emitted by a binary black hole merger is of the order $10^2$ Hz. The first detected binary black hole merger \cite{abbott2016observation} had a maximum frequency of 250 Hz for example. On the other hand, the lowest-frequency electromagnetic radiation typically observed by astronomers is of order 10 MHz \cite{lofar} five orders of magnitude higher. These frequencies are much too different to be directly simulated with our current setup. However, we can expand upon work of previous sections that took the gravitational and electromagnetic wave frequencies as $10$ m$^{-1}$ $\approx$ $3\times10^9$ Hz, and decrease the frequency of the gravitational wave. From this, we can deduce the changing behaviour of the frequency shift of the electromagnetic wave and postulate an idea for what should happen in the realistic case, albeit indirectly. We note that a decreasing gravitational wave frequency leads to a decreasing strain frequency as well.

\begin{figure}[H]
    \centering
    \begin{subfigure}[t]{0.5\textwidth}
        \centering
        \includegraphics[height=5.5cm]{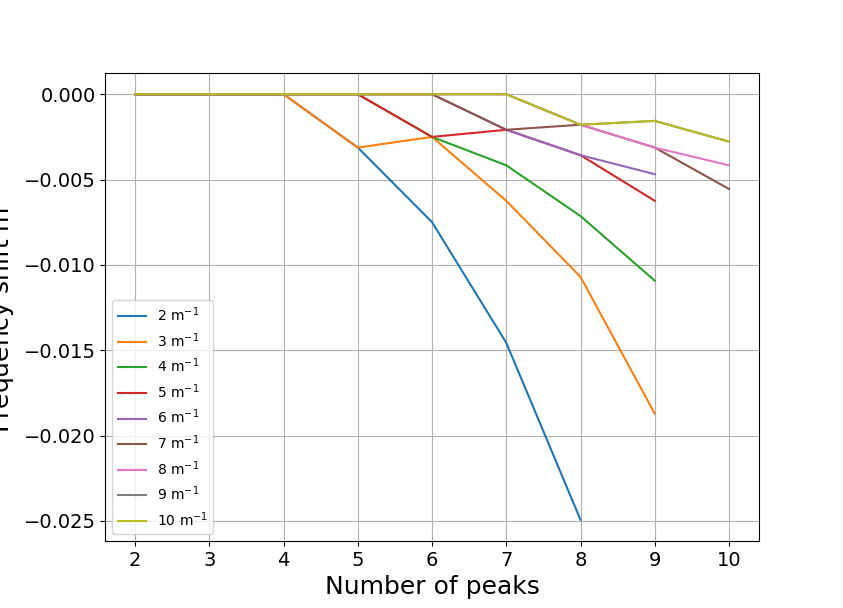}
        \caption{Frequency shift for boundary conditions Eq.~(\ref{eq:GWEMW_Train_BCs}) with $a = 16$, $f_{\phi} = 10$ m$^{-1}$ and $f_{\Psi} = 2, 3, ..., 10$ m$^{-1}$.}
    \end{subfigure}%
    ~ 
    \begin{subfigure}[t]{0.4\textwidth}
        \centering
        \includegraphics[height=5.5cm]{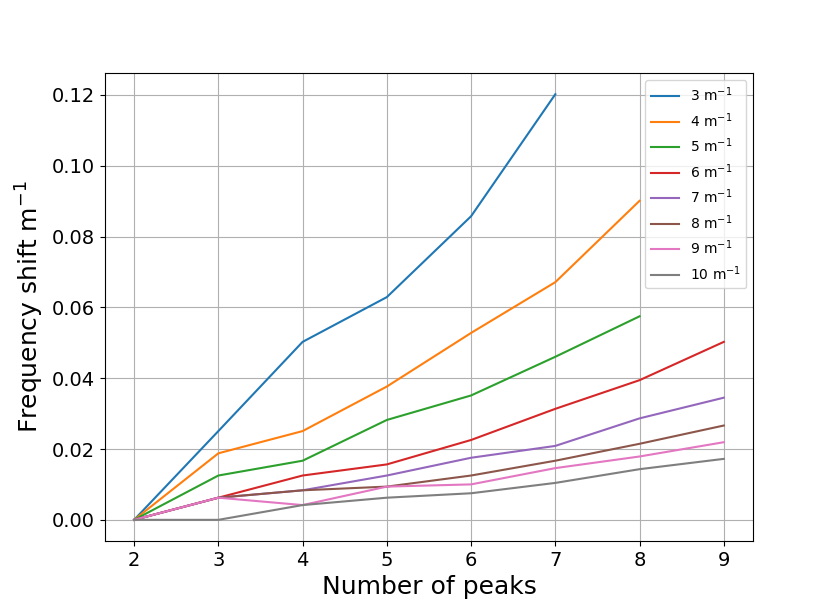}
        \caption{Frequency shift for boundary conditions Eq.~(\ref{eq:GWEMW_Train_BCs2}) with $a = 16$, $f_{\phi} = 10$ m$^{-1}$ and $f_{\Psi} = 3, 4, ..., 10$ m$^{-1}$.}
    \end{subfigure}
    \caption{The frequency shift of the electromagnetic wave $\phi_0$ as a function of the number of peaks used in its determination at $z = -1/2$ for (a) colliding gravitational and electromagnetic waves and (b) where the electromagnetic wave propagates together with a gravitational wave.}\label{fig:freq_shift}
\end{figure}

We first analyse the time dependency of the frequency shift of the electromagnetic wave following its collision with a gravitational wave, and then the same but where the electromagnetic wave has a gravitational counterpart. Using a spatial resolution of 6401 grid points and the boundary conditions given by Eq.~(\ref{eq:GWEMW_Train_BCs}) and Eq.~(\ref{eq:GWEMW_Train_BCs2}) respectively with $f_{\Psi} = 2, 4, \ldots, 10$ m$^{-1}$ and $f_{\phi} = 10$ m$^{-1}$, we calculate the induced frequency shift of the electromagnetic wave $\phi_0$ at $z = -1/2$ by using a successively larger number of periods. The resulting frequency shift is shown in Fig.~\ref{fig:freq_shift}. An increasingly negative frequency shift is observed when more peaks are analysed for the collision between a gravitational wave with an electromagnetic waves.  Conversely, an increasingly positive frequency shift is observed when more peaks are analysed for the collision between a gravitational wave with an electromagnetic and gravitational wave. This is what one would expect delving further into the scattering region given the future curvature singularity.

This brings up an important point: How much of the effects we are seeing are due to the future curvature singularity pathology exhibited through our plane symmetry assumption, and how much are through the scattering process? To shed some light on this, we note that as the gravitational wave frequency decreases towards more physical values, less peaks need to be considered to resolve a shift in the frequency of the electromagnetic wave.  That is, we do not have to penetrate quite so far into the scattering region and thus remain further away from the curvature blow-up instigated by the plane-symmetric regime. This suggests that in the limit of physical gravitational wave frequencies, these frequency shifts can be attributed to the wave collision itself rather than the curvature effects. It is true that as we decrease the gravitational wave's frequency, for a fixed amplitude we are increasing its area and therefore undoubtedly creating larger curvature changes which may lead to the curvature singularity appearing in a shorter proper time. This is something that is very difficult to rule out in our current paradigm, and we mention this in the discussion in Sec.~\ref{sec:summary}. However, below we note that the time delays and frequency shifts are comparable to the order-of-magnitude expectations of \cite{detweiler_pulsar_1979}.


\begin{figure}[H]
    \centering
    \begin{subfigure}[t]{0.45\textwidth}
        \centering
        \includegraphics[height=5.5cm]{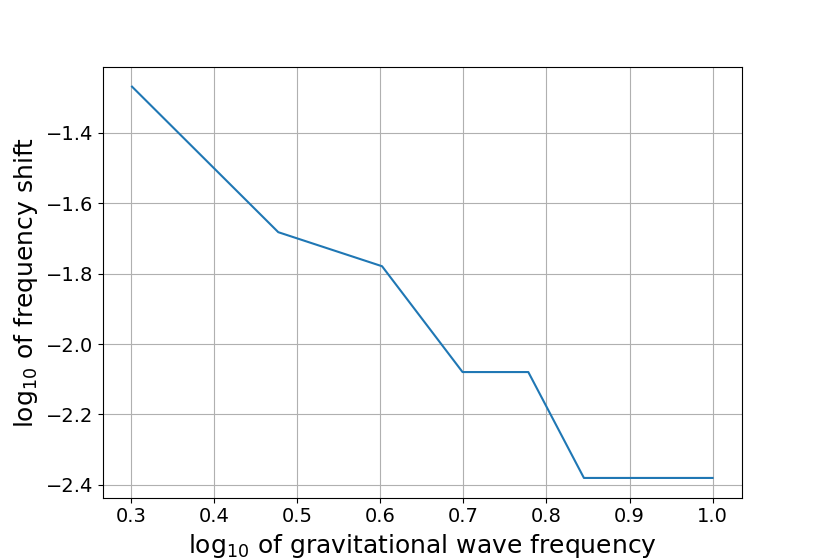}
        \caption{$\log_{10}$ of the absolute magnitude of the frequency shift for boundary conditions Eq.~(\ref{eq:GWEMW_Train_BCs}) with $a = 16$, $f_{\phi} = 10$ m$^{-1}$ and $f_{\Psi} = 2, 3,..., 10$ m$^{-1}$, depicted on a log-log plot base 10.}
    \end{subfigure}%
    ~ 
    \begin{subfigure}[t]{0.45\textwidth}
        \centering
        \includegraphics[height=5.5cm]{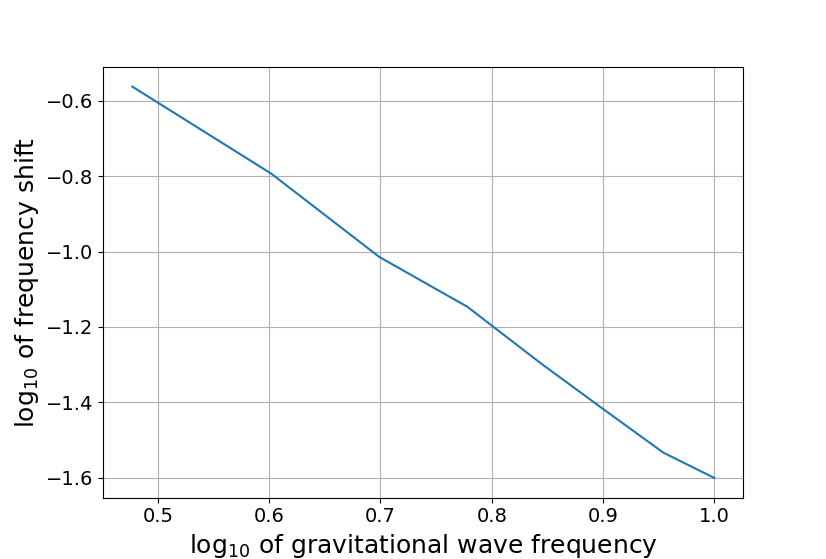}
        \caption{$\log_{10}$ of the absolute magnitude of the frequency shift for boundary conditions Eq.~(\ref{eq:GWEMW_Train_BCs2}) with $a = 16$, $f_{\phi} = 10$ m$^{-1}$ and $f_{\Psi} = 3, 4, ..., 10$ m$^{-1}$, depicted on a log-log plot base 10.}
    \end{subfigure}
    \caption{The magnitude of the frequency shift of the electromagnetic wave $\phi_0$ for colliding gravitational and electromagnetic waves at $z = -1/2$, $1.0 < t < 1.4$ with increasing gravitational wave frequency.}\label{fig:freq_norm}
\end{figure}

However, Fig.~\ref{fig:freq_shift} does not take into account the time of arrival of each peak of $\phi_0$ at $z = -1/2$, and `as a result the individual trends for each electromagnetic wave frequency cannot be compared directly.  To remedy this, we consider the region of spacetime in the scattering region along $z = -1/2$ and for $1.0 < t < 1.4$, for the same boundary conditions and wave frequencies used above.  This places us far enough into the scattering region that the frequency shift of the electromagnetic wave $\phi_0$ is non-negligible, without being too close to any curvature singularities which may alter the behaviour of $\phi_0$.  The resulting frequency shift as calculated by applying a fast Fourier transform to $\phi_0$ between the first and last peaks within this time period is depicted in Fig.~\ref{fig:freq_norm}.  Importantly, it is seen that as the frequency of the gravitational wave decreases, and thus the area of this gravitational wave increases, the magnitude of the frequency shift of the electromagnetic wave in the scattering region increases. This indicates that the large frequency difference between realistic gravitational and electromagnetic waves would result in a even bigger shift in frequency than what we have presented here.  Moreover, the use of a log-log plot indicates that the frequency shift again behaves like a power law. For boundary conditions Eq.~(\ref{eq:GWEMW_Train_BCs}) the frequency shift has a best fit of $f - f_0 \approx -0.20f_{\Psi}^{-1.90}$, while for boundary conditions Eq.~(\ref{eq:GWEMW_Train_BCs2}) it has a best fit of $f - f_0 \approx 2.45f_{\Psi}^{-1.99}$, where $f_0=10$ m$^{-1}$ is the initial electromagnetic wave frequency before scattering.

\begin{figure}[H]
    \centering
    \begin{subfigure}[t]{0.45\textwidth}
        \centering
        \includegraphics[height=5.5cm]{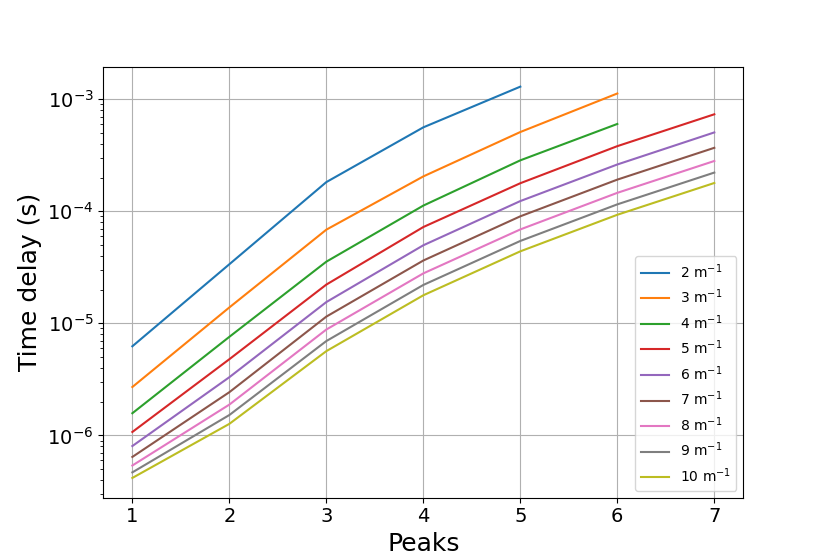}
        \caption{Time delay for boundary conditions Eq.~(\ref{eq:GWEMW_Train_BCs}) with $a = 16$, $f_{\phi} = 10$ m$^{-1}$ and $f_{\Psi} = 2, 3,..., 10$ m$^{-1}$}
    \end{subfigure}%
    ~ 
    \begin{subfigure}[t]{0.45\textwidth}
        \centering
        \includegraphics[height=5.5cm]{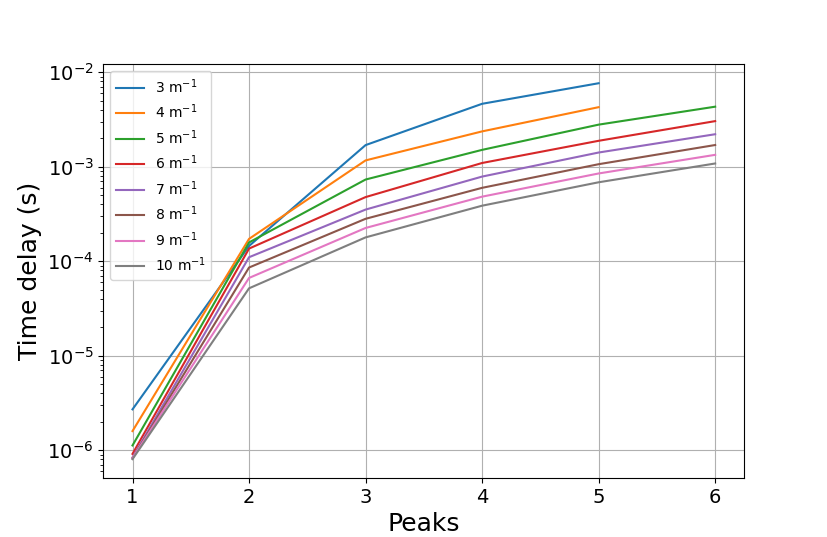}
        \caption{Time delay for boundary conditions Eq.~(\ref{eq:GWEMW_Train_BCs2}) with $a = 16$, $f_{\phi} = 10$ m$^{-1}$ and $f_{\Psi} = 3, 4, ..., 10$ m$^{-1}$}
    \end{subfigure}
    \caption{The magnitude of the time delay of the electromagnetic wave $\phi_0$ for colliding gravitational and electromagnetic waves at successive electromagnetic wave peaks as a function of the gravitational wave's frequency.}\label{fig:time_peaks}
\end{figure}

\begin{figure}[H]
    \centering
    \begin{subfigure}[t]{0.5\textwidth}
        \centering
        \includegraphics[height=5.5cm]{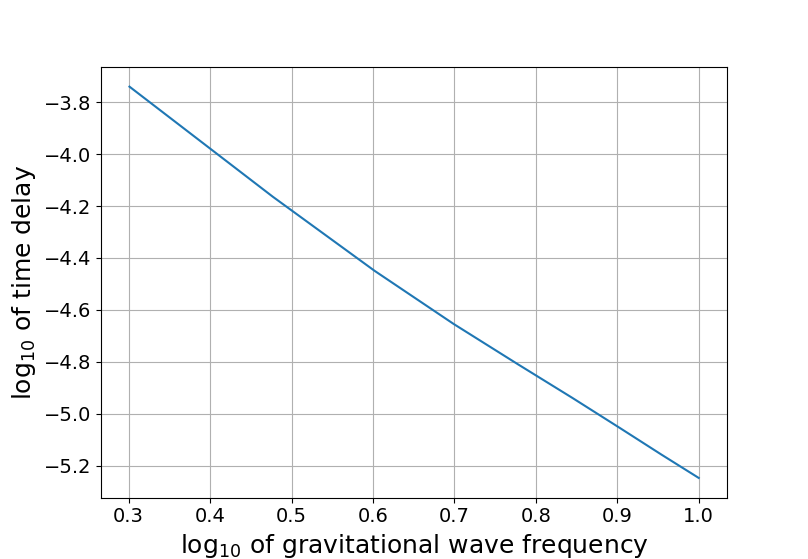}
        \caption{$\log_{10}$ of the time delay for boundary conditions Eq.~(\ref{eq:GWEMW_Train_BCs}) with $a = 16$, $f_{\phi} = 10$ m$^{-1}$ and $f_{\Psi} = 2, 3, ..., 10$ m$^{-1}$, depicted on a log-log plot base 10.}
    \end{subfigure}%
    ~ 
    \begin{subfigure}[t]{0.4\textwidth}
        \centering
        \includegraphics[height=5.5cm]{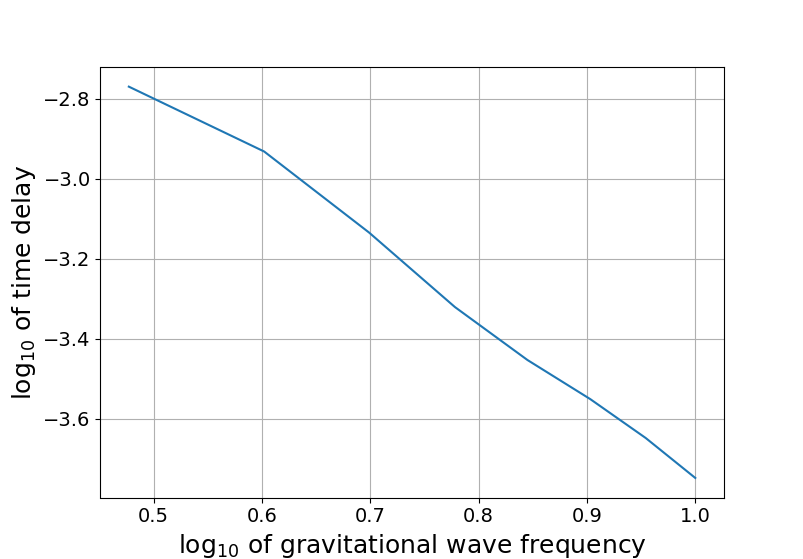}
        \caption{$\log_{10}$ of the time delay for boundary conditions Eq.~(\ref{eq:GWEMW_Train_BCs2}) with $a = 16$, $f_{\phi} = 10$ m$^{-1}$ and $f_{\Psi} = 3, 4, ..., 10$ m$^{-1}$, depicted on a log-log plot base 10.}
    \end{subfigure}
    \caption{The magnitude of the time delay of the third electromagnetic wave peak for colliding gravitational and electromagnetic waves as a function of the gravitational wave's frequency. }\label{fig:time_peaks_third_peak}
\end{figure}

The magnitude of the time delay of the incoming electromagnetic wave for successive wave peaks is depicted in Fig.~\ref{fig:time_peaks} for a range of gravitational wave frequencies.  As before, the time delay increases in magnitude both as the frequency of the gravitational wave increases, and as we penetrate further into the scattering region. The time delay of the third peak, occurring before the curvature blows up too significantly, is plotted against the gravitational wave frequency in Fig.~\ref{fig:time_peaks_third_peak}.  Again, the time delay scales like a power law with a best fit of $\tau \approx 0.001f_{\Psi}^{-2.32}$ for boundary conditions Eq.~(\ref{eq:GWEMW_Train_BCs}), and a best fit of $\tau \approx -0.012f_{\Psi}^{-1.77}$ for boundary conditions Eq.~(\ref{eq:GWEMW_Train_BCs2}). 


In a seminal paper by \cite{detweiler_pulsar_1979} which details how pulsar timing measurements can be used to search for gravitational waves, the TOA delay is estimated at an order-of-magnitude level as $\sim A/f_\Psi$ (their equation 29), with $A$ corresponding to the dimensionless strain of the gravitational wave. Via this time delay between adjacent peaks of the electromagnetic wave, the frequency of the electromagnetic wave would become $f\sim f_0(1 + A f_0/f_\Psi)^{-1}$. This corresponds to a frequency shift of $f-f_0 \sim - f_0 \tilde{A}/(1+\tilde{A})$ where $\tilde{A} = A f_0/f_\Psi$. For $\tilde{A}\ll1$, the frequency shift reduces to $f-f_0 \sim - f_0^2 A/f_\Psi$, and for $\tilde{A}\gg 1$, $f\sim f_\Psi/A$. Here, in the non-linear regime, we note that there is a steeper dependence on the gravitational wave frequency than expected from the simple argument in \cite{detweiler_pulsar_1979}, being closer to an inverse square. Nonetheless, the order of magnitude of the frequency shift and time delay are numerically similar to the simple estimate. The exception is the case where the electromagnetic wave is accompanied by a gravitational wave, which results in a frequency shift in the opposite direction and hence is not the result of anything resembling this simple time-delay argument. 

Further, it is noted that in the above we fixed the electromagnetic wave's amplitude to be $\mathcal{O}(1)$ so that the interactions are well resolved numerically. A small calculation shows that the absolute value of the $z$-component of the Poynting vector $S^z$ is $\frac{c^5}{4\pi G}|\phi_0|^2\;\mathrm{W}\ \mathrm{m}^{-2}$ when $\phi_2=0$. This shows that if one takes the electromagnetic wave amplitude to be $\mathcal{O}(10^n)$ then is $|S^z| = \mathcal{O}(10^{2n+51})$ $\mathrm{W}\ \mathrm{m}^{-2}$, a very large quantity. One then is interested to see how the frequency shift scales as the electromagnetic wave amplitude decreases from the value taken in this section. Limited by the plane pathologies with increasing $|\phi_0|$ and numerical resolvability with decreasing $|\phi_0|$, the frequency shift is tabulated in Tab.~\ref{tab:Poynting} for a small window of varying $|\phi_0|$ that is not prohibited by these issues. The boundary conditions are the same as Eq.~\eqref{eq:GWEMW_Train_BCs} with $a=64$ but varying $|\phi_0|$. The power law fit for the frequency shift is $0.00183|\phi_0|^{2.424}\ \mathrm{m}^{-1}$.

\begin{table}[!h]
    \begin{center}
        \begin{tabular}{|c||c|c|c|c|} 
             \hline
             $|\phi_0|$ & 1 & 1.5 & 2 & 2.5 \\ [0.5ex]
             \hline\hline
             Freq. Shift in $m^{-1}$ & 0.002083 & 0.004165 & 0.01041 & 0.016634 \\ 
             \hline
        \end{tabular}
    \end{center}
    \caption{The frequency shift of the electromagnetic wave $\phi_0$, with initial frequency $10$ m$^{-1}$ along the timelike geodesic $z=-1/2$ when scattered by a gravitational wave $\Psi_4$ with an initial frequency of $10$ m$^{-1}$ for a variety of electromagnetic wave amplitudes $|\phi_0|$.}
    \label{tab:Poynting}
\end{table}

Extrapolating to lower electromagnetic wave amplitudes gives frequency shifts of $10^{-13}\ \mathrm{m}^{-1}$ for $|S^z|=10^{43}\ \mathrm{W}\ \mathrm{m}^{-2}$ and $10^{-49}\ \mathrm{m}^{-1}$ for $|S^z|=10^{14}\ \mathrm{W}\ \mathrm{m}^{-2}$. These energy fluxes are, respectively, roughly associated with the collapse of a 10 solar mass black hole at the center a core collapse supernova, and the Eddington luminosity of a quasar. The former is close to the upper limit of energy fluxes allowed in the modern Universe, and the latter is realistic for the considerations of observability we discuss throughout this work. However, we expect that the strong scaling with $|\phi_0|$ should break for $|\phi_0| \ll 1$, since the frequency shifts and time delays in the low-amplitude regime have no dependence on $|\phi_0|$ (see \cite{detweiler_pulsar_1979} and the discussion above).

\section{Discussion and summary}\label{sec:summary}

In this work we utilized the Friedrich-Nagy gauge for the Einstein equations, a well posed IBVP formulation, fully coupled to Maxwell's equations as a means to study local, non-linear scattering properties of gravitational and electromagnetic plane waves. Given an initial space-time, which was always taken to be the Minkowski space-time, and boundary conditions for the ingoing waves, this framework was used to numerically approximate the IBVP. Maximally dissipative boundary conditions were implemented which, neglecting reflections of the outgoing modes, leave the boundary conditions completely arbitrary. Although Friedrich and Nagy prove wellposedness in vacuum (but without symmetry assumptions), it is immediately seen here in plane symmetry that when coupled to the electromagnetic field the system remains symmetric hyperbolic and the constraints propagate. Constraint propagation was also confirmed numerically and in Sec.~\ref{sec:emulationofexactsolns} numerical checks of correctness were made against exact solutions describing both gravitational and electromagnetic waves, with convergence found to these at the correct order.

As non-linear scattering effects of gravitational and electromagnetic waves are largely unexplored outside of the existing exact solutions, and interesting in their own right, our first aim was to uncover scattering properties in situations without a corresponding exact solution. The results of this investigation are presented in Sec.~\ref{sec:nonlinearbehaviour}, where we experimented with a large variety of boundary conditions for both the gravitational and electromagnetic waves. A detailed analysis of how the initial polarization of the waves affected parameters after scattering like the Weyl scalar curvature invariants, wave profiles and subsequent polarizations was presented. Specific, interesting features were found and commented upon that do not appear in the current literature of exact solutions. These included situations where the Weyl scalar curvature invariants nearly cancelled completely after scattering, while in other situations grew to a similar order as the gravitational waves, propagated with them after scattering and emulated their wave profile. New situations outside the scope of existing exact solutions that were explored include collisions of smoothed impulsive electromagnetic waves, pairs of smoothed impulsive gravitational waves and `gravitational spotlights', gravitational waves accompanied by an electromagnetic counterpart.

The scattering framework was then used to investigate whether non-linear scattering, such as what would occur near compact binary mergers, could affect electromagnetic wave observations. Although working in plane symmetry, which generically results in pathologies such as future curvature singularities, this should still give a good indication as to the non-linear effects at play. As such, we looked at two major observable changes: a time delay and a frequency shift. It was noted that the specific gauge used throughout this paper is the Gau\ss\;gauge, which is ideal for measuring the strain between two timelike free-falling observers. This was used as a measure of the gravitational wave strength required to induce the computed time delays and frequency shifts of the electromagnetic wave.

An important finding was that the direction of frequency shift changed depending on whether the electromagnetic wave had a gravitational counterpart or not. This corresponds to the proper length of the spatial computational domain increasing or decreasing. When the electromagnetic wave did not have a gravitational counterpart, the frequency shifted in the positive direction, when it did, it shifted negatively. This was also reflected in the time of arrival of the electromagnetic wave to propagate from one boundary to the other. Without the presence of a gravitational counterpart, the time was delayed, but with, it was advanced. 

It was found that the frequency shifted by up to $\mathcal{O}(10^{-1})$ m$^{-1}$, corresponding to $\mathcal{O}(10^7)$ Hz in units of $s^{-1}$ for radiation with an initial frequency of $\mathcal{O}(10^9)$ Hz. The corresponding strain to induce this shift was of the order unity. The time delay in relativistic units was found up to $\mathcal{O}(10^{-2})$ m, which equates to a change of the order $\mathcal{O}(10^{-11})$ seconds. Taken at face value, frequency shifts of this order are observable in principle with current instrumentation, while time delays of this order are not. However, for realistic gravitational wave frequencies, both the time delay and frequency shift are expected to be substantially larger.

To explore this further we investigated the change in frequency shift when we collided an electromagnetic wave of $10$ m$^{-1}$ with a gravitational wave of frequencies as low as $2$ m$^{-1}$. This was done when the electromagnetic wave was and was not accompanied by a gravitational wave. It was found that, whether the electromagnetic wave has a gravitational counterpart or not, the lower the frequency of the gravitational wave, the larger the absolute frequency shift. In the weak-field limit these quantities are expected to scale with the frequency of the gravitational wave $f_\Psi$ as $f_\Psi^{-1}$ \cite{detweiler_pulsar_1979}, whereas in the nonlinear regime over a limited range of frequencies we found scalings closer to, but not exactly, $f_\Psi^{-2}$. This gives tentative evidence that with a larger frequency discrepancy between gravitational and electromagnetic waves than we could resolve here, the time delays and frequency shifts will be larger by at least $f_\Psi^{-1}$.

The high-frequency end of the gravitational wave spectrum, as probed by LIGO, involves gravitational wave frequencies of $\mathcal{O}(10^2)$ Hz corresponding to mergers of stellar mass black holes. Meanwhile the lowest-frequency electromagnetic waves observed have frequencies of $\mathcal{O}(10^7)$ Hz, meaning that realistically the ratio of the frequency of the electromagnetic wave to that of the gravitational wave will be at minimum $\mathcal{O}(10^5)$. Note that the important quantity is not this ratio per se, but rather the product of this ratio with the dimensionless strain in the case of the frequency shift, and $A/f_\Psi$ in the case of the time delay. The large values of these quantities may be offset by the large dependence we have found on the amplitude of the electromagnetic wave. More realistic lower-amplitude waves will have reduced frequency shifts and time delays, but it is unclear by how much because we do not know the electromagnetic wave amplitude at which the dependence on the electromagnetic wave amplitude will be attenuated.

Given the potentially-large changes in frequency and time of arrival of electromagnetic waves as the result of an interaction with a gravitational wave, there may be a rich range of astrophysical phenomena where this effect is observable. A promising example is the coalescence of two black holes in the presence of accretion disks around one or both black holes, or the binary itself. Since the source of illumination is local, a non-negligible fraction of the radiation on the far side of the disks may interact with the gravitational waves propagating away from Earth. As the gravitational waves propagate outwards, $A f_0/f_\Psi$ decreases from $> 10^5$ towards 0, which would produce an electromagnetic transient that begins at very low frequency and evolves continuously up towards $f_0$. The transient will remain at low frequencies until $A f_0/f_\Psi \sim 1$, corresponding to the time it takes for the gravitational waves to propagate $\sim f_0/f_\Psi$ gravitational radii, and provided the disk is still emitting radiation at that radius. Given that the mergers will occur at cosmological distances, it is likely that it will be easier to observe these effects in supermassive black hole mergers, the components of which are likely to possess accretion disks, than in stellar mass binaries which have no ready source of accreting material.

Prior to coalescence, a binary will produce a continuous stream of gravitational waves with more modest strains, which may yet have an observable effect on the electromagnetic radiation. Compact binaries in the Galaxy emitting in the LISA band are expected to be so numerous that they will be difficult to distinguish. Light from background stars that happen to be closely aligned on the sky with a foreground compact object binary may therefore experience detectable time delays and frequency shifts. A number of these binaries are already identified electromagnetically as the LISA verification binaries \cite{kupfer_2018}, so our results may be tested by observing stars behind the subset of these binaries that are close to coalescence.

Although throughout this paper we were careful to mention and take into account the pathologies induced by the plane symmetry assumption, it is desirable to move to a regime without any symmetry assumptions. This is the subject of future work. It would be of great interest to see how the results stemming from this toy model port to the more realistic scenario without plane symmetry. Although our results were for a large electromagnetic wave amplitude and associated Poynting vector, higher resolution runs in the case without symmetry assumptions would give more concrete predictions about more realistic frequency shifts and time delays. Further, a general numerical implementation of the Friedrich-Nagy IBVP could potentially be useful to explore other areas of research, such as Ellis' finite infinity, whereby boundary conditions must be incorporated and easily altered. 

\ack
We acknowledge the use of the gravity computer cluster at the University of Canterbury during this project.

\appendix

\section{Subsidiary system}\label{app:CP}
We prove that the constraints propagate analytically by showing that the time derivative of each constraint is a linear combination of the constraints themselves
\begin{subequations}
	\begin{align}
(1/A)\del _tC_1 & = \left(F + \bar{F}+\mu + \bar{\mu } +6 \rho \right)C_1 + \bar{\sigma }C_3 + \sigma \bar{C_3} - 2  \bar{\varphi}_1C_5 - 2 \varphi_1\bar{C_5},\\
(1/A)\del _tC_2 & = \left(-F - \bar{F}+\mu + \bar{\mu } +6 \rho' \right)C_2 + \bar{\sigma }'C_4 + \sigma' \bar{C_4} - 2  \bar{\varphi}_1C_5 - 2 \varphi_1\bar{C_5},\\
(1/A)\del _tC_3 & = \left(4 \sigma + \bar{\sigma}'\right)C_1 - \sigma C_2 + \left(3 F -\bar{F}+ \mu + \bar{\mu }+4 \rho -\rho'\right)C_3 + \rho \bar{C_4},\\
(1/A)\del _tC_4 & = - \sigma' C_1 + \left(4 \sigma' + \bar{\sigma}\right)C_2 + \rho' \bar{C_3} + \left(-3 F +\bar{F}+ \mu + \bar{\mu }- \rho +4\rho'\right)C_4,\\(1/A)\del _tC_5 & = 2\varphi_1 \left(C_1+C_2\right) +  \left(\bar{\mu }+\mu +2 \rho' +2 \rho \right)C_5.
\end{align}
\end{subequations}
Further, as these are all ordinary differential equations, the subsidiary system contains no spatially-propagating modes and we are left with a wellposed IBVP in the same sense as Friedrich and Nagy's vacuum case \cite{friedrich1999initial}.

\section*{References}

\bibliographystyle{elsarticle-num} 
\bibliography{refs}

\begin{thebibliography}{10}
\expandafter\ifx\csname url\endcsname\relax
  \def\url#1{\texttt{#1}}\fi
\expandafter\ifx\csname urlprefix\endcsname\relax\def\urlprefix{URL }\fi
\expandafter\ifx\csname href\endcsname\relax
  \def\href#1#2{#2} \def\path#1{#1}\fi

\bibitem{friedrich1999initial}
H.~Friedrich, G.~Nagy, The initial boundary value problem for {E}instein's vacuum field equation, Communications in Mathematical Physics 201~(3) (1999) 619--655.

\bibitem{frauendiener2014numerical}
J.~Frauendiener, C.~Stevens, B.~Whale, Numerical evolution of plane gravitational waves in the {F}riedrich--{N}agy gauge, Physical Review D 89~(10) (2014) 104026.

\bibitem{frauendiener2021can}
J.~Frauendiener, J.~Hakata, C.~Stevens, Can gravitational waves halt the expansion of the universe?, Universe 7~(7) (2021) 228.

\bibitem{khan1971scattering}
K.~Khan, R.~Penrose, Scattering of two impulsive gravitational plane waves, Nature 229~(5281) (1971) 185--186.

\bibitem{nutku1977colliding}
Y.~Nutku, M.~Halil, Colliding impulsive gravitational waves, Physical Review Letters 39~(22) (1977) 1379.

\bibitem{griffiths1975colliding}
J.~Griffiths, Colliding gravitational and electromagnetic waves, Physics Letters A 54~(4) (1975) 269--270.

\bibitem{bell1974interacting}
P.~Bell, P.~Szekeres, Interacting electromagnetic shock waves in general relativity, General Relativity and Gravitation 5 (1974) 275--286.

\bibitem{penrose1984spinors}
R.~Penrose, W.~Rindler, Spinors and space--time: {V}olume 1, {T}wo-spinor calculus and relativistic fields, Vol.~1, Cambridge University Press, 1984.

\bibitem{griffiths2016colliding}
J.~B. Griffiths, Colliding plane waves in general relativity, Courier Dover Publications, 2016.

\bibitem{chandrasekhar1984nutku}
S.~Chandrasekhar, V.~Ferrari, On the {N}utku--{H}alil solution for colliding impulsive gravitational waves, Proceedings of the Royal Society of London. A. Mathematical and Physical Sciences 396~(1810) (1984) 55--74.

\bibitem{doulis2019coffee}
G.~Doulis, J.~Frauendiener, C.~Stevens, B.~Whale, {COFFEE}--{A}n {MPI}--parallelized {P}ython package for the numerical evolution of differential equations, SoftwareX 10 (2019) 100283.

\bibitem{strand1994summation}
B.~Strand, Summation by parts for finite difference approximations for d/dx, Journal of Computational Physics 110~(1) (1994) 47--67.

\bibitem{carpenter1994time}
M.~H. Carpenter, D.~Gottlieb, S.~Abarbanel, Time--stable boundary conditions for finite-difference schemes solving hyperbolic systems: methodology and application to high-order compact schemes, Journal of Computational Physics 111~(2) (1994) 220--236.

\bibitem{Griffiths1975}
J.~B. Griffiths, \href{https://www.sciencedirect.com/science/article/pii/0375960175902546}{Colliding gravitational and electromagnetic waves}, Physics Letters A 54~(4) (1975) 269--270.
\newblock \href {https://doi.org/https://doi.org/10.1016/0375-9601(75)90254-6} {\path{doi:https://doi.org/10.1016/0375-9601(75)90254-6}}.
\newline\urlprefix\url{https://www.sciencedirect.com/science/article/pii/0375960175902546}

\bibitem{abbott2023population}
R.~Abbott, T.~D. Abbott, F.~Acernese, K.~Ackley, C.~Adams, N.~Adhikari, R.~X. Adhikari, et~al., \href{https://link.aps.org/doi/10.1103/PhysRevX.13.011048}{Population of merging compact binaries inferred using gravitational waves through gwtc-3}, Phys. Rev. X 13 (2023) 011048.
\newblock \href {https://doi.org/10.1103/PhysRevX.13.011048} {\path{doi:10.1103/PhysRevX.13.011048}}.
\newline\urlprefix\url{https://link.aps.org/doi/10.1103/PhysRevX.13.011048}

\bibitem{callister2024}
T.~A. {Callister}, W.~M. {Farr}, {Parameter-Free Tour of the Binary Black Hole Population}, Physical Review X 14~(2) (2024) 021005.
\newblock \href {http://arxiv.org/abs/2302.07289} {\path{arXiv:2302.07289}}, \href {https://doi.org/10.1103/PhysRevX.14.021005} {\path{doi:10.1103/PhysRevX.14.021005}}.

\bibitem{Background}
G.~Agazie, A.~Anumarlapudi, A.~M. Archibald, Z.~Arzoumanian, P.~T. Baker, B.~Bécsy, L.~Blecha, A.~Brazier, P.~R. Brook, S.~Burke-Spolaor, J.~A. Casey-Clyde, M.~Charisi, S.~Chatterjee, T.~Cohen, J.~M. Cordes, N.~J. Cornish, F.~Crawford, H.~T. Cromartie, K.~Crowter, M.~E. DeCesar, P.~B. Demorest, T.~Dolch, B.~Drachler, E.~C. Ferrara, W.~Fiore, E.~Fonseca, G.~E. Freedman, E.~Gardiner, N.~Garver-Daniels, P.~A. Gentile, J.~Glaser, D.~C. Good, K.~Gültekin, J.~S. Hazboun, R.~J. Jennings, A.~D. Johnson, M.~L. Jones, A.~R. Kaiser, D.~L. Kaplan, L.~Z. Kelley, M.~Kerr, J.~S. Key, N.~Laal, M.~T. Lam, W.~G. Lamb, T.~J.~W. Lazio, N.~Lewandowska, T.~Liu, D.~R. Lorimer, J.~Luo, R.~S. Lynch, C.-P. Ma, D.~R. Madison, A.~McEwen, J.~W. McKee, M.~A. McLaughlin, N.~McMann, B.~W. Meyers, C.~M.~F. Mingarelli, A.~Mitridate, C.~Ng, D.~J. Nice, S.~K. Ocker, K.~D. Olum, T.~T. Pennucci, B.~B.~P. Perera, N.~S. Pol, H.~A. Radovan, S.~M. Ransom, P.~S. Ray, J.~D. Romano, S.~C. Sardesai, A.~Schmiedekamp, C.~Schmiedekamp, K.~Schmitz, L.~Schult, B.~J. Shapiro-Albert, X.~Siemens, J.~Simon, M.~S. Siwek, I.~H. Stairs, D.~R. Stinebring, K.~Stovall, A.~Susobhanan, J.~K. Swiggum, S.~R. Taylor, J.~E. Turner, C.~Unal, M.~Vallisneri, S.~J. Vigeland, H.~M. Wahl, C.~A. Witt, O.~Young, \href{https://dx.doi.org/10.3847/2041-8213/acf4fd}{The {NANOGrav} 15 yr data set: Search for anisotropy in the gravitational-wave background}, The Astrophysical Journal Letters 956~(1) (2023) L3.
\newblock \href {https://doi.org/10.3847/2041-8213/acf4fd} {\path{doi:10.3847/2041-8213/acf4fd}}.
\newline\urlprefix\url{https://dx.doi.org/10.3847/2041-8213/acf4fd}

\bibitem{Pulsars}
G.~Hobbs, A.~Archibald, Z.~Arzoumanian, D.~Backer, M.~Bailes, N.~D.~R. Bhat, M.~Burgay, S.~Burke-Spolaor, D.~Champion, I.~Cognard, W.~Coles, J.~Cordes, P.~Demorest, G.~Desvignes, R.~D. Ferdman, L.~Finn, P.~Freire, M.~Gonzalez, J.~Hessels, A.~Hotan, G.~Janssen, F.~Jenet, A.~Jessner, C.~Jordan, V.~Kaspi, M.~Kramer, V.~Kondratiev, J.~Lazio, K.~Lazaridis, K.~J. Lee, Y.~Levin, A.~Lommen, D.~Lorimer, R.~Lynch, A.~Lyne, R.~Manchester, M.~McLaughlin, D.~Nice, S.~Oslowski, M.~Pilia, A.~Possenti, M.~Purver, S.~Ransom, J.~Reynolds, S.~Sanidas, J.~Sarkissian, A.~Sesana, R.~Shannon, X.~Siemens, I.~Stairs, B.~Stappers, D.~Stinebring, G.~Theureau, R.~van Haasteren, W.~van Straten, J.~P.~W. Verbiest, D.~R.~B. Yardley, X.~P. You, \href{https://dx.doi.org/10.1088/0264-9381/27/8/084013}{The international pulsar timing array project: using pulsars as a gravitational wave detector}, Classical and Quantum Gravity 27~(8) (2010) 084013.
\newblock \href {https://doi.org/10.1088/0264-9381/27/8/084013} {\path{doi:10.1088/0264-9381/27/8/084013}}.
\newline\urlprefix\url{https://dx.doi.org/10.1088/0264-9381/27/8/084013}

\bibitem{cortes_ALMA_2023}
P.~Cortes, C.~Vlahakis, A.~Hales, J.~Carpenter, B.~Dent, S.~Kameno, R.~Loomis, B.~Vila~Vilaro, A.~Biggs, A.~Miotello, R.~Rosen, F.~Stoehr, K.~Saini, {{ALMA Cycle}} 10 {{Technical Handbook}} (Apr. 2023).
\newblock \href {https://doi.org/10.5281/zenodo.7822943} {\path{doi:10.5281/zenodo.7822943}}.

\bibitem{vogt1994}
S.~S. {Vogt}, S.~L. {Allen}, B.~C. {Bigelow}, L.~{Bresee}, B.~{Brown}, T.~{Cantrall}, A.~{Conrad}, M.~{Couture}, C.~{Delaney}, H.~W. {Epps}, D.~{Hilyard}, D.~F. {Hilyard}, E.~{Horn}, N.~{Jern}, D.~{Kanto}, M.~J. {Keane}, R.~I. {Kibrick}, J.~W. {Lewis}, J.~{Osborne}, G.~H. {Pardeilhan}, T.~{Pfister}, T.~{Ricketts}, L.~B. {Robinson}, R.~J. {Stover}, D.~{Tucker}, J.~{Ward}, M.~Z. {Wei}, {HIRES: the high-resolution echelle spectrometer on the Keck 10-m Telescope}, in: D.~L. {Crawford}, E.~R. {Craine} (Eds.), Instrumentation in Astronomy VIII, Vol. 2198 of Society of Photo-Optical Instrumentation Engineers (SPIE) Conference Series, 1994, p. 362.
\newblock \href {https://doi.org/10.1117/12.176725} {\path{doi:10.1117/12.176725}}.

\bibitem{antoniadis2023second}
J.~Antoniadis, S.~Babak, A.-S.~B. Nielsen, C.~Bassa, A.~Berthereau, M.~Bonetti, E.~Bortolas, P.~Brook, M.~Burgay, R.~Caballero, et~al., The second data release from the european pulsar timing array-i. the dataset and timing analysis, Astronomy \& Astrophysics 678 (2023) A48.

\bibitem{antoniadis2023second2}
J.~Antoniadis, P.~Arumugam, S.~Arumugam, S.~Babak, M.~Bagchi, A.-S.~B. Nielsen, C.~Bassa, A.~Bathula, A.~Berthereau, M.~Bonetti, et~al., The second data release from the european pulsar timing array-ii. customised pulsar noise models for spatially correlated gravitational waves, Astronomy \& Astrophysics 678 (2023) A49.

\bibitem{antoniadis2023second3}
J.~Antoniadis, P.~Arumugam, S.~Arumugam, S.~Babak, M.~Bagchi, A.-S.~B. Nielsen, C.~Bassa, A.~Bathula, A.~Berthereau, M.~Bonetti, et~al., The second data release from the european pulsar timing array-iii. search for gravitational wave signals, Astronomy \& Astrophysics 678 (2023) A50.

\bibitem{abbott2016observation}
B.~P. Abbott, R.~Abbott, T.~Abbott, M.~Abernathy, F.~Acernese, K.~Ackley, C.~Adams, T.~Adams, P.~Addesso, R.~X. Adhikari, et~al., Observation of gravitational waves from a binary black hole merger, Physical review letters 116~(6) (2016) 061102.

\bibitem{lofar}
M.~P. van Haarlem, M.~W. Wise, A.~Gunst, G.~Heald, J.~P. McKean, J.~W. Hessels, A.~G. de~Bruyn, R.~Nijboer, J.~Swinbank, R.~Fallows, et~al., Lofar: The low-frequency array, Astronomy \& astrophysics 556 (2013) A2.

\bibitem{detweiler_pulsar_1979}
S.~Detweiler, Pulsar timing measurements and the search for gravitational waves, Astrophysical Journal, Part 1, vol. 234, Dec. 15, 1979, p. 1100-1104. 234 (1979) 1100--1104.

\bibitem{kupfer_2018}
T.~Kupfer, V.~Korol, S.~Shah, G.~Nelemans, T.~Marsh, G.~Ramsay, P.~Groot, D.~Steeghs, E.~Rossi, Lisa verification binaries with updated distances from gaia data release 2, Monthly Notices of the Royal Astronomical Society 480~(1) (2018) 302--309.

\end{thebibliography}
\end{document}